\documentclass[graybox]{svmult}

\usepackage{type1cm}        
\usepackage{makeidx}         
\usepackage{graphicx}        
\usepackage{multicol}        
\usepackage[bottom]{footmisc}

\usepackage{newtxtext}       %
\usepackage{newtxmath}       
\usepackage{hyperref}

\def\snn{\mbox{$\sqrt{s_{_{\rm NN}}}$}}

\newcommand{ \bv }{{\bf v}}

\newcommand{ \bS }{{\bf S}}

\newcommand{ \bomega }{{\boldsymbol \omega}}

\newcommand{\di}{{\rm d}}

\def\wT{{\widehat T}}
\def\wj{{\widehat j}}

\def\wrhol{{\widehat\rho_{\rm LE}}}

\newcommand{\tr}{{\rm tr}}

\newcommand{\omegav}{\boldsymbol{\omega}}

\newcommand{\p}{{\rm p}}

\newcommand{\be}{\begin{equation}}
\newcommand{\ee}{\end{equation}}                                                                               
\newcommand{\bea}{\begin{eqnarray}}
\newcommand{\eea}{\end{eqnarray}}

\makeindex             

\begin{document}

\title*{Vorticity and Polarization in Heavy Ion Collisions: Hydrodynamic Models}
\author{Iurii Karpenko}
\institute{Iurii Karpenko \at Czech Technical University in Prague, B\v rehov\'a 7, 11519 Prague 1, Czech Republic, \email{yu.karpenko@gmail.com}}
%
%
\maketitle
\abstract{ Fluid dynamic approach is a workhorse for modelling collective dynamics in relativistic heavy-ion collisions. The approach has been successful in describing various features of the momentum distributions of hadrons produced in the heavy-ion collisions, such as $p_T$ spectra, flow coefficients $v_n$ etc. As such, the description of the phenomenon of polarization of $\Lambda$ hyperons in heavy-ion collisions has to be incorporated into the hydrodynamic approach. We start this chapter by introducing different definitions of vorticity in relativistic fluid dynamics. Then we present a derivation of the polarization of spin 1/2 fermions in the relativistic fluid. The latter is directly applied to compute the spin polarization of the $\Lambda$ hyperons, which are produced from the hot and dense medium, described with fluid dynamics. It is followed by a review of the existing calculations of global or local polarization of $\Lambda$ hyperons in different hydrodynamic models of relativistic heavy-ion collisions. We particularly focus on the explanations of the collision energy dependence of the global $\Lambda$ polarization from the different hydrodynamic models, the polarization component in the beam direction as well as on the origins of the global and local $\Lambda$ polarization.}

\section{Introduction: vorticities in a fluid}
Heavy-ion collisions at ultra-relativistic energies create a strongly
interacting system characterized by extremely high temperature and
energy density. For a large fraction of its lifetime the system shows
strong collective effects and can be described by relativistic hydrodynamics. 
In particular, the large elliptic flow observed in such collisions, indicates 
that the created quasi-macroscopic system is strongly coupled, and has an extremely low viscosity to entropy
ratio. From the very success of the hydrodynamic description, 
one can also conclude that the system might possess an extremely high
vorticity, likely the highest ever made under the laboratory conditions.

\begin{figure}
\begin{center}
\includegraphics[keepaspectratio, width=0.8\columnwidth]{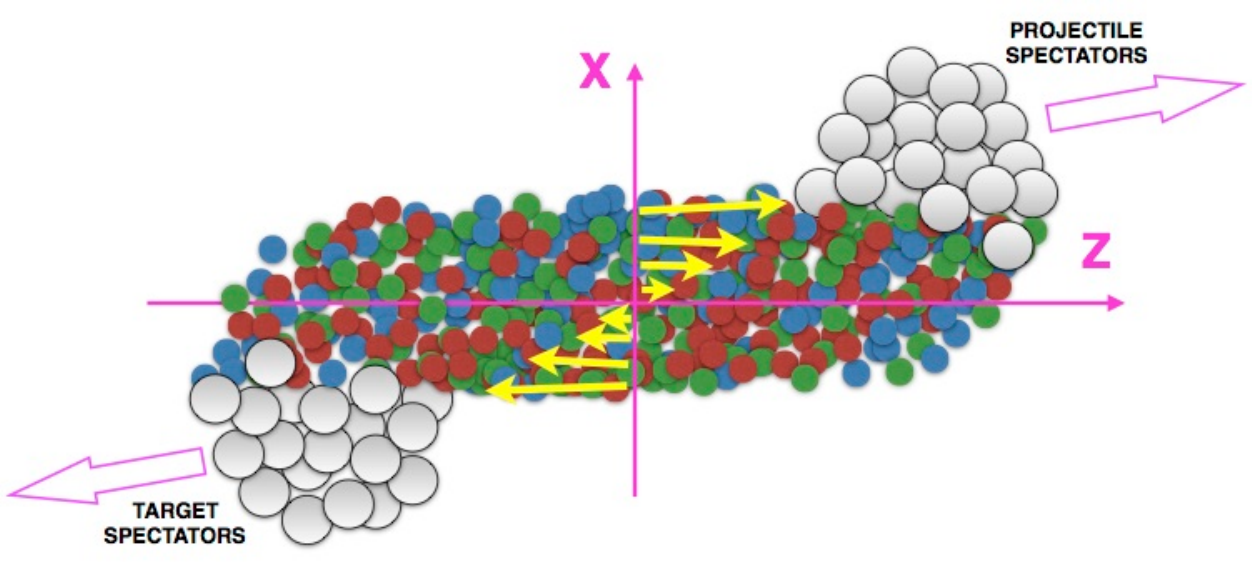}
\end{center}
  \caption{
    Schematic view of the collision. Arrows indicate the flow velocity
  field.
  The $+\hat{y}$ direction is out of the page; both
  the orbital angular momentum and the magnetic field point
  into the page.}
  \label{fig:collision}
\end{figure}

A simple estimate of the non-relativistic vorticity, defined as 
\be\label{nrv}
 \bomega =\frac{1}{2}\, \nabla \times \bv,
\ee
\footnote{sometimes the vorticity is defined without the factor
$1/2$; we use the definition that gives the vorticity of the fluid
rotating as a whole with a constant angular velocity $\Omega$, to be
$\omega=\Omega$} 
can be made based on a very schematic picture of the collision
depicted in Fig.~\ref{fig:collision}. As the projectile and target
spectators move in opposite direction with the velocity close to the
speed of light, the $z$ component of the collective velocity in the
system close to the projectile spectators and that close to the target
spectators are expected to be different. Assuming that this difference
is a fraction of the speed of light, e.g.  0.1 (in units of the speed
of light), and that the transverse size of the system is about 5~fm,
one concludes that the vorticity in the system is of the order
$0.02\,{\rm fm^{-1}}\approx 10^{22}\,{\rm s}^{-1}$.  

Unlike in classical hydrodynamics, where vorticity is the curl of the velocity field 
${\bf v}$, several vorticities can be defined in relativistic hydrodynamics which 
can be useful in different applications (for more on that, we refer the reader to \cite{Becattini:2015ska}). We list and discuss the different vorticity definitions as follows:

\subparagraph{The kinematical vorticity} The kinematical vorticity is defined as:
\be\label{kinvort}
  \omega_{\mu\nu} = \frac{1}{2} (d_\nu u_\mu - d_\mu u_\nu) = 
  \frac{1}{2} (\partial_\nu u_\mu - \partial_\mu u_\nu)
\ee
where $d_\mu$ is a covariant derivative (different from the ordinary derivative $\partial_\mu$) and $u$ is the four-velocity field. This tensor includes both the acceleration $A$ and the relativistic extension of the angular velocity pseudo-vector $\omega_\mu$ in the usual decomposition of an antisymmetric tensor field into a polar and 
pseudo-vector fields:
\bea\label{decomp}
   \omega_{\mu\nu} &=& \epsilon_{\mu\nu\rho\sigma} \omega^\rho u^\sigma
   + \frac{1}{2} (A_\mu u_\nu - A_\nu u_\mu)  \nonumber\\ 
   A_\mu &=& 2 \omega_{\mu\nu} u^\nu = u^\nu d_\nu u_\mu \equiv D u_\mu  \nonumber\\ 
   \omega_\mu &=& -\frac{1}{2} \epsilon_{\mu\rho\sigma\tau} \,\omega^{\rho\sigma} 
   u^\tau
\eea
where $\epsilon_{\mu\nu\rho\sigma}$ is the Levi-Civita symbol.
Using of the transverse (to $u$) projector:
$$
 \Delta^{\mu\nu}\equiv g^{\mu\nu} - u^{\mu}u^{\nu},
$$
and the usual definition of the orthogonal derivative
$$
  \nabla_\mu \equiv  \Delta^\alpha_\mu d_\alpha = d_\mu - u_\mu D,
$$
where $D=u^\alpha d_\alpha$ is a so-called co-moving derivative, it is convenient to define also a transverse kinematical 
vorticity as:
\be
  \omega^{\Delta}_{\mu\nu} = \Delta_{\mu\rho} \Delta_{\nu\sigma} 
  \omega^{\rho\sigma} = \frac{1}{2}(\nabla_\nu u_\mu - \nabla_\mu u_\nu)
\ee
Using the above definition in the decomposition (\ref{decomp}) it can be shown that:
\be
 \omega^{\Delta}_{\mu\nu} = \epsilon_{\mu\nu\rho\sigma} \omega^\rho u^\sigma
\ee
that is $\omega^\Delta$ is the tensor formed with the angular velocity vector only.
As we will show in the next subsection, only $\omega^\Delta$ shares the ``conservation''
property of the classical vorticity for an ideal barotropic fluid.

\subparagraph{The T-vorticity}

This is defined as:
\be\label{tvort}
  \Omega_{\mu\nu} = \frac{1}{2} \left[ \partial_\nu (T u_\mu) - \partial_\mu 
  (T u_\nu) \right]
\ee
and it is particularly useful for a relativistic uncharged fluid, such as the QCD
plasma formed in nuclear collisions at very high energy. This is because from the 
basic thermodynamic relations when the temperature is the only independent 
thermodynamic variable, the ideal relativistic equation of motion $(\varepsilon + p) 
A_\mu = \nabla_\mu p$ can be recast in the simple form:
\be\label{carter}
  u^\mu \Omega_{\mu\nu} = \frac{1}{2} (T A_\nu - \nabla_\nu T)  = 0 
\ee     

The above (\ref{carter}) is also known as Carter-Lichnerowicz equation \cite{gourg} for 
an ideal uncharged fluid and it entails conservation properties which do not hold for the 
kinematical vorticity. This can be better seen in the the language of differential forms, 
rewriting the definition of the T-vorticity as the exterior derivative of a the vector 
field (1-form) $Tu$, that is $\Omega = \mathbf{d} (Tu)$. Indeed, the eq.~(\ref{carter})
implies - through the Cartan identity - that the Lie derivative of $\Omega$ along 
the vector field $u$ vanishes, that is
\be\label{cartan}
  {\cal L}_u \,\Omega = u \cdot {\bf d} \Omega + {\bf d} (u \cdot \Omega) = 0
\ee
because $\Omega$ is itself the external derivative of the vector field $T u$ and
${\bf d}{\bf d} = 0$. The eq.~(\ref{cartan}) states that the T-vorticity is conserved
along the flow and, thus, if it vanishes at an initial time it will remain so at
all times. This can be made more apparent by expanding the Lie derivative definition
in components:
\be\label{lie}
 ({\cal L}_u \,\Omega)^{\mu\nu} = D \Omega^{\mu\nu} - \partial_\sigma u^\mu
 \Omega^{\sigma\nu} - \partial_\sigma u^\nu \Omega^{\sigma\mu} = 0
\ee 
The above equation is in fact a differential equation for $\Omega$ precisely showing 
that if $\Omega=0$ at the initial time then $\Omega\equiv 0$. Thereby, the T-vorticity
has the same property as the classical vorticity for an ideal barotropic fluid, such 
as the Kelvin circulation theorem, so the integral of $\Omega$ over a surface enclosed 
by a circuit comoving with the fluid will be a constant.

One can write the relation between the T-vorticity and the kinematical vorticity by expanding
the definition (\ref{tvort}):
$$
  \Omega_{\mu\nu} = \frac{1}{2}\left[ (\partial_\nu T) \, u_\mu - (\partial_\mu T) \,
  u_\nu \right] + T \omega_{\mu\nu}
$$
implying that the double-transverse projection of $\Omega$:
$$
  \Delta_{\mu\rho}\Delta_{\nu\sigma} \Omega^{\rho\sigma} \equiv \Omega^\Delta_{\mu\nu}
  = T \omega_{\mu\nu}^\Delta 
$$
Hence, the tensor $\omega^\Delta$ shares the same conservation properties of $\Omega^\Delta$,
namely it vanishes at all times if it is vanishing at the initial time. Conversely, 
the mixed projection of the kinematical vorticity:
$$
  u^\rho \omega_{\rho\sigma} \Delta^{\sigma\nu} = \frac{1}{2} A_{\sigma}
$$
does not. It then follows that for an ideal uncharged fluid with $\omega^\Delta=0$
at the initial time, the kinematical vorticity is simply:
\be\label{omeglong}
  \omega_{\mu\nu} = \frac{1}{2} ( A_\mu u_\nu - A_\nu u_\mu)
\ee
%

\subparagraph{The thermal vorticity}

The thermal vorticity is defined as \cite{Becattini:2013vja}:
\be\label{thvort}
  \varpi_{\mu\nu} = \frac{1}{2} (\partial_\nu \beta_\mu - \partial_\mu \beta_\nu)   
\ee
where $\beta$ is the temperature four-vector. This vector is defined as $(1/T) u$
once a four-velocity $u$, that is a hydrodynamical frame, is introduced, but it
can also be taken as a primordial quantity to define a velocity through $u \equiv
\beta/\sqrt{\beta^2}$ \cite{becaframe}. The thermal vorticity features two 
important properties: it is adimensional in natural units (in cartesian coordinates)
and it is the actual constant vorticity at the global equilibrium with rotation 
\cite{Becattini:2012tc} for a relativistic system, where $\beta$ is a Killing vector field 
whose expression in Minkowski spacetime is $\beta_\mu = b_\mu + \varpi_{\mu\nu} x^\nu$ 
being $b$ and $\varpi$ constant. In this case the magnitude of thermal vorticity is 
- with the natural constants restored - simply $\hbar \omega/k_BT$ where $\omega$ is 
a constant angular velocity. In general, (replacing $\omega$ with the classical 
vorticity defined as the curl of a proper velocity field) it can be readily realized 
that the adimensional thermal vorticity is a tiny number for most hydrodynamical 
systems, though it can be significant for the plasma formed in relativistic nuclear 
collisions.

\section{Polarization of particles in the fluid}
Particles produced in relativistic heavy ion collisions are expected to be polarized 
in peripheral collisions because of angular momentum conservation. At finite impact 
parameter, the QGP has a finite angular momentum perpendicular to the reaction plane 
and some fraction thereof may be converted into spin of final state hadrons. Therefore, 
measured particles may show a finite mean {\em global} polarization along the angular
momentum direction. In a fluid at local thermodynamic equilibrium, the polarization
can be calculated by using the principle of quantum statistical mechanics, that is
assuming that the spin degrees of freedom are at local thermodynamical equilibrium 
at the hadronization stage, much the same way as the momentum degrees of freedom.

The crucial role in the calculation of the polarization for the fluid produced in 
relativistic heavy ion collisions is played by the density operator. For a system 
at Local Thermodynamic Equilibrium (LTE), this reads \cite{becaframe}:
\be\label{gencov}
  \wrhol = (1/Z) \exp \left[- \int_\Sigma \di\Sigma_\mu  \left( \wT^{\mu\nu} \beta_\nu 
- \zeta \wj^\mu \right) \right]
\ee
where $\beta = (1/T)u$ is the four-temperature vector, $\wT$ the stress-energy tensor, $\wj$
a conserved current - like the baryon number - and $\zeta=\mu/T$. The mean value 
of a local operator $\widehat O(x)$ (such as, for instance the stress-energy tensor 
$\wT$, or the current $\wj$) at LTE:
\be
  O(x) = \tr ( \wrhol \widehat O(x) )
\ee
and if the fields $\beta$,$\zeta$ vary significantly over a distance which is much 
larger than the typical microscopic length (indeed the {\em hydrodynamic limit}), 
then they can be Taylor expanded in the density operator starting from the point 
$x$ where the mean value $O(x)$ is to be calculated. The leading terms in the 
exponent of (\ref{gencov}) then become \cite{becaframe}:
\begin{align}\label{ltedensop}
 \wrhol \simeq \frac{1}{Z_{\rm LE}}\exp \left[- \beta_\nu(x) \widehat P^\nu + 
 \xi(x) \widehat Q  - \frac{1}{4} (\partial_\nu \beta_\lambda(x) - \partial_\lambda 
 \beta_\nu(x)) \widehat J_x^{\lambda\nu} \right. \nonumber\\
 \left. + \frac{1}{2}(\partial_\nu \beta_\lambda(x) 
 + \partial_\lambda \beta_\nu(x)) \,\widehat L^{\lambda\nu}_x + \nabla_\lambda \xi(x) 
 \, \widehat d^\lambda_x \right].
\end{align}
where the last two terms with the shear tensor and the gradient of $\zeta$ are 
dissipative and vanish at equilibrium. The $\nabla_\lambda$ operator stands for:
$$
  \nabla_\lambda = \partial_\lambda - u_\lambda u \cdot \partial
$$
as usual in relativistic hydrodynamics. The term which is responsible for a non-vanishing
polarization is the one involving the angular momentum-boosts operators $\widehat J_x$.

The polarization of particles in a fluid at LTE can in principle be obtained by 
calculating matrices like:
$$
 W_{\sigma,\sigma^\prime} = \tr (\wrhol a^\dagger(p)_\sigma a(p^\prime)_{\sigma^\prime})
$$
where $a(p)_\sigma$ are the destruction operators of final state particles of four-momentum
$p$ and $\sigma$ is the spin state index. Nevertheless, the exact calculation of 
$W$ is a difficult one even with the expansion of $\wrhol$ and the mean polarization
was obtained in ref.~\cite{becaspin} by means of a different method, involving
the spin tensor and an  {\em ansatz} about the form of the covariant Wigner function at
LTE (see also \cite{xnwang2}). As a result, the mean spin vector of $1/2$ particles 
with four-momentum $p$, turns out to be:

\be\label{eq:Pixp}
  S^\mu(x,p)= - \frac{1}{8m} (1-f(x,p)) \epsilon^{\mu\rho\sigma\tau} p_\tau \varpi_{\rho\sigma}
\ee
where $f(x,p) = (1+\exp[\beta(x) \cdot p - \mu(x) Q/T(x)] +1)^{-1}$ is the Fermi-Dirac 
distribution and $\varpi(x)$ is the {\em thermal vorticity}, that is:
\be\label{thvort}
   \varpi_{\mu\nu} = -\frac{1}{2} \left( \partial_\mu \beta_\nu - \partial_\nu 
   \beta_\mu \right)
\ee

In hydrodynamic picture of heavy ion collisions, particles with a given momentum are produced across entire particlization hypersurface (see the next Subsection for details). Therefore to calculate the relativistic mean spin vector of a given particle species with given momentum, one has to integrate the above expression over the 
particlization hypersurface $\Sigma$ \cite{becaspin}:
\be\label{eq-Pip}
 S^\mu(p)=\frac{\int d\Sigma_\lambda p^\lambda f(x,p) S^\mu(x,p)}{\int d\Sigma_\lambda 
 p^\lambda f(x,p)}
\ee
With the expression for $S^\mu$, Eq.~\ref{eq-Pip} can be expanded as follows:
\be\label{basic}
 S^\mu(p)=  - \frac{1}{8m} \epsilon^{\mu\rho\sigma\tau} p_\tau 
 \frac{\int d\Sigma_\lambda p^\lambda f(x,p) (1-f(x,p))\varpi_{\rho\sigma}}
{\int d\Sigma_\lambda p^\lambda f(x,p)}
\ee 
The mean (i.e.\ momentum average) spin vector of all particles of given species can be expressed as:
\begin{equation}\label{eq-Pi}
  S^\mu =\frac{1}{N} \int\frac{\di^3 \p}{p^0} \int d\Sigma_\lambda p^\lambda f(x,p) S^\mu(x,p)
\end{equation}
where $N=\int\frac{\di^3\p}{p^0}\int d\Sigma_\lambda p^\lambda f(x,p)$ is the average number of 
particles produced at the particlization surface.

In the experiment, the $\Lambda$ polarization is measured in its rest frame, therefore one can derive 
the expression for the mean polarization vector in the rest frame from Eq.~(\ref{eq-Pi}) taking into account 
Lorentz invariance of most of the terms in it:
\begin{align}\label{eq-mean-S}
  S^{*\mu} = \frac{1}{N} \int \frac{\di^3 \p}{p^0} \int d\Sigma_\lambda p^\lambda f(x,p) 
   S^{*\mu}(x,p)
\end{align}
where asterisk denotes a quantity in the rest frame of particle.

The Eqs.~\ref{basic} and \ref{eq-mean-S} have been used in all numerical calculations of polarization, either based on the hydrodynamic model discussed in the following sub-sections, or transport approaches (next Section) and a good agreement with the data is observed. A crucial feature of Eq.~\ref{basic}, and more in general of this effect, is that it predicts an almost
equal polarization of particles and anti-particles (if quantum statistics effect
are not important) for it is a statistical thermodynamic effect driven by local
equilibration and not by an external C-odd field like the electromagnetic field.
This distinctive feature is confirmed - modulo small deviations - by the experimental measurements 
described in Chapter 10.

\subparagraph{Non-relativistic limit of Eq. \ref{eq:Pixp}}
It is instructive to check that the eq.~(\ref{eq:Pixp}) yields, in the non-relativistic and global
equilibrium limit, the formulae obtained in the first part of this Section. First of all, 
at low momentum, in eq.~(\ref{eq:Pixp}) one can keep only the term corresponding to $\tau=0$
and $p_0 \simeq m$, so that $S^0 \simeq 0$ and:
\be
  S^\mu(x,p) \simeq - \epsilon^{\mu\rho\sigma 0}\frac{1-f(x,p)}{8} \varpi_{\rho\sigma}
\ee
Then, the condition of global equilibrium makes the thermal vorticity field constant and equal to the 
ratio of a constant angular velocity $\omegav$ and a constant temperature $T$ \cite{Becattini:2012tc}
that is:
\be
 - \frac{1}{2} \epsilon^{ijk0} \varpi_{jk} = \frac{1}{T_0} \omega^i
\ee
Finally, in the Boltzmann statistics limit $1 - n_F \simeq 1$ and one finally gets the spin 
3-vector as:
\be
 \bS(x,p) \simeq \frac{1}{4} \frac{\omegav}{T}
\ee

\section{Hydrodynamic modelling of heavy ion collisions}

Let us start the section by outlining the established paradigm of hydrodynamic modelling of heavy ion collisions. Hydrodynamic approximation is not used to describe all stages of a heavy ion collision; instead, a multi-component approach is generally adoped in the field. The approach also reflects different dominant physics processes, which happen at different stages of the heavy ion collision.

The fist stage of heavy ion collision comprises the primary nucleon-nucleon scatterings, which - at top RHIC or LHC energy take less than a fraction of fm/c, due to a strong Lorentz contraction of the incoming nuclei. At this stage, a dense parton (at lower energies - hadron) system is formed. Within the first fermi/c, the system is assumed to reach enough degree of local equilibration, so that the subsequent evolution is described by relativistic hydrodynamics of ideal or viscous fluid.

The modelling of the next, hydrodynamic stage of collision became more sophisticated over the last decades. Successful interpretation of the early results from heavy ion collisions at the RHIC collider within the hydrodynamic picture and the associated discovery of the nearly-perfect fluid at RHIC led to a boom in hydrodynamic modelling. The simulations evolved from 1+1D to 2+1D ideal fluid to 2+1D and 3+1D viscous fluid approximation.

\begin{figure}
\includegraphics[height=130pt]{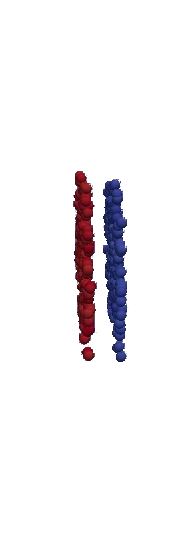}
\includegraphics[height=130pt]{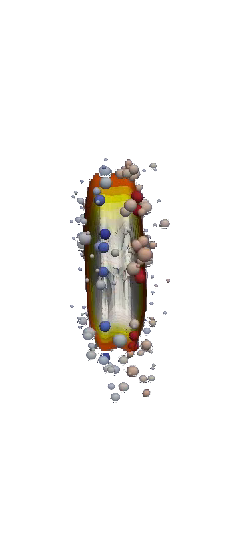}
\includegraphics[height=130pt]{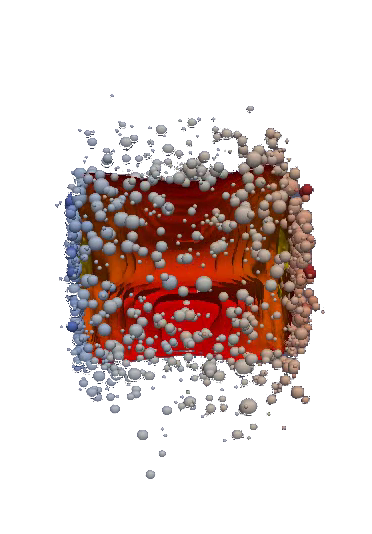}
\includegraphics[height=130pt]{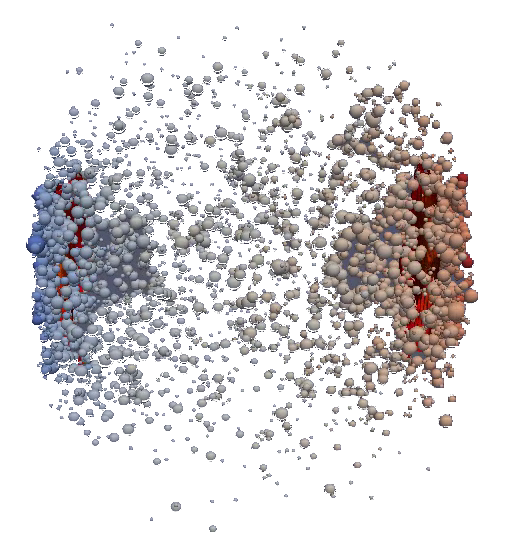}
 \caption{Different stages of relativistic heavy ion collision. From left to right: (1) two Lorentz-contracted nuclei right before the collision, (2) formation of dense, hydrodynamically expanding matter at around 1-2 fm/c after the collision, (3) hydrodynamic expansion of the dense core, surrounded by hadronic corona (the particles on the plot represent individual hadrons), (4) final-state hadronic interactions and decoupling of the fireball. Images taken from an animation by MADAI\protect\footnotemark}
\end{figure}

\footnotetext{\url{http://madai.phy.duke.edu/indexaae2.html?page_id=503}}

As the fireball expands, its density decreases, and so the mean free path of the constituents of the medium becomes larger. When the mean free path becomes comparable to the size of the fireball, the hydrodynamic picture does not apply anymore. At this point, a so-called particlization (see \cite{Huovinen:2012is} for more details) takes place: the fluid medium decouples into particles (hadrons). In the state-of-the-art models, the process of particlization typically takes place at a hyper-surface of constant temperature or constant local rest frame energy density. Such 3-dimensional hyper-surface in 4-dimensional space-time is reconstructed from a full hydrodynamic solution evolved till large enough time. To convert the fluid dynamic degrees of freedom to hadrons, a Cooper-Frye prescription, first introduced in \cite{Cooper:1974mv}, is often used.
Practically speaking, in some of the hydrodynamic models discussed below, the spin polarization is calculated at the hypersurface of particlization, using Eqs.~\ref{eq-Pip},\ref{eq-Pi}, or \ref{eq-mean-S} - which are nothing more but modified Cooper-Frye formulas. The latter makes the computation relatively strightforward. Other hydrodynamic models take a simpler way to compute the spin polarization on a hypersurface of constant proper time $\tau=const$, even if it does not coincide with the hypersurface of particlization in the model.

However, the cross-sections of hadron scatterings are still not small right after the particlization. Therefore, both inelastic scatterings - which change the composition of hadrons in the event - and elastic reactions which only change hadron's momenta, take place. An effective moment when the inelastic reaction cease is known as a {\it chemical freeze-out}, whereas the moment where also elastic scatterings cease is known as a {\it kinetic freeze-out}. As those processes happen gradually, the post-hydrodynamic phase is often modelled using a hadronic cascade, sometimes called a hadronic afterburner.

\section{Hydrodynamic calculations at $\snn=7\dots62$~GeV}

Hydrodynamic modelling of heavy ion collisions at very high energies, such as $\snn=200$~GeV at RHIC or $\snn=2.76, 5.02$~TeV at the LHC, can be numerically simplified by taking into account a strong Lorentz contraction of the colliding nuclei. This practically means that, as long as observables in the central rapidity slice $y\approx 0$ are concerned, such as transverse momentum distributions of produced hadrons, their flow coefficients $v_n(p_T)$, the initial state for the hydrodynamic expansion at $t=t_0$ can be approximated by a thin disk with thickness $z_0\approx t_0$, with initial longitudinal flow $v_z=z/t_0$. The hydrodynamic solution will then have a symmetry with respect to Lorentz boosts in the longitudinal direction. Then the dynamics in the longitudinal direction can be integrated out analytically, leaving only the transverse expansion to the numerics. Likewise, widely used initial state models, such as CGC (Color Glass Condensate), IP-Glasma, Monte Carlo Glauber, only evaluate the initial energy/entropy density profiles in the direction, transverse to the beam axis.

Modelling of relativistic heavy-ion collisions at collision energies $\snn$ from few to a hundred GeV is more challenging as compared to the high-energy regime.
Many of the hydrodynamical and hybrid models used to model
collisions at top RHIC and LHC energies are not directly applicable
to collisions at the lower energies. The simplifying approximations of boost invariance and zero net-baryon density are not valid, and different
kinds of non-equilibrium effects play a larger role.

At such collision energies, the colliding nuclei do not resemble thin disks because of a weaker Lorentz contraction; also, the partonic models of the initial state (CGC, IP-Glasma) gradually lose their applicability in this regime. The longitudinal boost invariance is not a good approximation anymore, therefore one needs to simulate a 3-dimensional hydrodynamic expansion.

Historically, the first full-fledged hydrodynamic model applied to study the polarization of hadrons - $\Lambda$ hyperons - in heavy-ion collisions at $\snn=7\dots62$~GeV is UrQMD+vHLLE model \cite{Karpenko:2015xea}. The second hydrodynamic calculation of $\Lambda$ polarization for this collision energy range was performed in PICR model \cite{Xie:2017upb}. More recently, $\Lambda$ polarization was calculated in three-fluid dynamics (3FD) model \cite{Ivanov:2019ern}. We proceed by describing the details of the abovementioned hydrodynamic models.

{\bf Initial states in the hydrodynamic calculations.}
In the UrQMD+vHLLE calculation, the UrQMD string/hadronic cascade is used to describe the primary
collisions of the nucleons, and to create the initial state of the
 hydrodynamical evolution. The two nuclei are initialized according to
 Woods-Saxon distributions and the initial binary interactions proceed
 via string or resonance excitations, the former process being dominant
 in ultrarelativistic collisions (including the BES collision
 energies). All the strings are fragmented into hadrons before the
 transition to fluid phase (fluidization) takes place, although not all
 hadrons are yet fully formed at that time, i.e., they do not yet have
 their free-particle scattering cross sections, and thus do not yet
 interact at all. The hadrons before conversion to fluid should not be
 considered physical hadrons, but rather marker particles to describe
 the flow of energy, momentum and conserved charges during the
 pre-equilibrium evolution of the system. The use of UrQMD to
 initialise the system allows us to describe some of the
 pre-equilibrium dynamics and dynamically generates event-by-event
 fluctuating initial states for hydrodynamical evolution.

The interactions in the pre-equilibrium UrQMD evolution are allowed
until a hypersurface of constant Bjorken proper time
$\tau_0=\sqrt{t^2-z^2}$ is reached, since the hydrodynamical code is
constructed using the Milne coordinates $(\tau,x,y,\eta)$, where $\tau
= \sqrt{t^2-z^2}$~\cite{Karpenko:2013wva}. The UrQMD evolution,
however, proceeds in Cartesian coordinates $(t,x,y,z)$, and thus
evolving the particle distributions to constant $\tau$ means evolving
the system until large enough time $t_{l}$ in such a way that the
collisional processes and decays are only allowed in the domain
$\sqrt{t^2-z^2} < \tau_0$. The resulting particles on $t = t_l$
surface are then propagated backwards in time to the $\tau = \tau_0$
surface along straight trajectories to obtain an initial state for the
hydrodynamic evolution.

The lower limit for the starting time of the hydrodynamic evolution depends on 
the collision energy according to
\begin{equation}
\tau_0=2R/\sqrt{(\snn/2m_N)^2-1}, \label{eqTau0}
\end{equation}
which corresponds to the average time, when two nuclei have passed
through each other, i.e., all primary nucleon-nucleon
collisions have happened. This is the earliest possible moment in
time, where approximate local equilibrium can be assumed.

To perform event-by-event hydrodynamics using fluctuating initial
conditions, every individual UrQMD event is converted to an initial state
profile. As mentioned, the hadron transport does not lead to an
initial state in full local equilibrium, and the thermalization of the
system at $\tau=\tau_0$ has to be artificially enforced.  The energy and
momentum of each UrQMD particle at $\tau_0$ is distributed to the
hydrodynamic cells $ijk$ assuming Gaussian density
profiles
\begin{align} \label{Gauss1}
\Delta P^\alpha_{ijk} & = P^\alpha \cdot C\cdot\exp\left(-\frac{\Delta x_i^2+\Delta y_j^2}{R_\perp^2}-\frac{\Delta\eta_k^2}{R_\eta^2}\gamma_\eta^2 \tau_0^2\right) \\
\Delta N^0_{ijk}&=N^0 \cdot C\cdot\exp\left(-\frac{\Delta x_i^2+\Delta y_j^2}{R_\perp^2}-\frac{\Delta\eta_k^2}{R_\eta^2}\gamma_\eta^2 \tau_0^2\right), \label{Gauss2}
\end{align}
where $\Delta x_i$, $\Delta y_j$, $\Delta \eta_k$ are the differences
between particle's position and the coordinates of the hydrodynamic
cell $\{i,j,k\}$, and $\gamma_\eta={\rm cosh}(y_p-\eta)$ is the
longitudinal Lorentz factor of the particle as seen in a frame moving
with the rapidity $\eta$. The normalization constant $C$ is calculated
from the condition that the discrete sum of the values of the Gaussian
in all neighboring cells equals one. The resulting $\Delta P^\alpha$
and $\Delta N^0$ are transformed into Milne coordinates and added
to the energy, momentum and baryon number in each cell. This procedure
ensures that in the initial transition from transport to hydrodynamics the
energy, momentum and baryon number are conserved.

\begin{figure}
\includegraphics[width=0.49\textwidth]{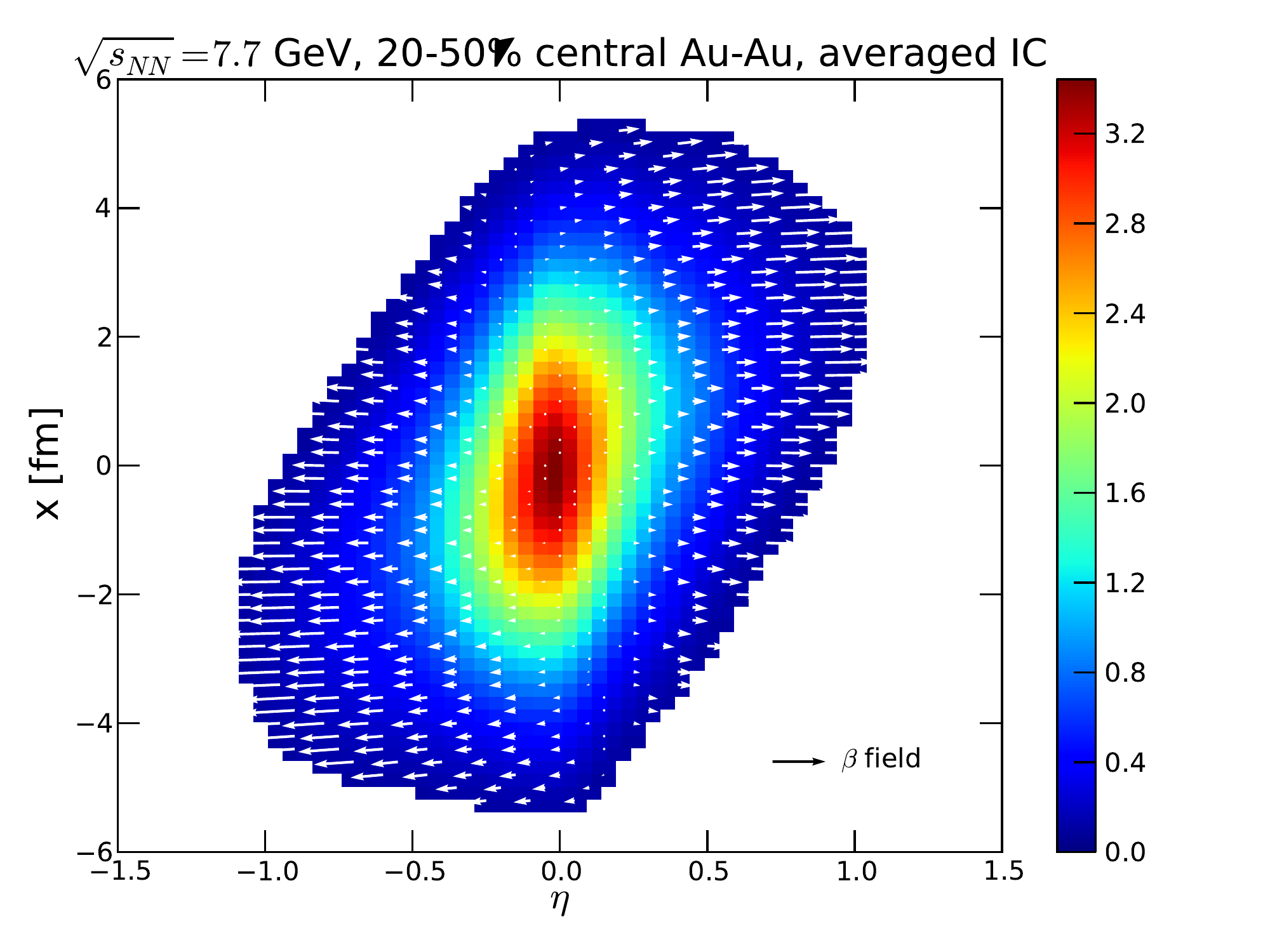}
\includegraphics[width=0.49\textwidth]{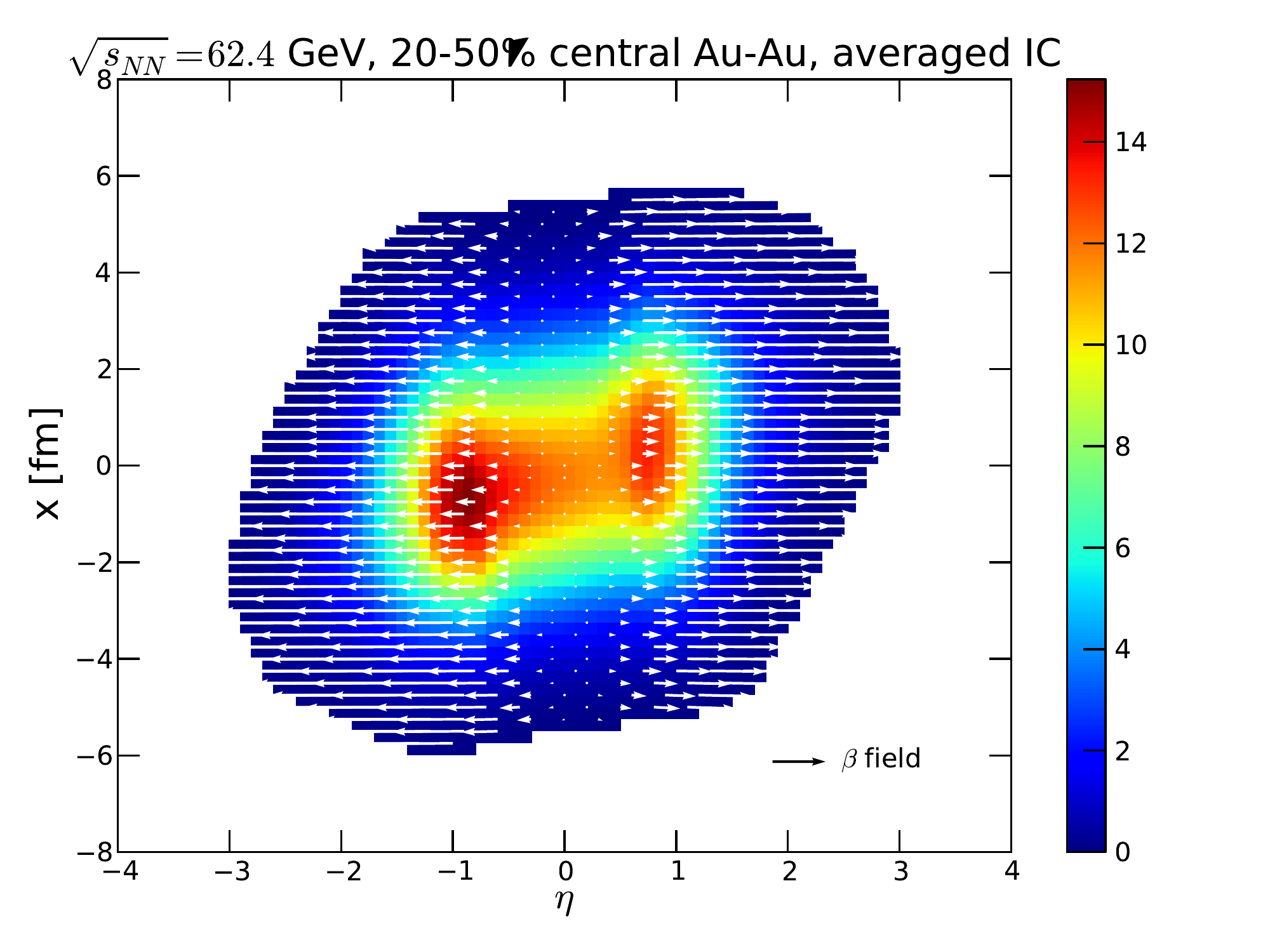}
\caption{Initial energy density profiles for hydrodynamic stage with arrows depicting initial beta field superimposed. The hydrodynamic evolutions start from averaged initial state corresponding to 20-50\% central Au-Au collisions at $\snn=7.7$ (top row) and $62.4$~GeV (bottom row).}\label{fig-ICxeta}
\end{figure}

\begin{figure*}
\includegraphics[width=0.47\textwidth]{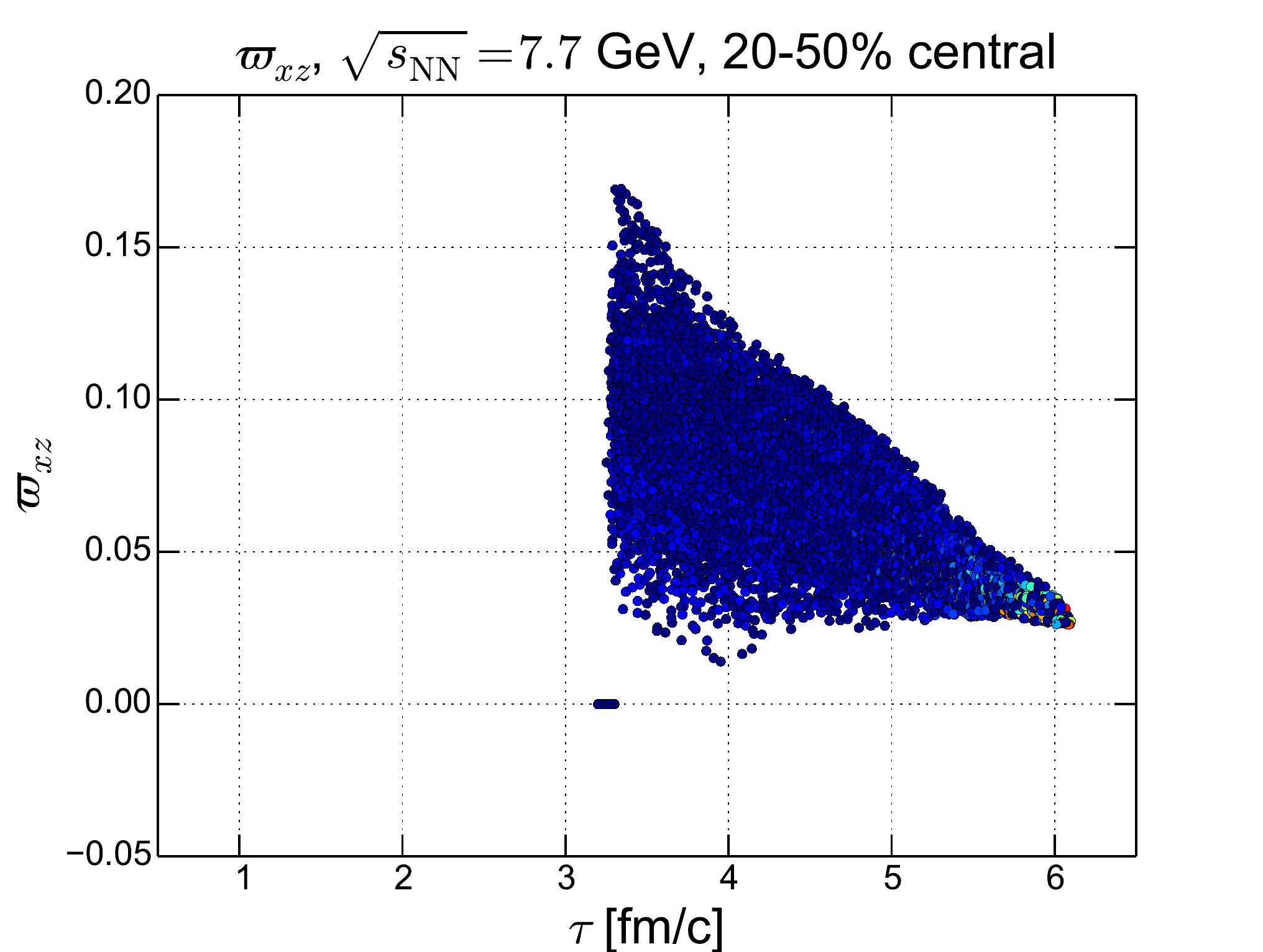}
\includegraphics[width=0.47\textwidth]{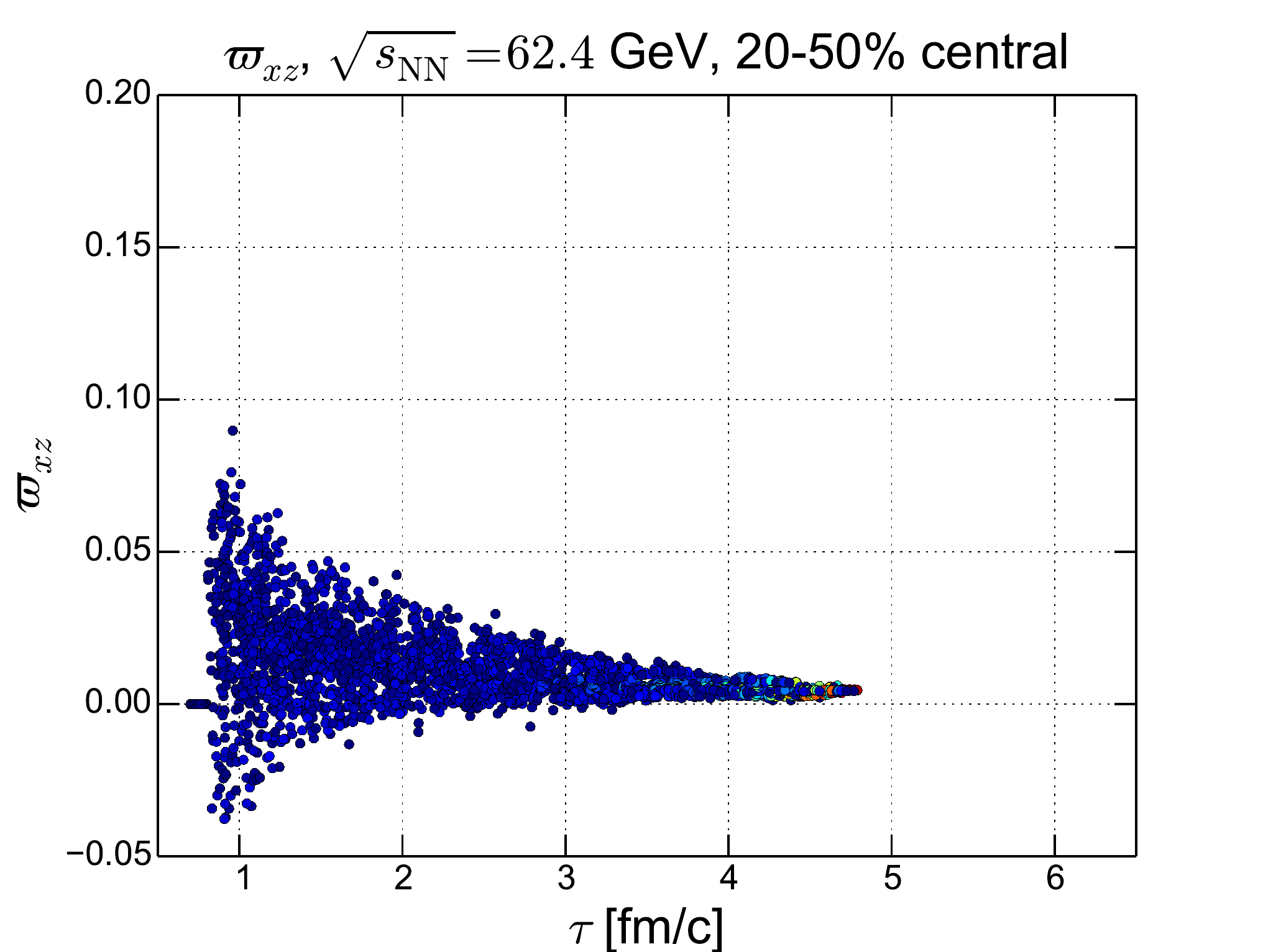}
\caption{Evolution of $\varpi_{xz}$ in midrapidity ($|y|<0.3$) slice of particlization surface, projected onto time axis (right column). The hydrodynamic evolutions start from averaged initial state corresponding to 20-50\% central Au-Au collisions at $\snn=7.7$ (top row) and $62.4$~GeV (bottom row).}\label{fig-omegaXZ-tau}
\end{figure*}

In Fig.~\ref{fig-omegaXZ-tau} the initial energy density profiles are visualized for two selected collision energies: $\snn=7.7$ and 62.4 GeV. To produce this figure, two single hydrodynamic calculations with averaged initial conditions from 100 initial UrQMD simulations each were run. At $\snn=62.4$~GeV, because of baryon transparency effect, the $x,z$ components of beta vector at midrapidity are small and do not have a regular pattern, therefore the distribution of $\varpi_{xz}$ in the hydrodynamic cells close to particlization energy density includes both positive and negative parts, as it is seen on the corresponding plot in the right column. At $\snn=7.7$~GeV, baryon stopping results in a shear flow structure, which leads to same (positive) sign of the $\varpi_{xz}$.

In PICR model, the physics picture of the initial state is a Yang-Mills field, stretched between Lorentz 
contracted streaks after impact \cite{Magas:2000jx}. Such initial state also produces torqued initial state of the fireball with finite angular momentum.

\begin{figure}
\includegraphics[width=1.0\textwidth]{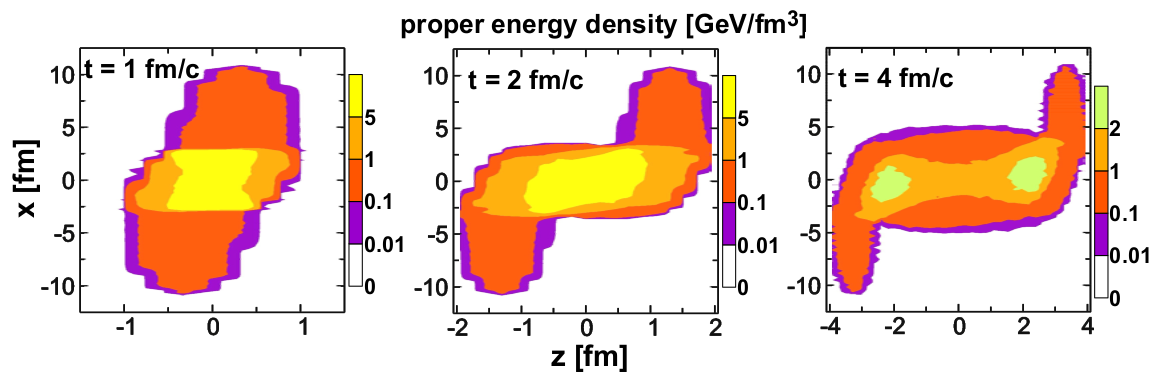}
\caption{Snapshots of the energy density profile in the $x-z$ plane in a 3-fluid dynamic simulation of a semi-central Au-Au collision at $\snn=19.6$~GeV. Plots are taken from \cite{Ivanov:2020wak}.}\label{fig-energy-density-3FD}
\end{figure}

The 3-fluid dynamic model is somewhat different from the UrQMD+vHLLE and PICR models. In the 3FD, the evolution starts with two nuclei right before the moment of impact, which are represented by two blobs of cold baryon-rich fluid \cite{Ivanov:2005yw}. The process of nucleus-nucleus collision is then modelled as an inter-penetration of the baryon-rich fluids, which leads to friction between the fluids. The fluids lose energy and momentum via the friction terms, which leads to a creation of the third fluid, which is baryon-free. Similarly to the UrQMD+vHLLE or PICR models, the friction between the baryon-rich fluids leads to a total energy density profile which is tilted in the $x-z$ plane, as shown in Fig.~\ref{fig-energy-density-3FD}. The friction also produces the velocity shear, which corresponds to a finite angular momentum of the participant system.

\textbf{Hydrodynamic stage} in UrQMD+vHLLE hybrid is simulated with a (3+1)-dimensional viscous hydrodynamical code \texttt{vHLLE}, which is described in full detail in Ref.~\cite{Karpenko:2013wva}. The code solves the local energy-momentum conservation equations:
\begin{align}
 d_\nu T^{\mu\nu}&=0, \label{hydro1}\\
 d_\nu N_{\rm B,Q}^\nu&=0, \label{hydro2}
\end{align}
where $N^\nu_{\rm B}$ and $N^\nu_{\rm Q}$ are the net baryon and electric charge
currents respectively, and we remind that $d_\nu$ denotes a covariant
derivative. The calculation\footnote{Typical grid spacing used in the calculations: $\Delta x=\Delta y=0.2$~fm, $\Delta\eta=0.05-0.15$ and timestep $\Delta\tau=0.05-0.1$~fm/c depending on the collision energy. A finer grid with $\Delta x=\Delta y=0.125$~fm was taken to simulate peripheral collisions.}
is done in Milne coordinates $(\tau,x,y,\eta)$, where $\tau = \sqrt{t^2-z^2}$ and
$\eta=1/2\ln[(t+z)/(t-z)]$.

In the Israel-Stewart framework of causal dissipative
hydrodynamics~\cite{Israel-Stewart}, the dissipative currents are
independent variables. For the calculations of $\Lambda$ polarization, 
$\zeta/s=0$ is set in the UrQMD+vHLLE calculation. The code works in the Landau frame, where
the energy diffusion flow is zero, and the baryon and charge
diffusion currents are neglected for simplicity, which is equivalent to zero heat conductivity. For
the shear stress evolution we choose the relaxation time 
$\tau_\pi = 5\eta/(Ts)$, the coefficient $\delta_{\pi\pi} = 4/3 \tau_\pi$, and
approximate all the other higher-order coefficients by zero. The following evolution equations are solved for the
shear-stress tensor $\pi^{\mu\nu}$:
\begin{equation}
\left<u^\gamma d_\gamma \pi^{\mu\nu}\right>
 =-\frac{\pi^{\mu\nu}-\pi_\text{NS}^{\mu\nu}}{\tau_\pi}
  -\frac 4 3 \pi^{\mu\nu}\partial_{;\gamma}u^\gamma, \label{evolutionShear}
\end{equation}
where the brackets denote the traceless and orthogonal to $u^\mu$ part
of the tensor and $\pi_\text{NS}^{\mu\nu}$ is the Navier-Stokes value
of the shear-stress tensor.

Another necessary ingredient for the hydrodynamic stage is the
equation of state (EoS) of the medium. In UrQMD+vHLLE, the chiral
model EoS \cite{Steinheimer:2010ib}, which features correct asymptotic
degrees of freedom, i.e., quarks and gluons in the high temperature and
hadrons in the low-temperature limits, crossover-type transition
between confined and deconfined matter for all values of $\mu_B$ and
qualitatively agrees with lattice QCD data at $\mu_B=0$.

Different from that, both PICR and 3FH models feature ideal fluid approximation. The hydrodynamic evolution is 
simulated with the Relativistic Particle-in-Cell (PICR) method. Both in the initial state and subsequent CFD simulation, a classic `Bag Model' EoS was applied: $P=c_0^2 e^2 - \frac43 B$, with constant $c_0^2 = \frac13$ and a fixed Bag constant $B$. The energy density takes the form:
$e=\alpha T^4+ \beta T^2 + \gamma + B $, where $\alpha$, $\beta$, $\gamma$ are constants arising from the degeneracy factors for (anti-)quarks and gluons. 

{\bf Final conditions for hydrodynamic stage.} In the UrQMD+vHLLE hybrid, the fluid-to-particle transition, or particlization, is performed using the conventional Cooper-Frye prescription \cite{Cooper:1974mv}. The Cooper-Frye prescription applied at a hypersurface of constant local rest frame energy density. The hadrons, generated at the particlization, are then re-scattered with the UrQMD cascade. This particlization hypersurface is reconstructed during the hydrodynamic evolution based on the criterion of a fixed energy density $\epsilon=\epsilon_\text{se}$ and using the Cornelius routine~\cite{Huovinen:2012is}. The default value for the particlization energy density is $\epsilon_{\rm sw}=0.5$~GeV/fm$^3$, which in the chiral model EoS corresponds to $T\approx 175$~MeV at $\mu_{\rm B}=0$.
At this energy density the crossover transition is firmly on the
hadronic side, but the density is still a little higher than the
chemical freeze-out energy density suggested by the thermal
models (for the topic of thermal models, we refer the reader to \cite{Becattini:2005xt}).

In PICR model, the particlization is set to happen at fixed time $t$ in the laboratory frame.

As given by the Cooper-Frye prescription, the hadron distribution on
each point of the hypersurface is
\begin{equation}
p^0 \frac{d^3N_i(x)}{d^3p} = d\Sigma_\mu p^\mu f(p\cdot u(x),T(x),\mu_i(x)).
 \label{CFp}
\end{equation}
The phase space distribution function $f$ is usually assumed to be the
one corresponding to a noninteracting hadron resonance gas in or close
to the local thermal equilibrium.

In a standard calculation in UrQMD+vHLLE, the Cooper-Frye formula (\ref{CFp}) is used as a probability density to sample ensembles of hadrons with Monte Carlo method. Further on, the sampled hadrons are passed to the UrQMD cascade to simulate inelastic and elastic interactions in the dilute post-hydrodynamic stage.
However, for the calculation of the polarization the Monte Carlo hadron sampling is replaced with a direct calculation based on the Eq.~(\ref{eq-Pip}), applied on particlization surfaces from event by event hydrodynamics. For that, one can realize that the formula for the mean polarization of spin 1/2 hadrons (\ref{basic}) looks similar to the Cooper-Frye formula, except for the factor $(1-f(x,p))\varpi_{\rho\sigma}$ under the integral.

\begin{table}
\vspace{10pt}
\begin{center}
\begin{tabular}{|l|l|l|l|l|}
\hline
 $\snn$~[GeV] & $\tau_0$~[fm/c] & $R_\perp$~[fm] & $R_\eta$~[fm] & $\eta/s$ \\ \hline
     7.7          &      3.2        &     1.4        &     0.5    &    0.2   \\ \hline
     8.8 (SPS)    &      2.83       &     1.4        &     0.5    &    0.2   \\ \hline
     11.5         &      2.1        &     1.4        &     0.5    &    0.2   \\ \hline
     17.3 (SPS)   &      1.42       &     1.4        &     0.5    &    0.15  \\ \hline
     19.6         &      1.22       &     1.4        &     0.5    &    0.15  \\ \hline
     27           &      1.0        &     1.2        &     0.5    &    0.12  \\ \hline
     39           &      0.9*        &     1.0        &     0.7    &    0.08  \\ \hline
     62.4         &      0.7*        &     1.0        &     0.7    &    0.08  \\ \hline
     200          &      0.4*        &     1.0        &     1.0    &    0.08  \\ \hline
 \end{tabular}
\caption{Collision energy dependence of the UrQMD+vHLLE parameters chosen to
  reproduce the experimental data in the RHIC BES range: $\snn=7.7-200$~GeV.}\label{tb-params}
\end{center}
\end{table}

The 3-fluid dynamic model has again a somewhat different final conditions for its hydrodynamic stage, as compared to UrQMD+vHLLE or PICR. The distributions of hadrons at the particlization are computed not with the Cooper-Frye but with Milekhin formula \cite{Milekhin}, and the criterion for the particlization is a fixed combined energy density of all 3 fluids in a given space-time point:
\begin{equation} \varepsilon_\text{tot}=\left(T^{00}_\text{proj}+T^{00}_\text{targ}+T^{00}_\text{B-free}\right)_\text{rest frame}<\varepsilon_\text{frz}\ ,\end{equation}
The 3-fluid dynamic model does not feature the final-state hadronic cascade, therefore the particlization in 3FD is the same as the freeze-out.

It is important to note that all the above-mentioned models had been tuned to reproduce the basic hadronic observables prior to the calculations of polarization. In particular, a reasonable reproduction of the experimental data - (pseudo)rapidity distributions, transverse momentum spectra and elliptic flow coefficients - has been achieved in UrQMD+vHLLE with the parameter values depending monotonically on the collision energy as it is shown in Table~\ref{tb-params}. This was obtained when the particlization energy density was fixed to $\epsilon_{\rm sw}=0.5$~GeV/fm$^3$ for the whole collision energy range. In 3FD, a typical choice of the freeze-out energy density to reproduce a broad set of experimental data is $\varepsilon_\text{frz}\simeq 0.2$~GeV/fm$^3$ \cite{Ivanov:2005yw}.

\subparagraph{Patterns of $\Lambda$ polarization in the UrQMD+vHLLE and PICR models}

\begin{figure}
\includegraphics[width=0.6\textwidth]{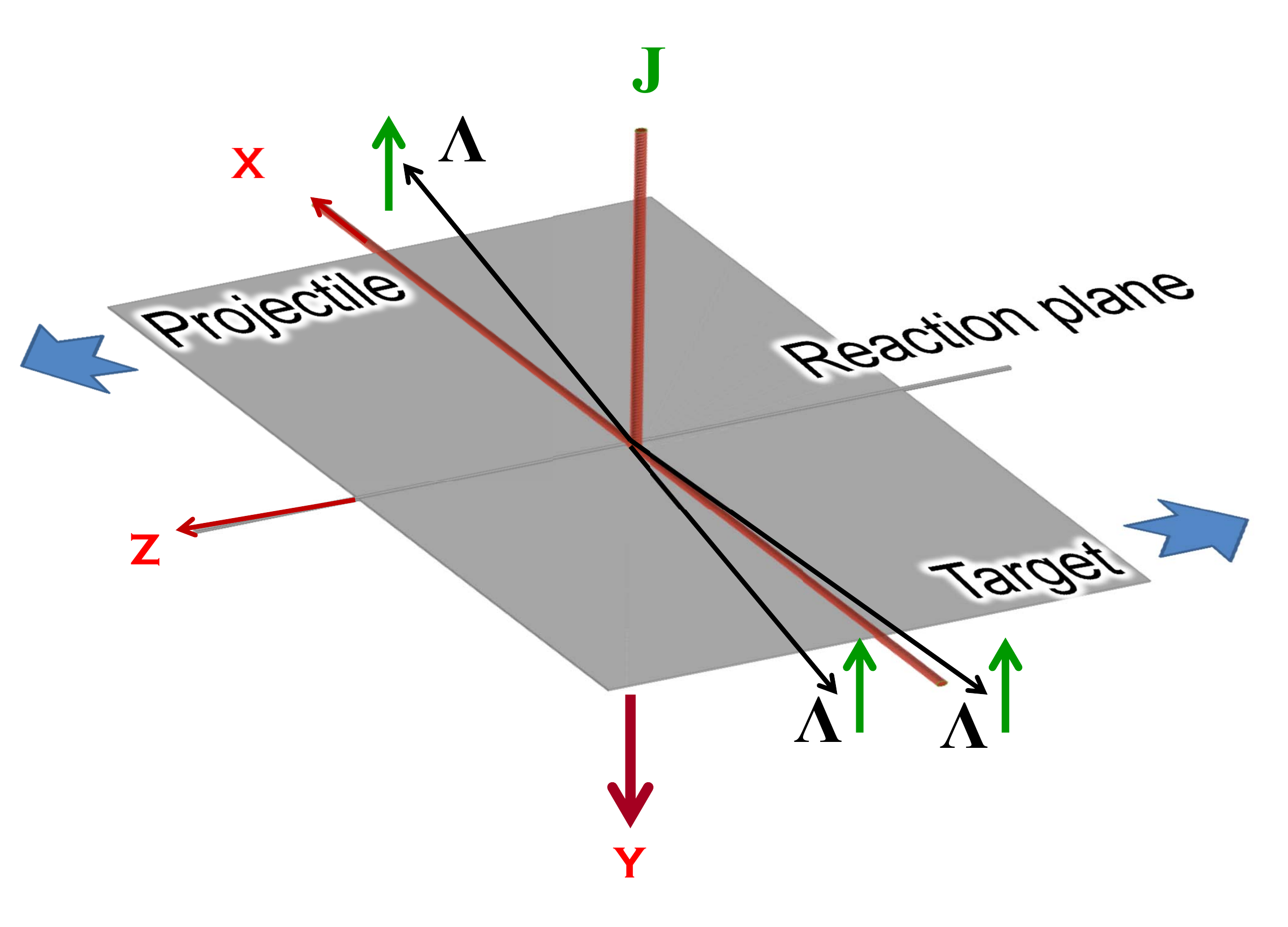}
\caption{Coordinate system used for the components of $\Lambda$ polarization vector. The reaction plane is $xz$, and $y$ coordinate is opposite to the vector of the global angular momentum of the system, which points upwards and perpendicular to the reaction plane. The plot is taken from \cite{Becattini:2013vja}.}\label{fig-coord}
\end{figure}

Before starting to refer to the components of the spin polarization vector of $\Lambda$ hyperons, it is worth to sketch the coordinate system used. The coordinate system is shown in Fig.~\ref{fig-coord}: the $x$ axis is parallel to the vector of the impact parameter of the heavy-ion collision, the $z$ axis is parallel to the beam direction, thus $xz$ is the so-called reaction plane. The $y$ axis points perpendicular to the reaction plane, and is directed opposite to the vector of the total angular momentum of the fireball.

\begin{figure}
\includegraphics[width=0.53\textwidth]{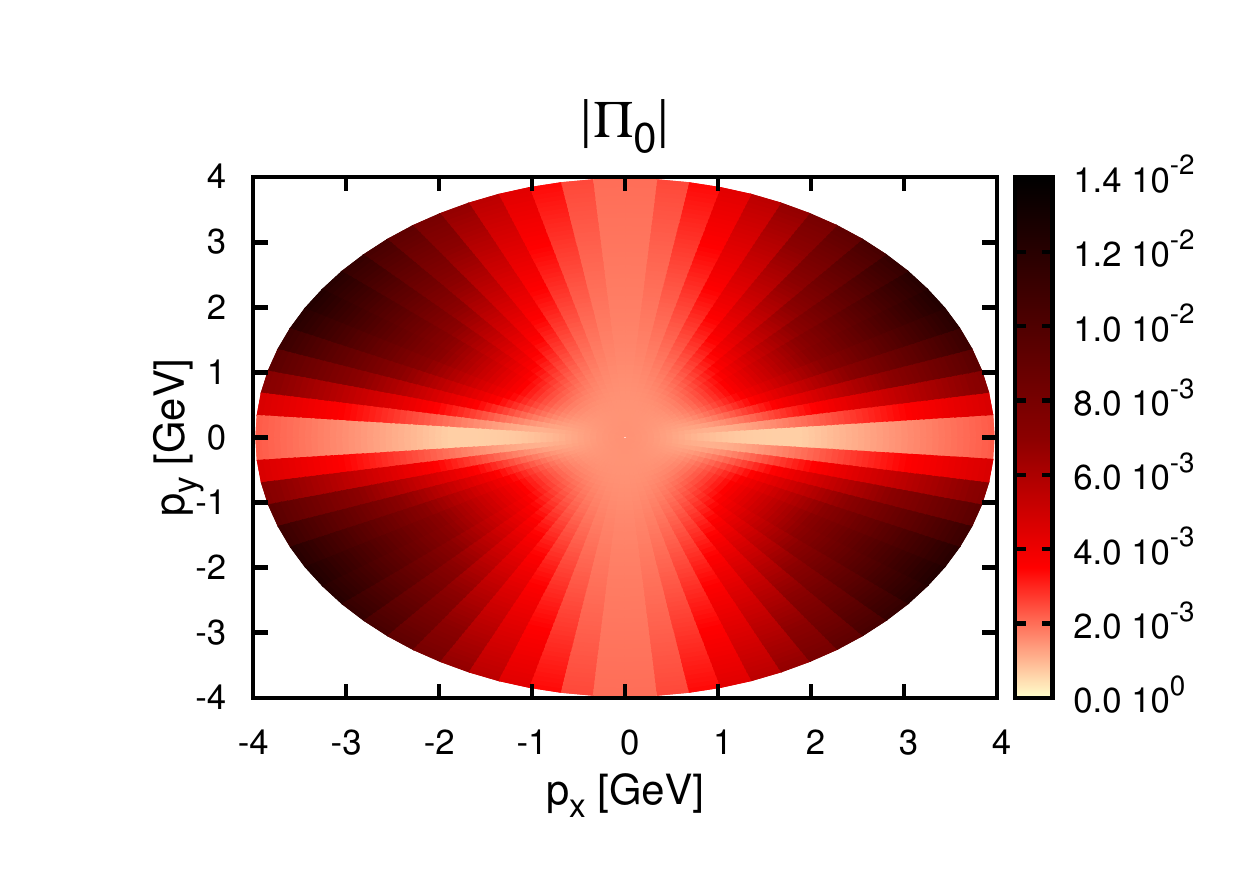}\hspace{-10pt}
\includegraphics[width=0.53\textwidth]{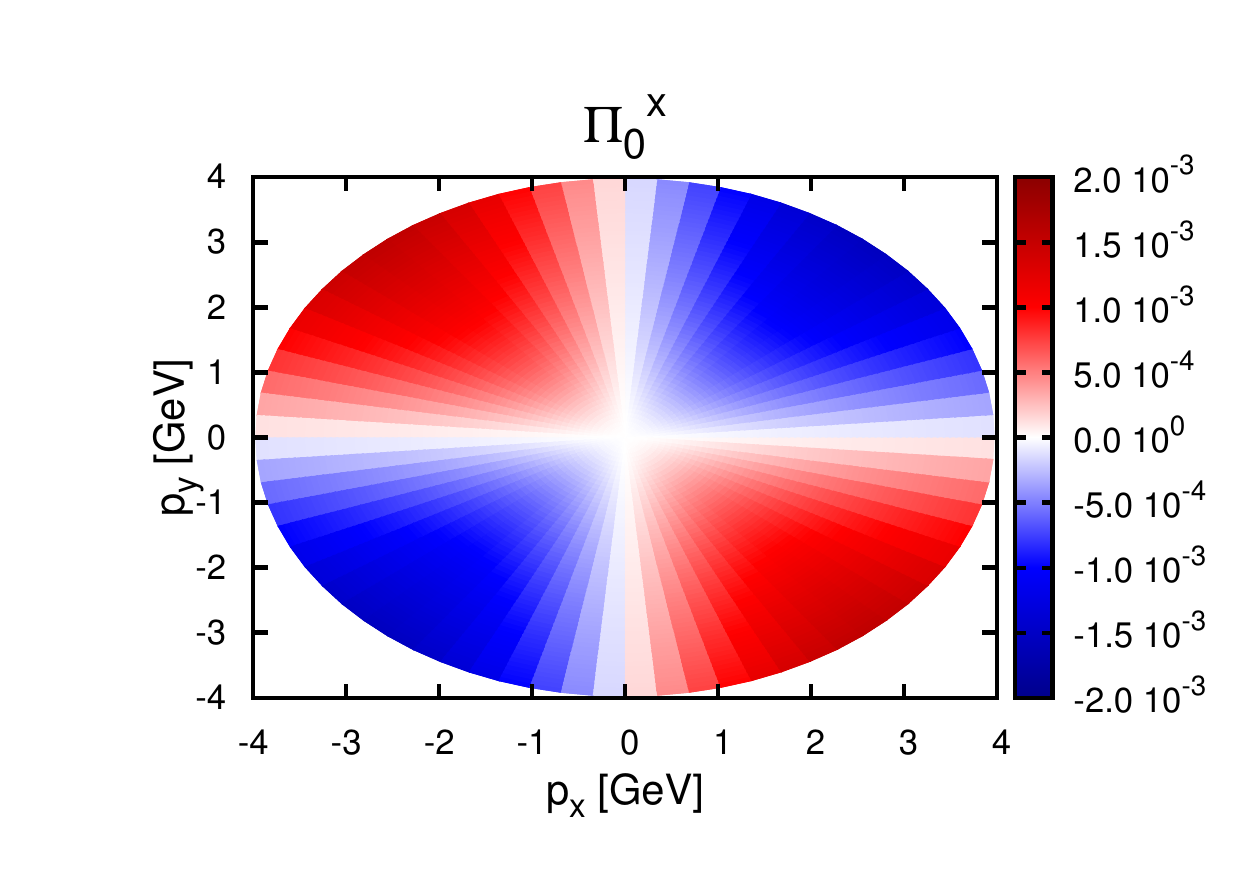}\\ \vspace{-10pt}
\includegraphics[width=0.53\textwidth]{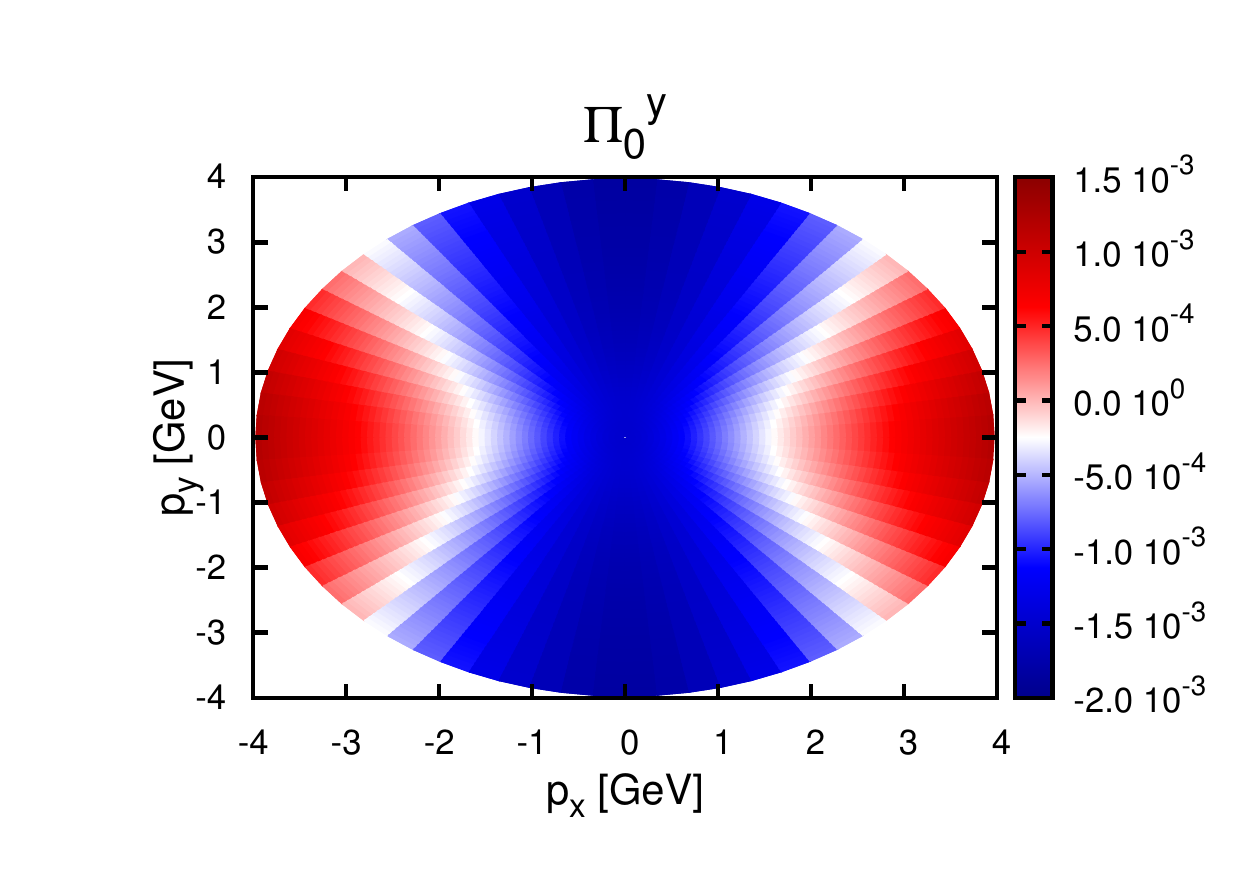}\hspace{-10pt}
\includegraphics[width=0.53\textwidth]{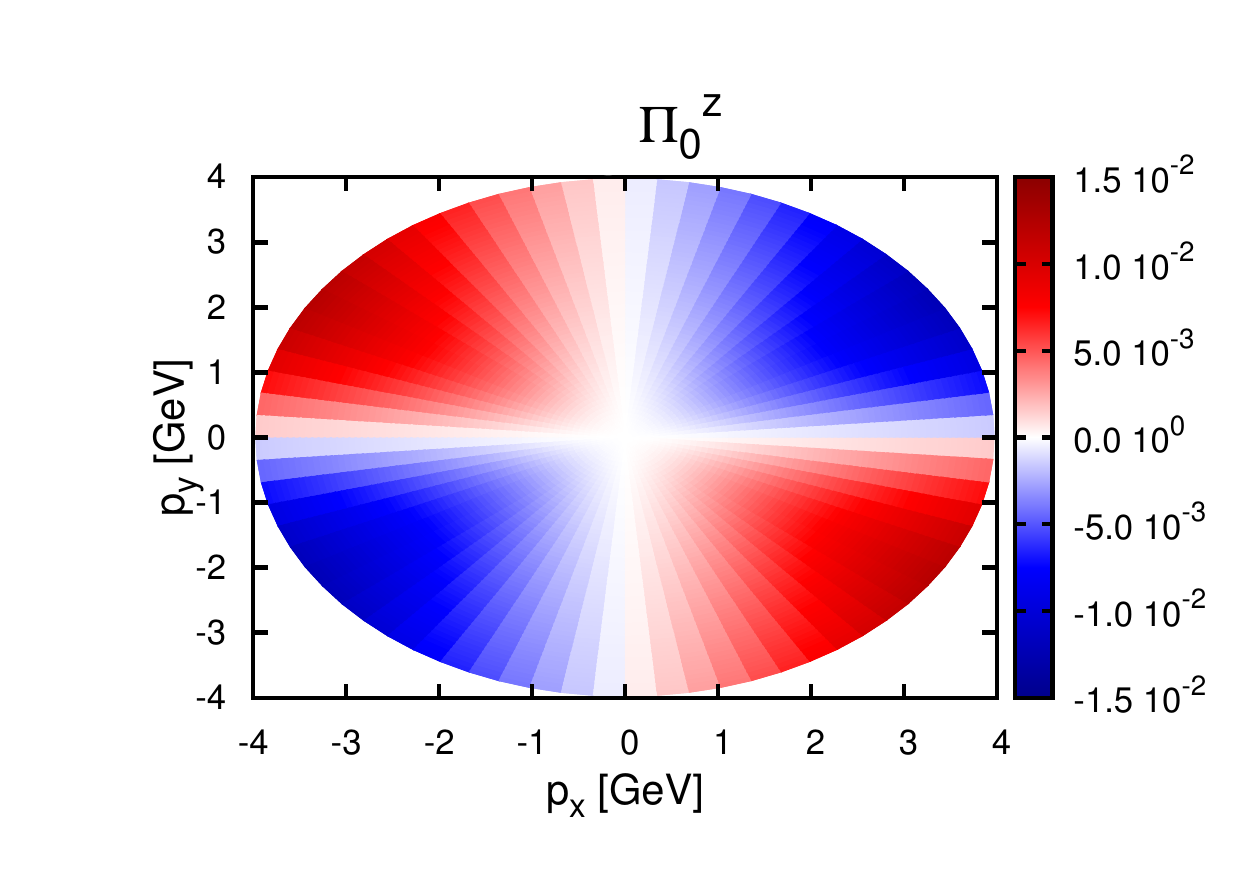}
\caption{Components and modulus of the $\Lambda$ polarization vector as a function of $p_T$ of the $\Lambda$, in ECHO-QGP simulation of Au-Au collisions with fixed impact parameter $b=11.6$~fm at $\snn=200$~GeV. The picture is taken from \cite{Becattini:2015ska}.}\label{fig-pxpypz-echoqgp}
\end{figure}

Already an early calculation \cite{Becattini:2015ska} of the $\Lambda$ polarization vector as a function of the $p_T$ of the $\Lambda$ at mid-rapidity performed with 3+1 dimensional hydrodynamic code ECHO-QGP \cite{DelZanna:2013eua} has shown quite an assorted pattern, see Fig.~\ref{fig-pxpypz-echoqgp}. The ECHO-QGP calculation has been made for Au-Au collisions at fixed impact parameter $b=11.6$~fm (corresponding to peripheral collisions) at the top RHIC energy $\snn=200$~GeV; hydrodynamic calculations for top RHIC and LHC energies will be discussed in a next sub-section.
At large transverse momenta and at $|p_x|=|p_y|$, the polarization vector component along the beam axis, $P^{z}$ (marked as $\Pi_0^z$ on Fig.~\ref{fig-pxpypz-echoqgp}, also marked as $P_{||}$ on some of the plots below) has the largest amplitude. The component along the impact parameter, $P^x$ (marked as $\Pi_0^x$ on Fig.~\ref{fig-pxpypz-echoqgp}, also marked as $P_{b}$ on some of the plots below) has a quadrupole pattern similar to $P^{z}$ but with a smaller amplitude. However, because of symmetry of the system, the ${\bf p}_T$ integrated $P^x$ and $P^z$ integrate out to zero, and the only nonzero component remaining is $P^y$ ($\Pi_0^y$ on Fig.~\ref{fig-pxpypz-echoqgp}), which is opposite to the direction of the total angular momentum $\bf J$ of the fireball.

\begin{figure*}
\includegraphics[width=\textwidth]{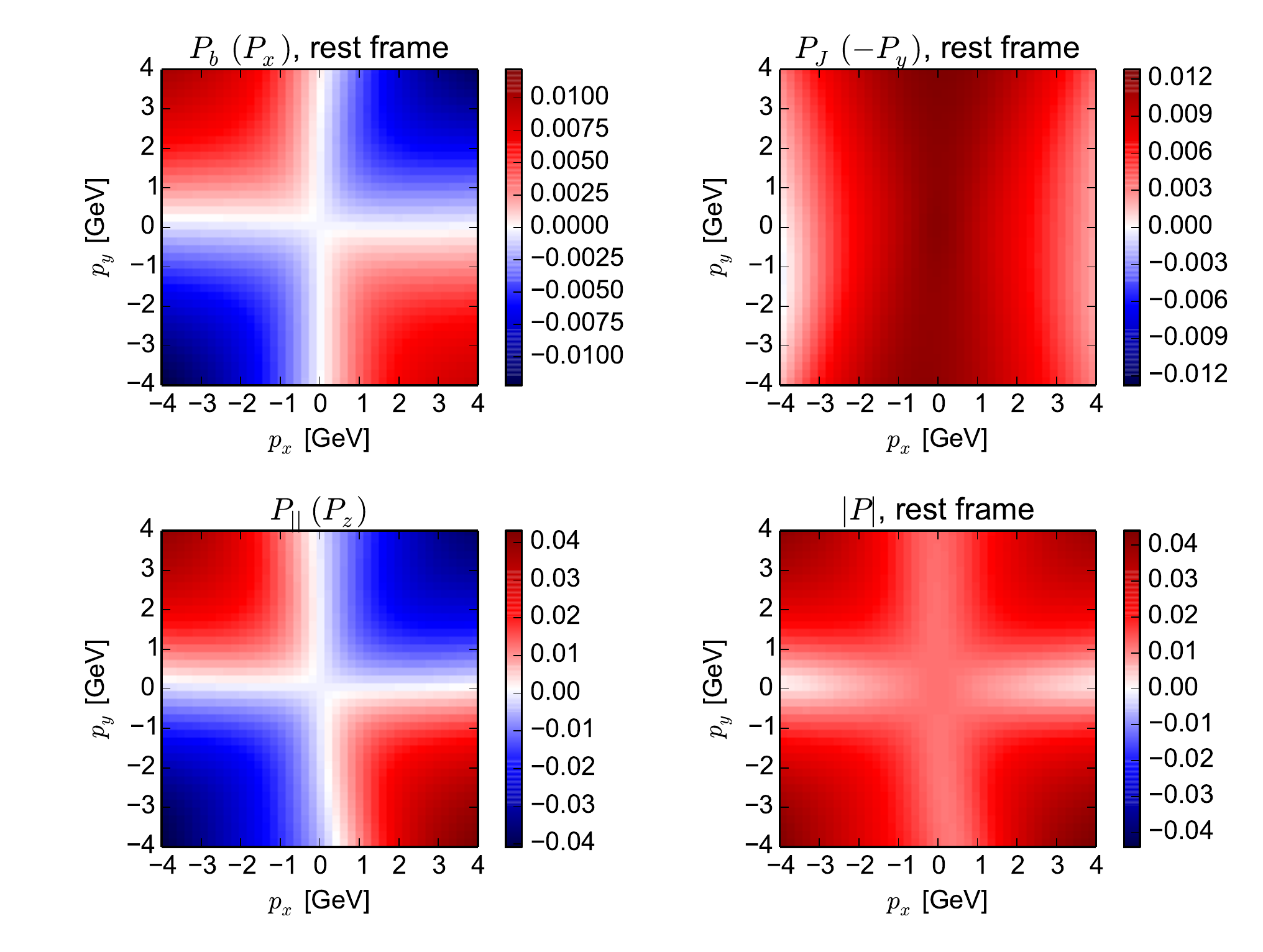}
\caption{Components of spin polarization vector of primary $\Lambda$ baryons produced at midrapidity 
in UrQMD+vHLLE calculation for 40-50\% central Au-Au collisions at $\snn=19.6$~GeV. 
The polarization is calculated in the rest frame of $\Lambda$.}\label{fig-196gev-4050}
\end{figure*}

A very similar patterns for the components of the polarization vector were observed later in the UrQMD+vHLLE calculations for the Beam Energy Scan energies. On Fig.~\ref{fig-196gev-4050} the transverse momentum dependence of the components of $\Lambda$ polarization vector is shown for 40-50\% central Au-Au collisions (impact parameter range $b=9.3-10.4$~fm) at collision energy $\snn=19.6$~GeV, which is located in the middle of the Beam Energy Scan range. 1000 event-by-event hydrodynamic simulations were executed, then the event-averaged denominator and numerator of eq.~\ref{eq-Pip} were computed as a function of $p_x$ and $p_y$, in order to produce Fig.~\ref{fig-196gev-4050}.

\begin{figure*}
\includegraphics[width=0.49\textwidth]{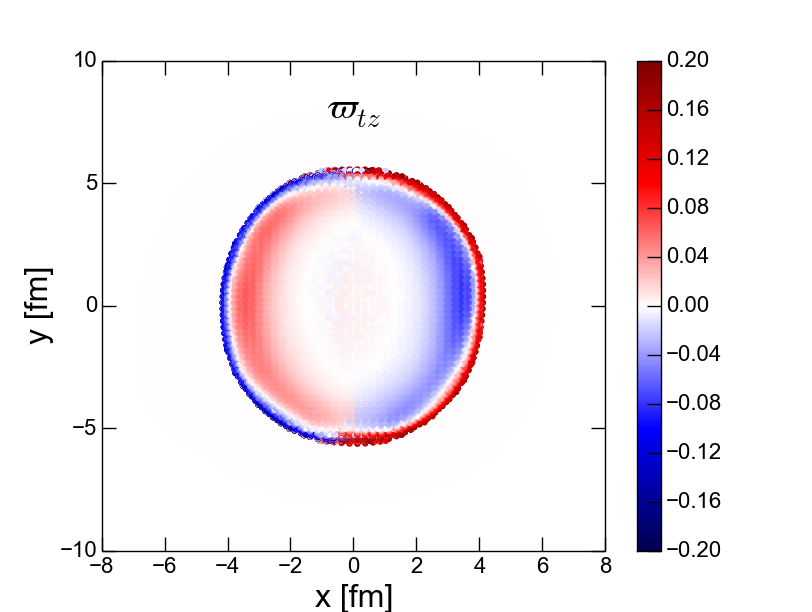}
\includegraphics[width=0.49\textwidth]{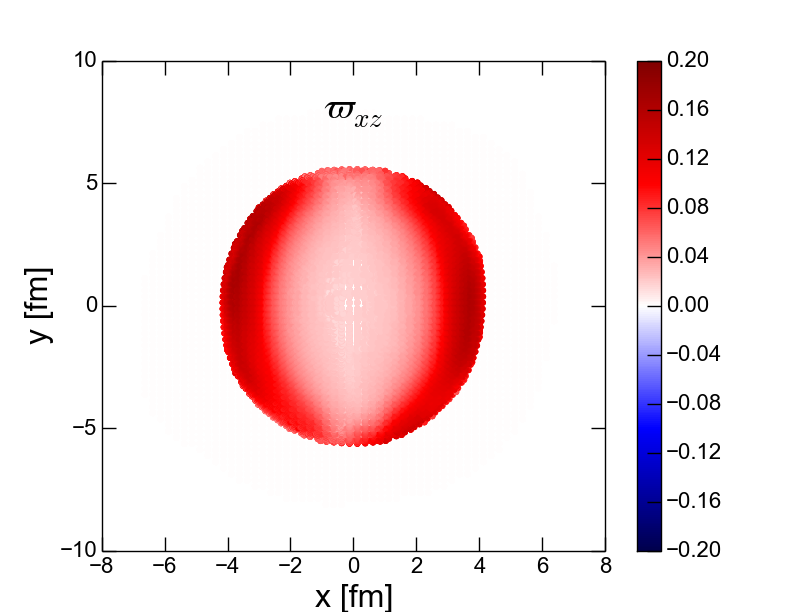}
\caption{Components of thermal vorticity $\varpi_{tz}$ (left) and $\varpi_{xz}$ (right) on the midrapidity slice of particlization hypersurface, projected on the $xy$ plane. The UrQMD+vHLLE calculation with an averaged initial state corresponds to 40-50\% central Au-Au collisions at $\snn=19.6$~GeV.}\label{fig-omega-196gev}
\end{figure*}

The polarization patterns in the $p_x p_y$ plane reflect the corresponding patterns of the components of thermal vorticity over the particlization hypersurface. In particular, it was found in \cite{Karpenko:2016jyx} that the leading contribution to $P^x$ stems from the term $\varpi_{tz}p_y$ in Eq.~\ref{eq:Pixp}. In turn, $\varpi_{tz}$, shown in left panel of Fig.~\ref{fig-omega-196gev} is a result of the interplay of $\partial_t \beta_z$ (acceleration of longitudinal flow and temporal gradients of temperature - conduction) and $\partial_z \beta_t$ (convection and conduction), according to eq.~(\ref{thvort}).
The $P^y$ component has a leading contribution from the term $\varpi_{xz}p_0$ (which is also the only non-vanishing contribution at $p_T=0$), and $\varpi_{xz}$ has a rather uniform profile over the midrapidity slice of the freezeout hypersurface, and the leading contribution to it comes from $\partial_x u_z$ (shear flow in $z$ direction).

\begin{figure}
\begin{center}
      \includegraphics[width=0.48\textwidth]{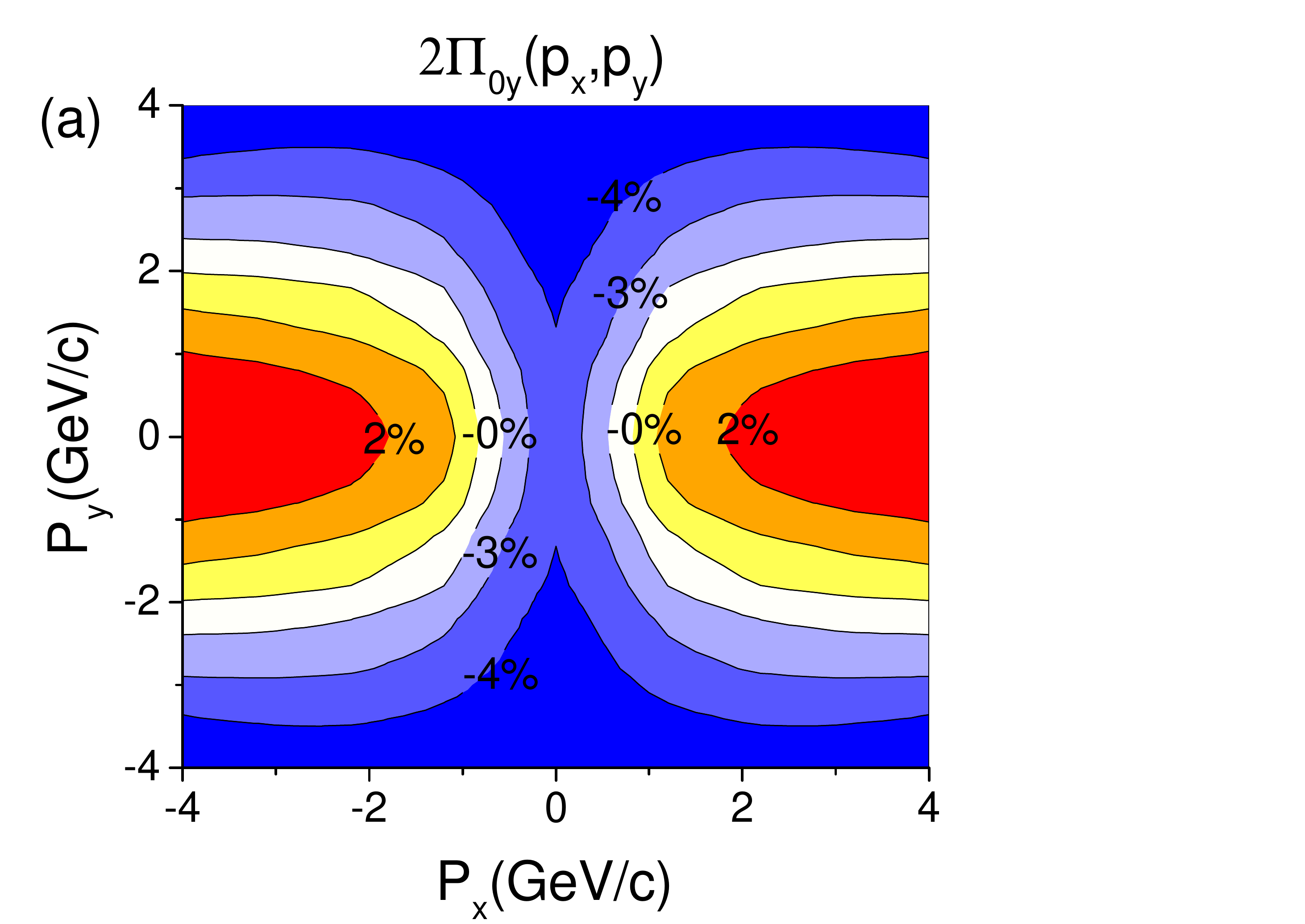}
      \includegraphics[width=0.48\textwidth]{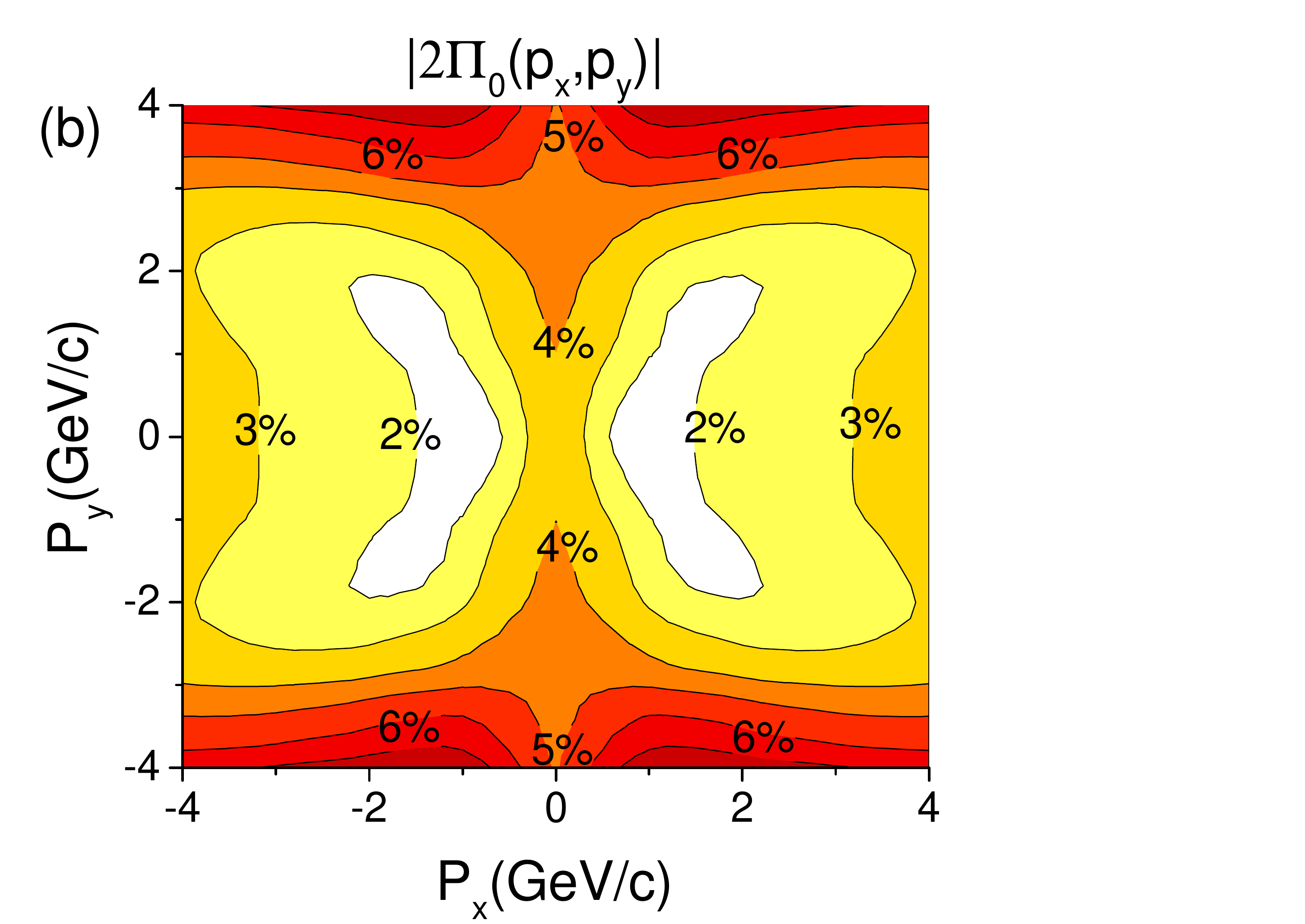}
\end{center}
\caption{
The $y$ component (left) and the modulus (right) of the $\Lambda$ 
polarization at $p_z=0$ in PICR model, 
for the Au+Au reaction at $\snn=11.5$ GeV.
The figure is in the frame of the $\Lambda$. The impact
parameter $b=0.7 b_m = 0.7\times 2R$, where $R$ is the radius of 
Au and $b_m=2R$ is the maximum value of $b$.
The freeze out time is $6.25 =(2.5 + 4.75)$ fm/c, including
2.5 fm/c for initial state and 4.75 fm/c for hydro-evolution.
}
\label{fig-Piy-PICR}
\end{figure}

The PICR calculation provides a transverse momentum pattern of the $y$ component of polarization ($P^y$), which is different from the UrQMD+vHLLE calculation, see the left panel of Figure~\ref{fig-Piy-PICR}. At the $p_y=0$ line, the polarization changes sign between large $|p_x|$ and zero $p_x$. As for now, it is not clear why the transverse momentum patterns are different in the two models.

\subparagraph{Centrality and collision energy dependence of the polarization}
\begin{figure}
\includegraphics[width=0.49\textwidth]{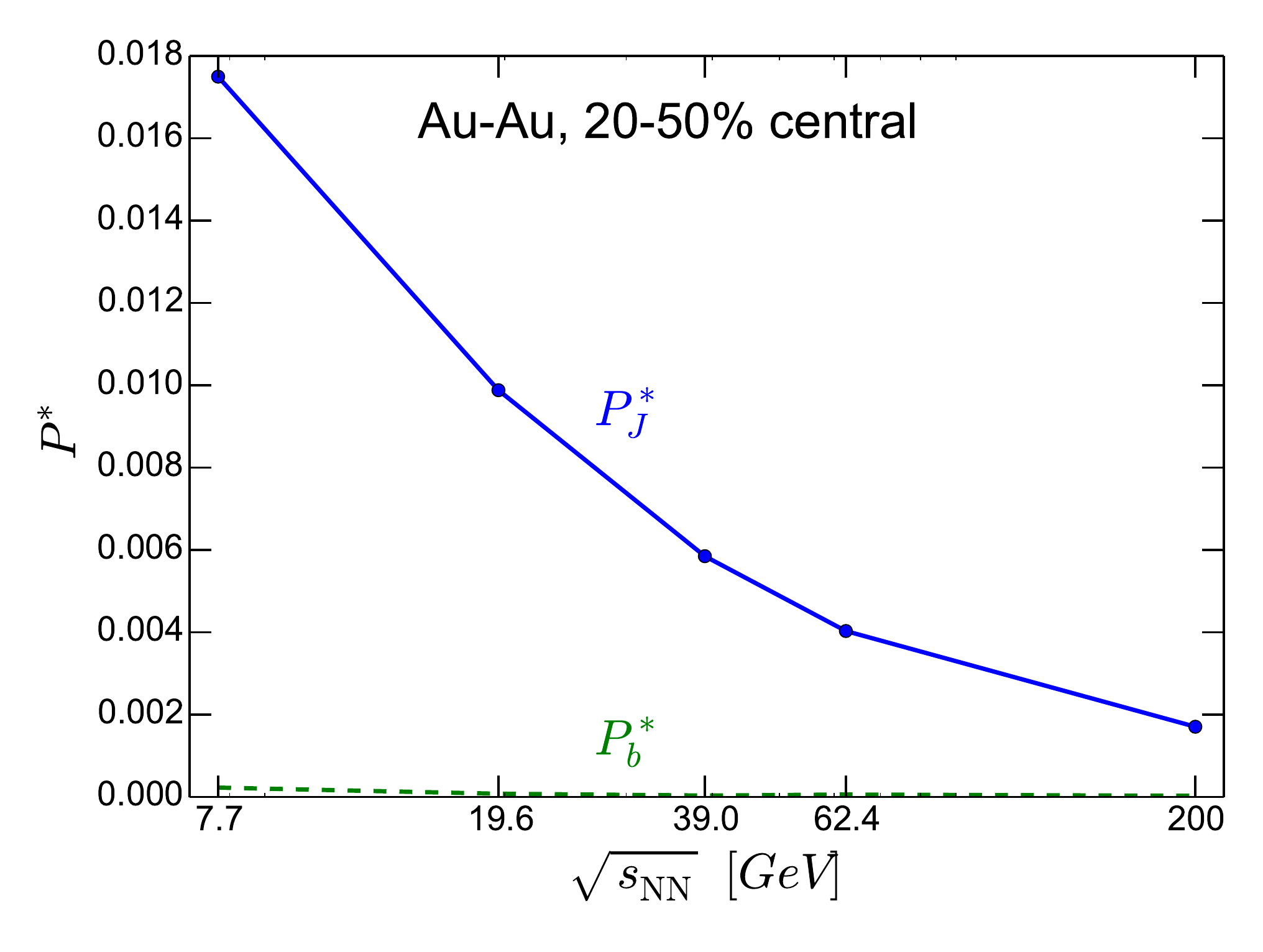}
\includegraphics[width=0.49\textwidth]{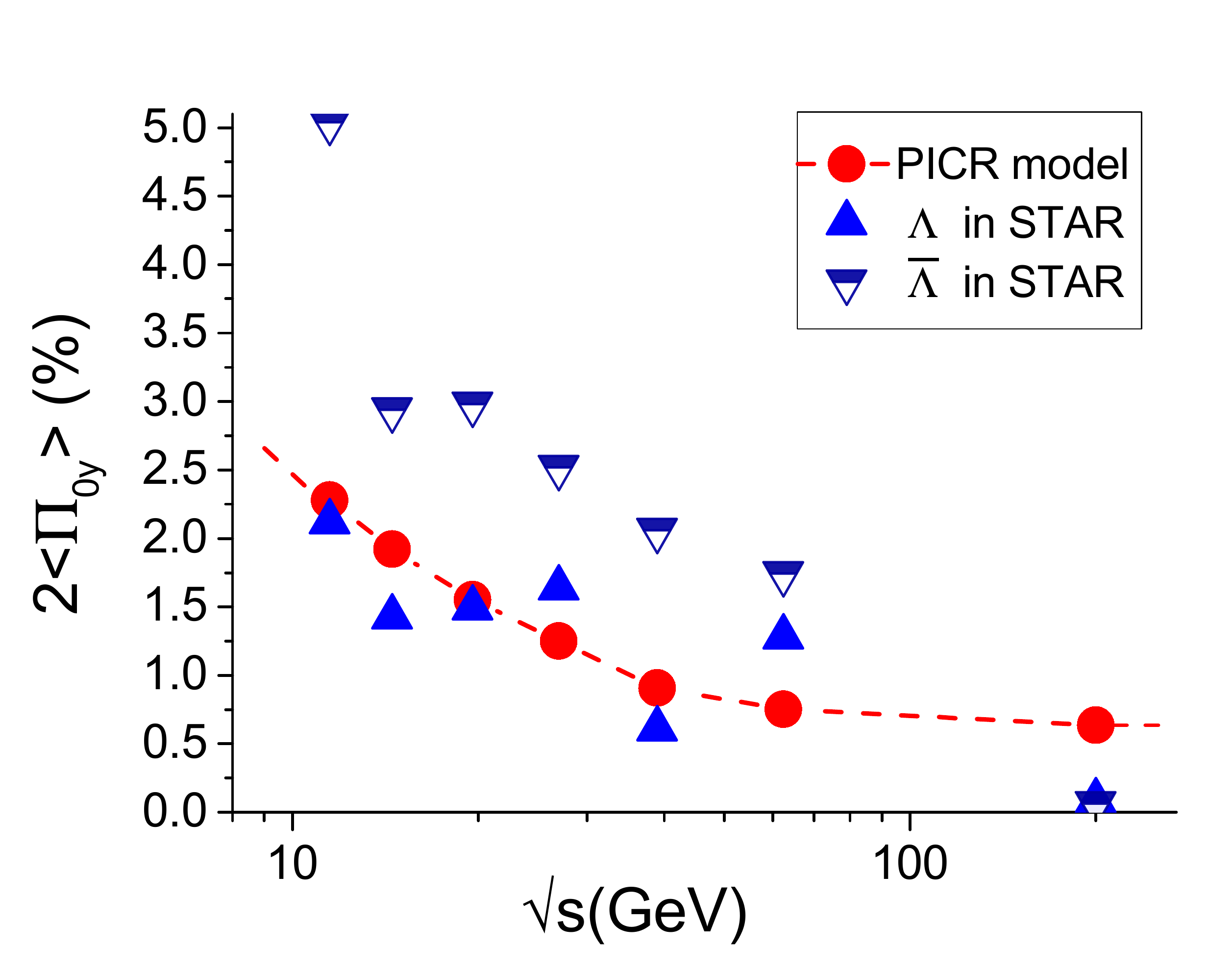}
\caption{Collision energy dependence of the $P_J\equiv P^y$ and $P_b\equiv P^x$ components of polarization vector of $\Lambda$, calculated in its rest frame, in UrQMD+vHLLE (left panel) and PICR (right panel) models for 20-50\% central Au-Au collisions.}\label{fig-Pixy-sqrts}
\end{figure}

\begin{figure}
\includegraphics[width=0.5\textwidth]{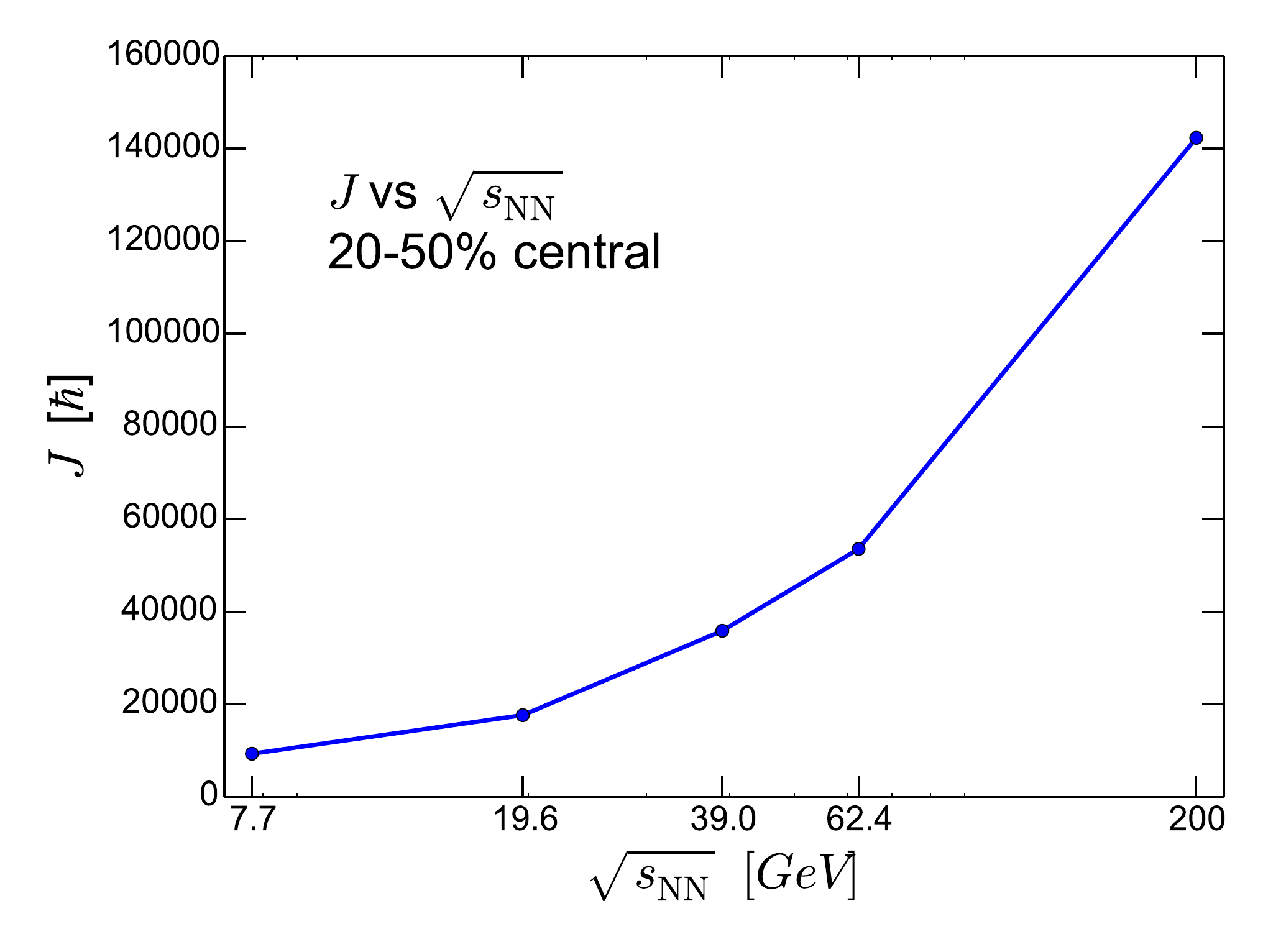}
\includegraphics[width=0.5\textwidth]{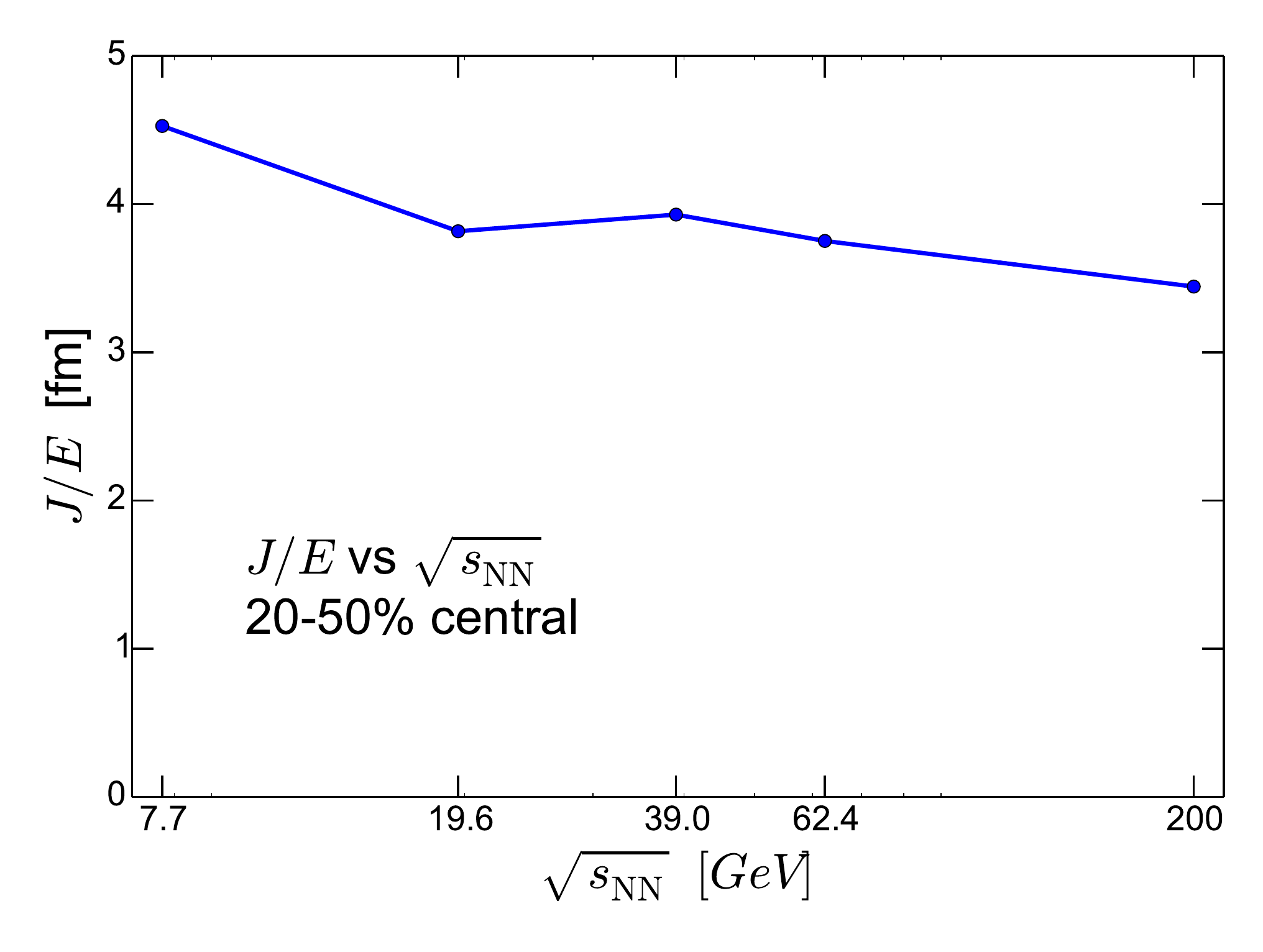}
\caption{Total angular momentum of the fireball (left) and total angular momentum scaled by total energy of the fireball (right) as a function of collision energy, in UrQMD+vHLLE calculation for 20-50\% central Au-Au collisions.}\label{fig-Jy-sqrts}
\end{figure}

Figure~\ref{fig-Pixy-sqrts} shows the collision energy dependence of the global polarization of $\Lambda$ in UrQMD+vHLLE and PICR models. To follow recent STAR measurements, in UrQMD+vHLLE calculation the 20-50\% centrality bin was constructed by correspondingly chosen range of impact parameters for the initial state UrQMD calculation. We observe that the polarization component along $\bf J$, the $P^y$ decreases by about one order of magnitude as collision energy increases from $\snn=7.7$~GeV to full RHIC energy, where it turns out to be consistent with an early calculation of the global hyperon polarization at the top RHIC energy in \cite{Becattini:2015ska}.
In the PICR calculation, the impact parameter $b_0=0.7$, which corresponds to centrality $c = 49\%$, was chosen to simulate the 20-50\% centrality bin. For comparison the data of $\Lambda$ and $\bar{\Lambda}$ polarization from STAR (RHIC) were inserted into the right panel of the Figuree~\ref{fig-Pixy-sqrts} with blue triangle symbols.

In the UrQMD+vHLLE calculation, the fall of the out-of-plane component $P^y$ is not directly related to a change in the out-of-plane component of total angular momentum of the fireball. In fact, the total angular momentum increases as the collision energy increases, which can be seen on top panel of Fig.~\ref{fig-Jy-sqrts}. However, the total angular momentum is not an intensive quantity like polarization, so, to have a better benchmark we took the ratio between the total angular momentum and the total energy, $J/E$, which is shown in the bottom panel of the same figure. Yet, one can see that the $J/E$ shows only a mild decrease as collision energy increases.
\begin{figure}
\includegraphics[width=0.5\textwidth]{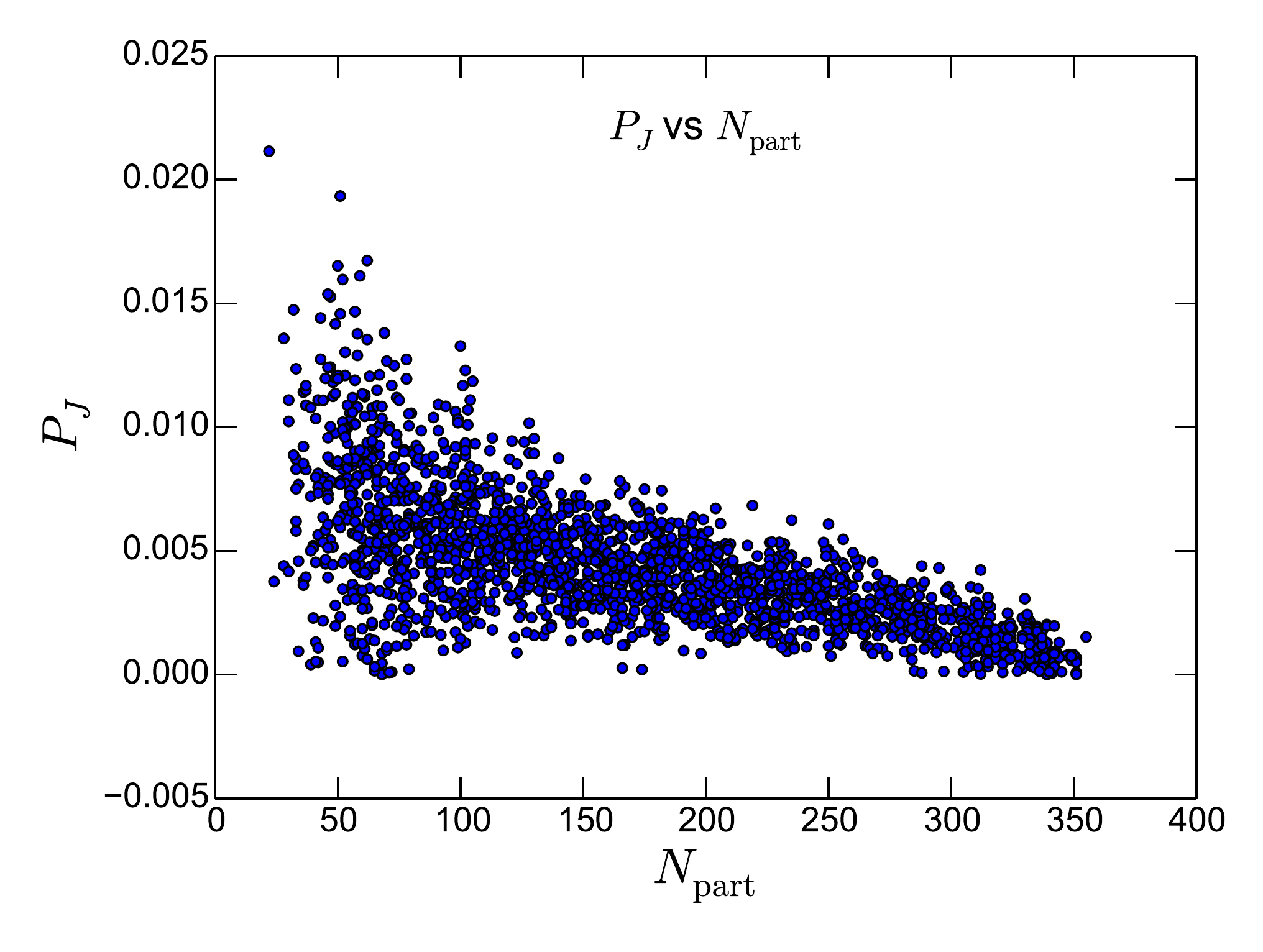}
\includegraphics[width=0.5\textwidth]{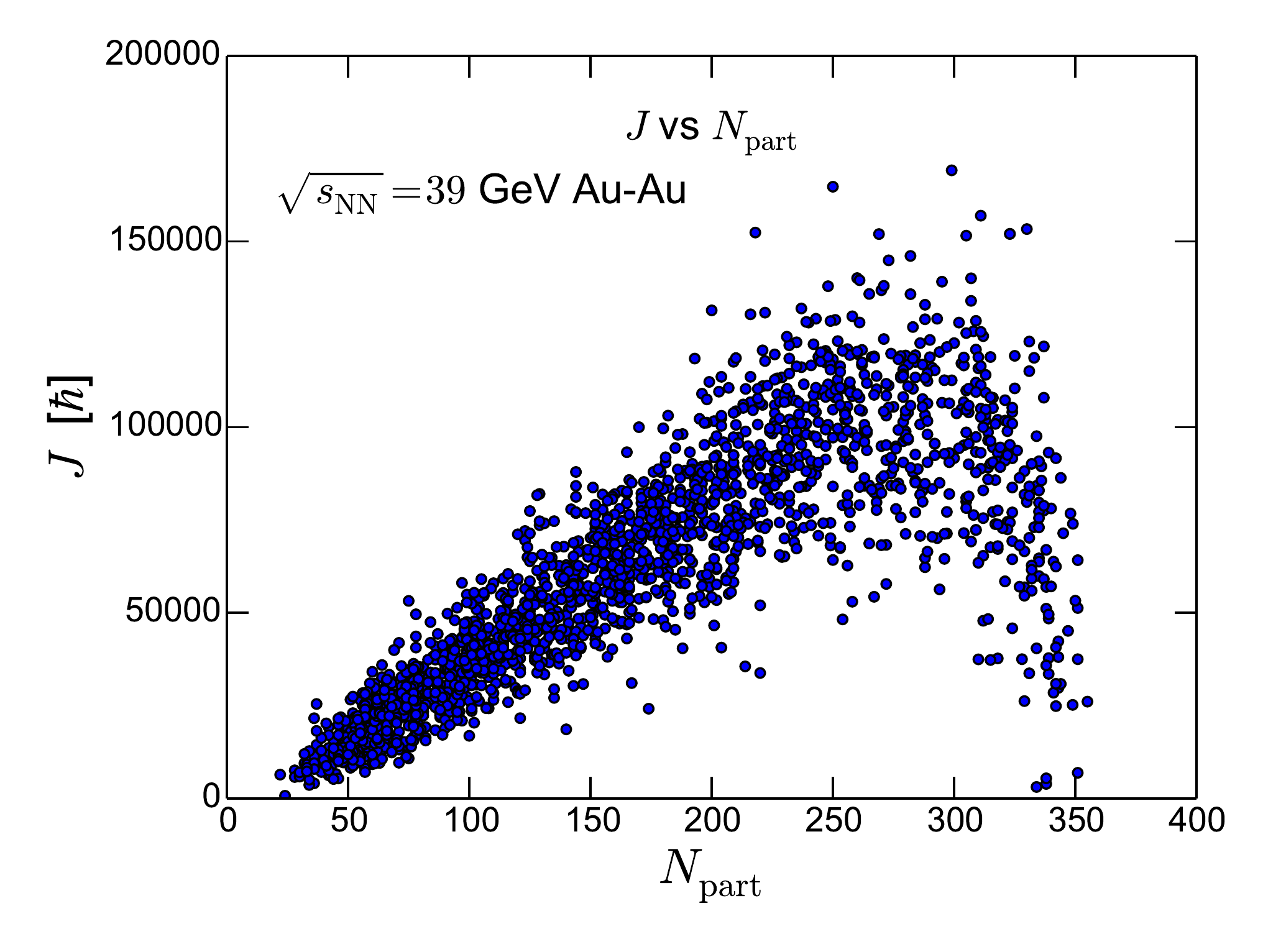}
\caption{Left: global $\Lambda$ polarization at mid-rapidity as a function of the number of participating nucleons $N_{\rm part}$ in the initial state. Each point represents one hydrodynamic configuration in an ensemble of 2000 event-by-event calculations for 0-50\% central Au-Au collisions at $\snn=39$~GeV. Right: out-of-plane component of initial angular momentum versus number of participating nucleons $N_{\rm part}$ in the same calculation.}\label{fig-Jy-npart}
\end{figure}

In Fig.~\ref{fig-Jy-npart} we show the distribution of the global polarization of $\Lambda$ as a function of centrality (i.e.\ $N_{\rm part}$), where each point corresponds to a hydrodynamic evolution with a given fluctuating initial condition characterized by $N_{\rm part}$; in the right panel one can see the corresponding distribution of total angular momentum $\bf J$. We observe that the total angular momentum distribution has a maximum at certain range of $N_{\rm part}$, and drops to zero for the most central events (where the impact parameter is zero) and most peripheral ones (where the system becomes small). In contrast to that, the polarization shows a steadily increasing trend towards peripheral collisions, where it starts to fluctuate largely from event to event because of smallness of the fireball, a situation where the initial state fluctuations start to dominate in the hydrodynamic stage.

\begin{figure}
\begin{center}
\includegraphics[width=0.5\textwidth]{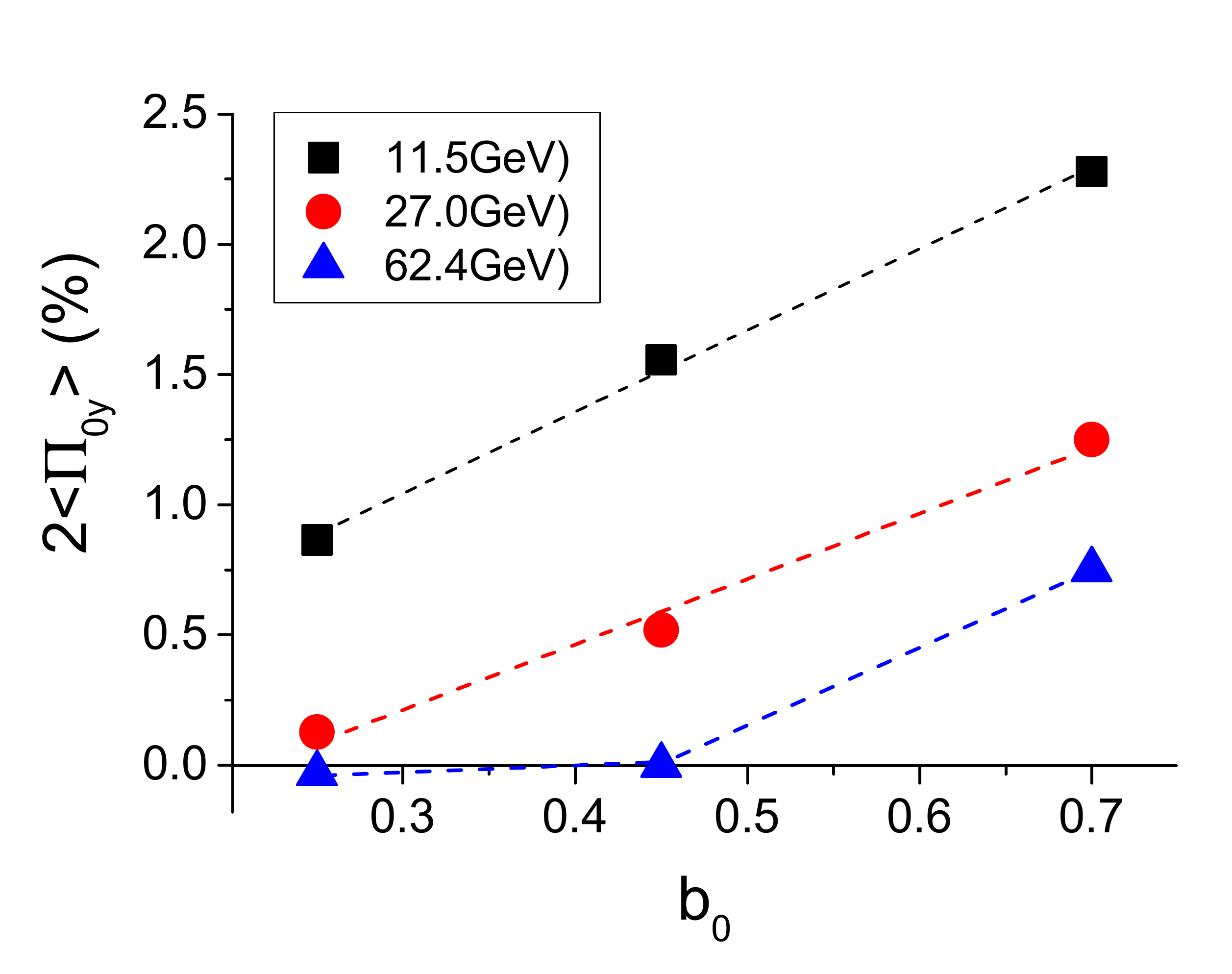}
\end{center}
\caption{The scaled impact parameter $b_0=b/(2R)$ dependence of global polarization in PICR model for 
$\snn=$11.5, 27.0 and 62.4 GeV.}\label{fig-impact-param-PICR}
\end{figure}

The overall trend of the impact parameter (and centrality) dependence of the global polarization is confirmed in the PICR calculation \cite{Xie:2017upb}, see Figure~\ref{fig-impact-param-PICR}. This figure shows the global polarization in Au+Au collisions as a function of ratio of impact parameter $b$ to Au's nuclear radius $R$, i.e. $b_0 = b/2R$. One could see that the polarization at different energies indeed approximately 
takes a linear increase with the increase of impact parameter, except for 62.4GeV due to the vanishing polarization signals at relatively central collisions. This linear dependence clearly indicates that 
the polarization in our model arises from the initial angular momentum. However, the polarization's linear dependence on $b$ is somewhat different from the angular momentum's quadratic dependence on $b$. This is because the angular momentum $L$ is an extensive quantity dependent on the system's mass, while 
the polarization $\Pi$ is an intensive quantity. 

\begin{figure}
\begin{center}
\includegraphics[width=0.5\textwidth]{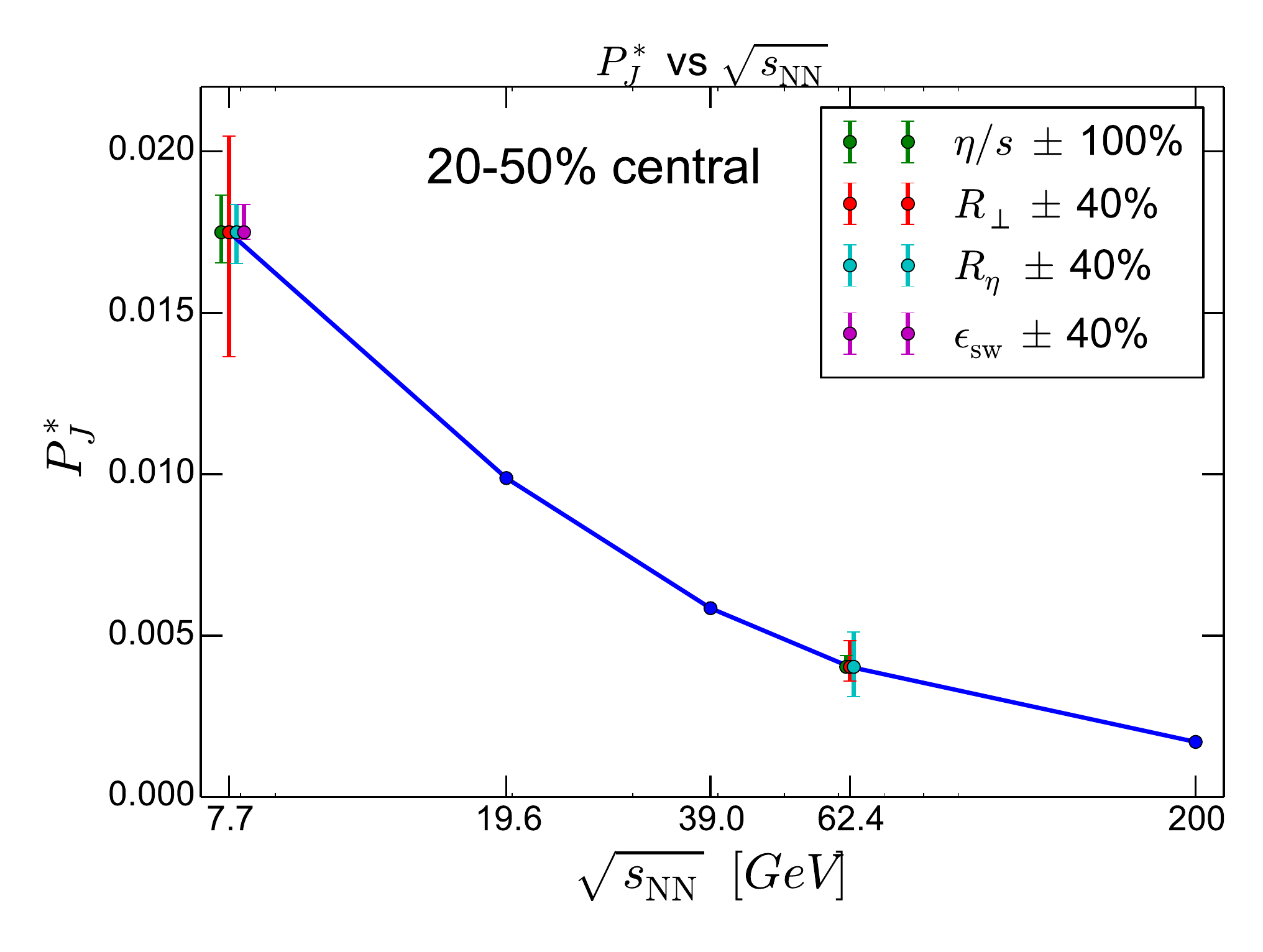}
\end{center}
\caption{Same as Fig.~\ref{fig-Pixy-sqrts} but with bands added, which correspond to variations of the model parameters.}\label{fig-Piy-errorbars}
\end{figure}

\begin{figure}
\begin{center}
\includegraphics[width=0.5\textwidth]{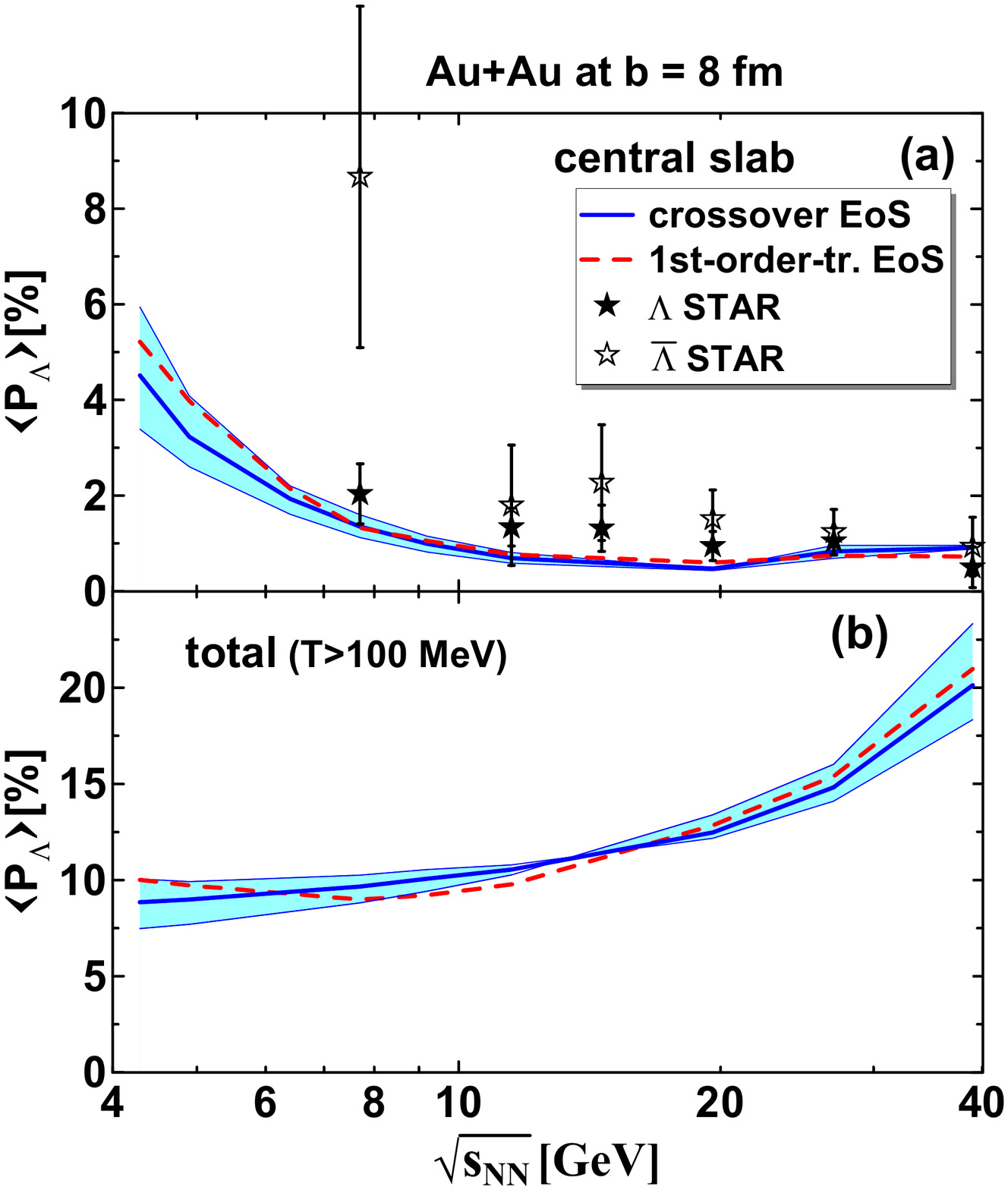}
\end{center}
\caption{Collision energy dependence of the components of polarization vector of $\Lambda$, calculated in its rest frame, in 3-fluid dynamics for 20-50\% central Au-Au collisions.}\label{fig-energy-dep-3fd}
\end{figure}

Finally, the energy dependence of $\Lambda$ polarization in the 3FD calculation is similar to the other two hydrodynamic models, see Fig.~\ref{fig-energy-dep-3fd}, top panel. Contrary to the decrease of the $\Lambda$ polarization at mid-rapidity with collision energy - a trend which will be discussed in the next sub-section - the polarization of all $\Lambda$ actually grows with the energy, as can be seen on the bottom panel of Fig.~\ref{fig-energy-dep-3fd}. The latter is explained in the 3FD calculation by the vorticity being pushed out to the fragmentation regions.

\begin{figure}
\begin{center}
\includegraphics[width=0.7\textwidth]{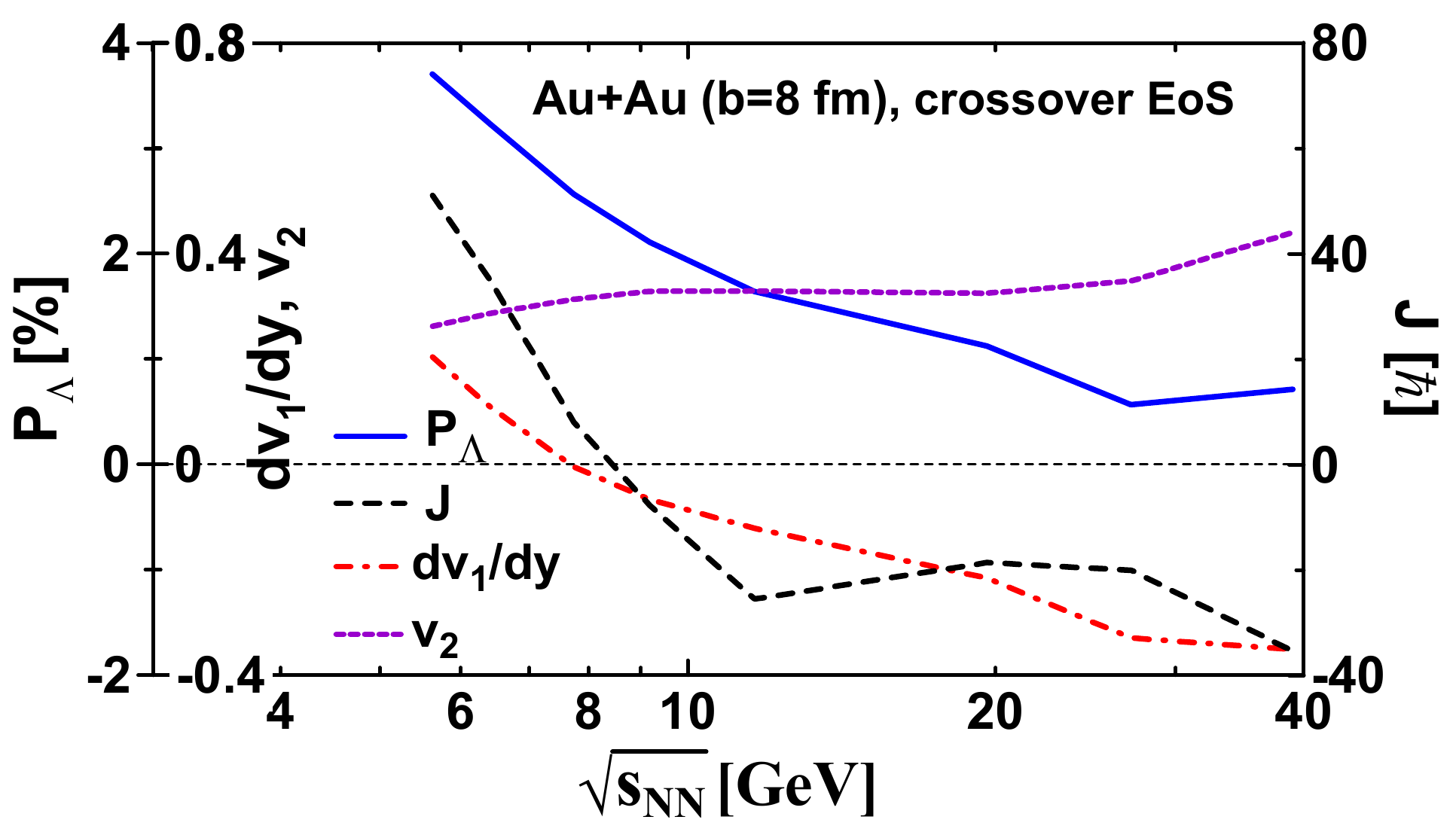}
\end{center}
\caption{$\Lambda$ polarization and angular momentum within space-time rapidity $|\eta|<0.5$ slab, the slope of directed flow $dv_1/dy$ and elliptic flow $v_2$ in 3FD calculations of Au-Au collisions with fixed impact parameter $b=8$~fm. The plot is taken from \cite{Ivanov:2020wak}.}\label{fig-p-j-v1-v2}
\end{figure}

The 3FD calculation further demonstrates that the $\Lambda$ polarization at the mid-spacetime rapidity slab of matter $|\eta|<0.5$ does not correlate with the total angular momentum of the slab, as shown in Fig.~\ref{fig-p-j-v1-v2}. In this calculation, the momentum of the slab changes sign around collision energy $\snn\approx 9$~GeV, whereas the polarization remains positive. Also, the polarization follows correlates neither with the slope of directed flow (which also changes sign around the same energy as the angular momentum), nor with the elliptic flow $v_2$.

{\bf Parameter dependence.} As it has been mentioned above, the parameters of the model are set to monotonically depend on collision energy in order to approach the experimental data for basic hadronic observables. The question may arise whether the collision energy dependence of $P^y$ is the result of an interplay of collision energy dependencies of the parameters. The UrQMD+vHLLE calculation argues that it is not the case: in Fig.~\ref{fig-Piy-errorbars} one can see how the $p_T$ integrated polarization component $P^y$ varies at two selected collision energies, $\snn=7.7$ and $62.4$~GeV, when the granularity of the initial state controlled by $R_\perp$, $R_\eta$ parameters, shear viscosity to entropy ratio of the fluid medium $\eta/s$ and particlization energy density $\epsilon_{\rm sw}$ change. It turns out that a variation of $R_\perp$ within $\pm 40\%$ changes $P^y$ by $\pm 20\%$, and a variation of $R_\eta$ by $\pm 40\%$ changes $P^y$ by $\pm 25\%$ at $\snn=62.4$~GeV only. The variations of the remaining parameters affect $P^y$ much less. We thus conclude that the observed trend in $p_T$ integrated polarization is robust with respect to variations of parameters of the model. 

\subparagraph{Discussion on the energy dependence}

Now we have to understand the excitation function of the $p_T$ integrated $P^y$ which is calculated in the hydrodynamic models. As it has been mentioned, $P^y$ at low momentum (which contributes most to the $p_T$ integrated polarization) has a dominant contribution proportional to $\varpi_{xz}p_0$. It turns out that the pattern and magnitude of $\varpi_{xz}$ over the particlization hypersurface do change with collision energy.

The latter is demonstrated in Fig.~\ref{fig-omegaXZ-tau}, for two selected collision energies. For this purpose, two single hydrodynamic calculations were performed with averaged initial conditions from 100 initial state UrQMD simulations each. At $\snn=62.4$~GeV, because of the baryon transparency effect\footnote{The phenomenon of baryon transparency describes transporting the baryon charge of the colliding nuclei to the forward and backward rapidities. Opposite to that, baryon stopping implies that the baryon charge from the colliding nuclei is {\it stopped} around mi-rapidity.}, the $x,z$ components of beta vector at midrapidity are small and do not have a regular pattern, therefore the distribution of $\varpi_{xz}$ in the hydrodynamic cells close to particlization energy density includes both positive and negative parts, as it is seen on the corresponding plot in the right column. At $\snn=7.7$~GeV, baryon stopping results in a shear flow structure, which leads to same (positive) sign of the $\varpi_{xz}$.

In the right column of Fig.~\ref{fig-omegaXZ-tau}, we plot the corresponding $\varpi_{xz}$ distributions over the particlization hypersurfaces projected on the proper time axis. Generally speaking, hydrodynamic evolution tends to dilute the initial vorticities. One can see that longer hydrodynamic evolution at $\snn=62.4$~GeV in combination with smaller absolute value of average initial vorticity results in factor 4-5 smaller average absolute vorticity at late times for $\snn=62.4$~GeV than for $\snn=7.7$~GeV. This results in corresponding difference in the momentum integrated polarization at these two energies, that is mostly determined by low-$p_T$ $\Lambda$ which are preferentially produced from the Cooper-Frye hypersurface at late times.

\begin{figure*}
\includegraphics[width=0.47\textwidth]{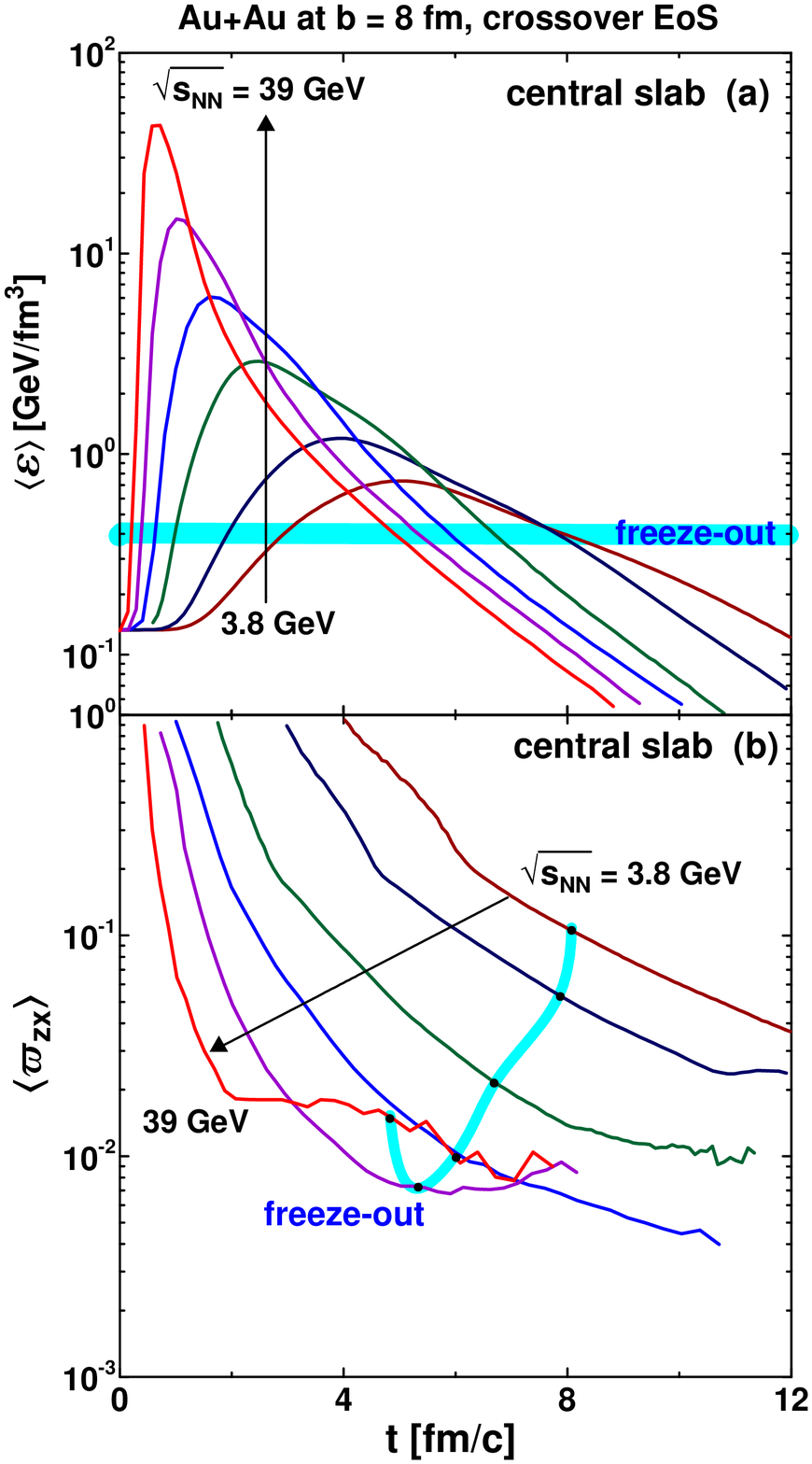}
\includegraphics[width=0.47\textwidth]{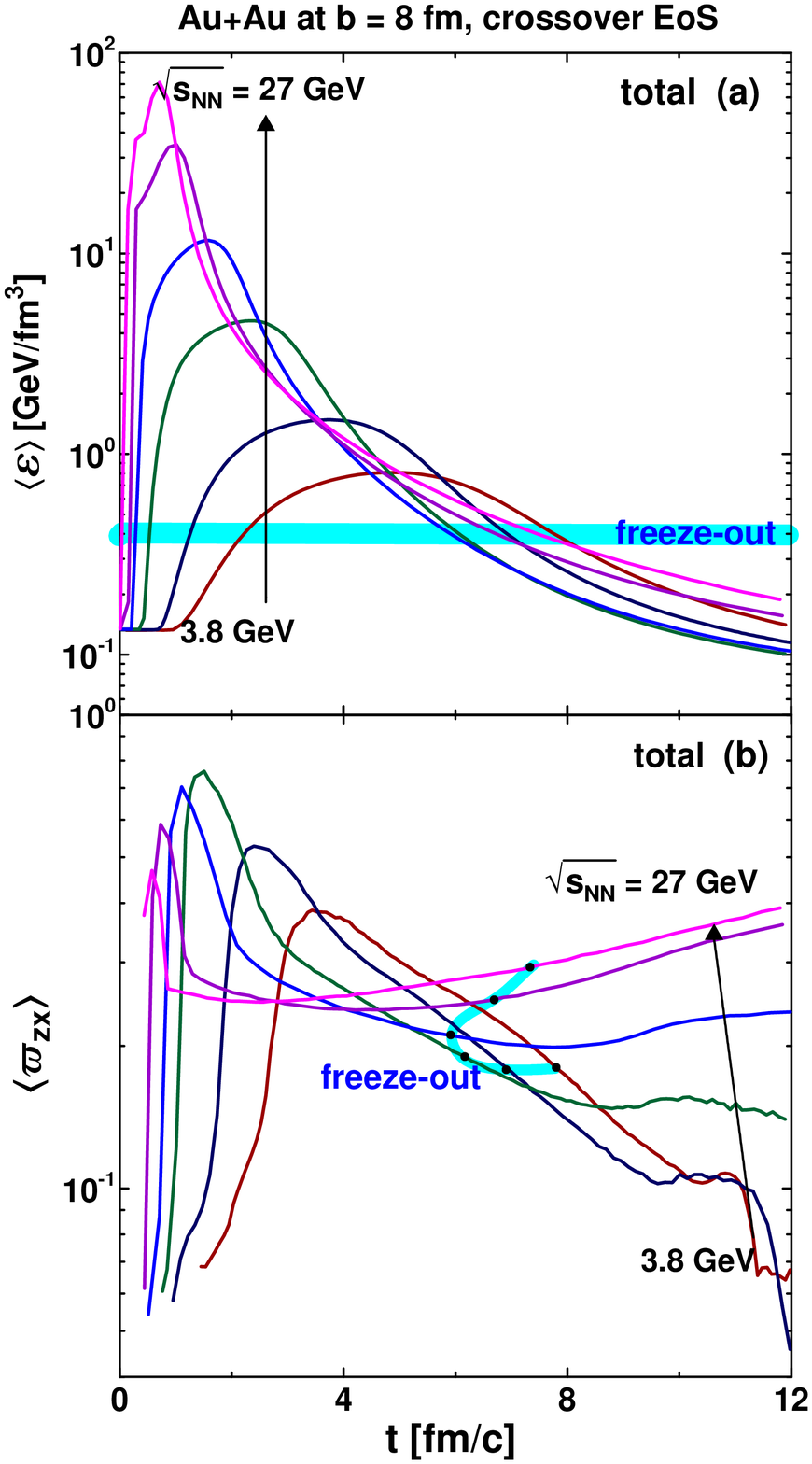}
\caption{Time evolution of the average energy density $\varepsilon$ (top) and average $\varpi_{zx}$ (bottom) in 3FD model \cite{Ivanov:2019ern}, in the central slab of the fireball corresponding to the mid-rapidity (left) and in the whole system (right). The different curves represent the 3-fluid dynamic evolutions for Au-Au collisions at different collision energies from $\snn=3.8$~GeV and up to $\snn=39$~GeV. The calculation is performed a with fixed impact parameter is $b=8$~fm, which approximately corresponds to 30\% centrality.}\label{fig-3fd-time-evol}
\end{figure*}

The explanation of the collision energy dependence of the $\Lambda$ polarization from 3FD is similar to the one above. The dominant effect is that the central-slab vorticity at the freezeout decreases with increasing collision energy because the vortical field is pushed out to the fragmentation regions. The evolution of the vortical field with collision energy and time at the central slab of the system is displayed in Fig.\ref{fig-3fd-time-evol}, bottom left. In addition to that, similarly to the UrQMD+vHLLE calculations, the $\varpi_{xz}$ decreases with time at any given collision energy, and the freeze-out time in 3FD actually decreases with collision energy, in the range $\snn=3.8\dots 27$~GeV.

To summarize: in hydrodynamic models, the effect of hyperon polarization emerges in noncentral collisions, where the angular momentum of the fireball is finite. However, the polarization does not (linearly) scale with the angular momentum of the system. The collision energy denendence in the hydrodynamic calculations is consistent with the experimental measurements by STAR collaboration in the Beam Energy Scan program at RHIC collider.

\section{Hydrodynamic calculations at $\snn=200$ and 2760~GeV}
\begin{figure}
\includegraphics[width=0.5\textwidth]{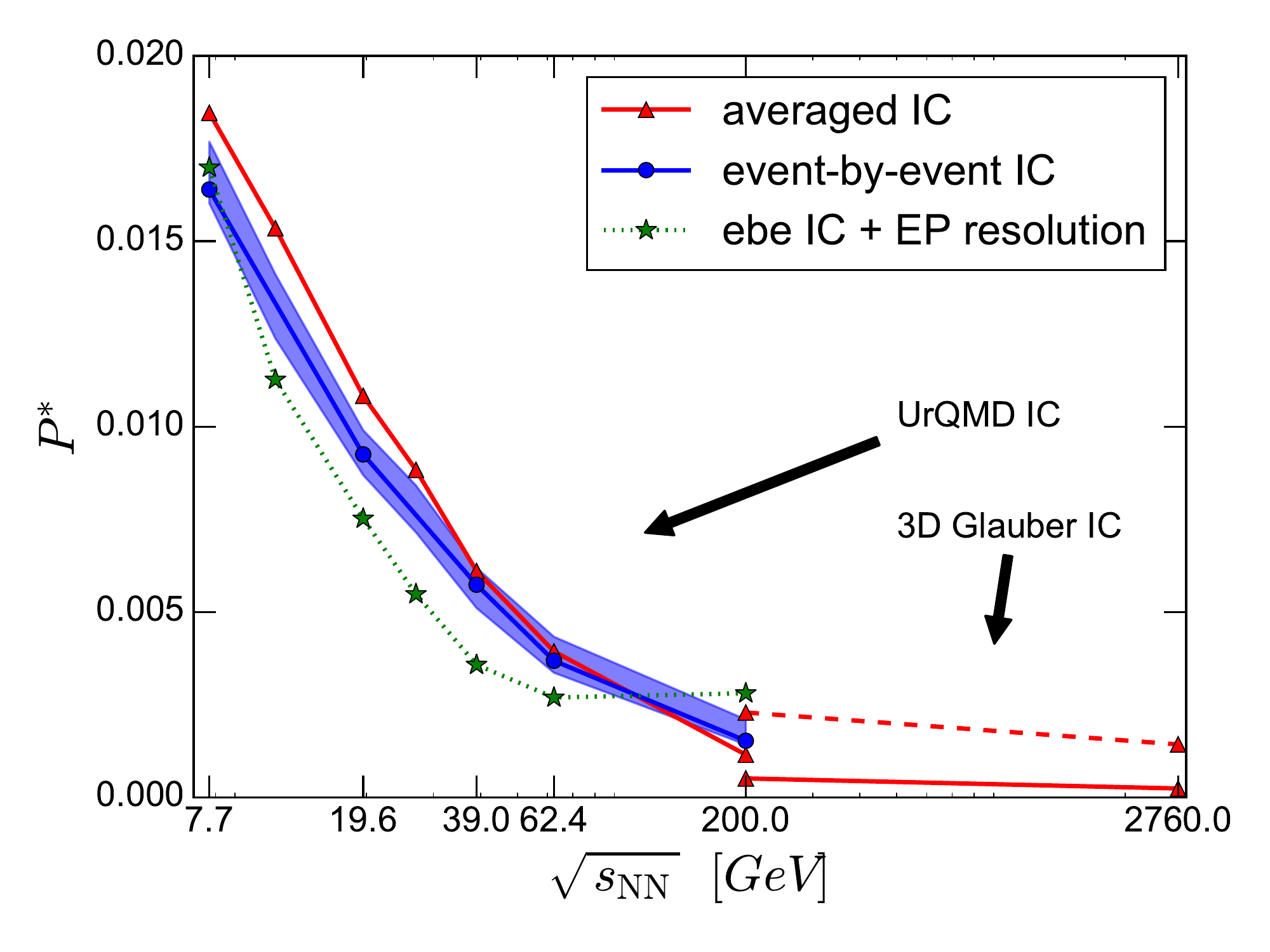}
\includegraphics[width=0.5\textwidth]{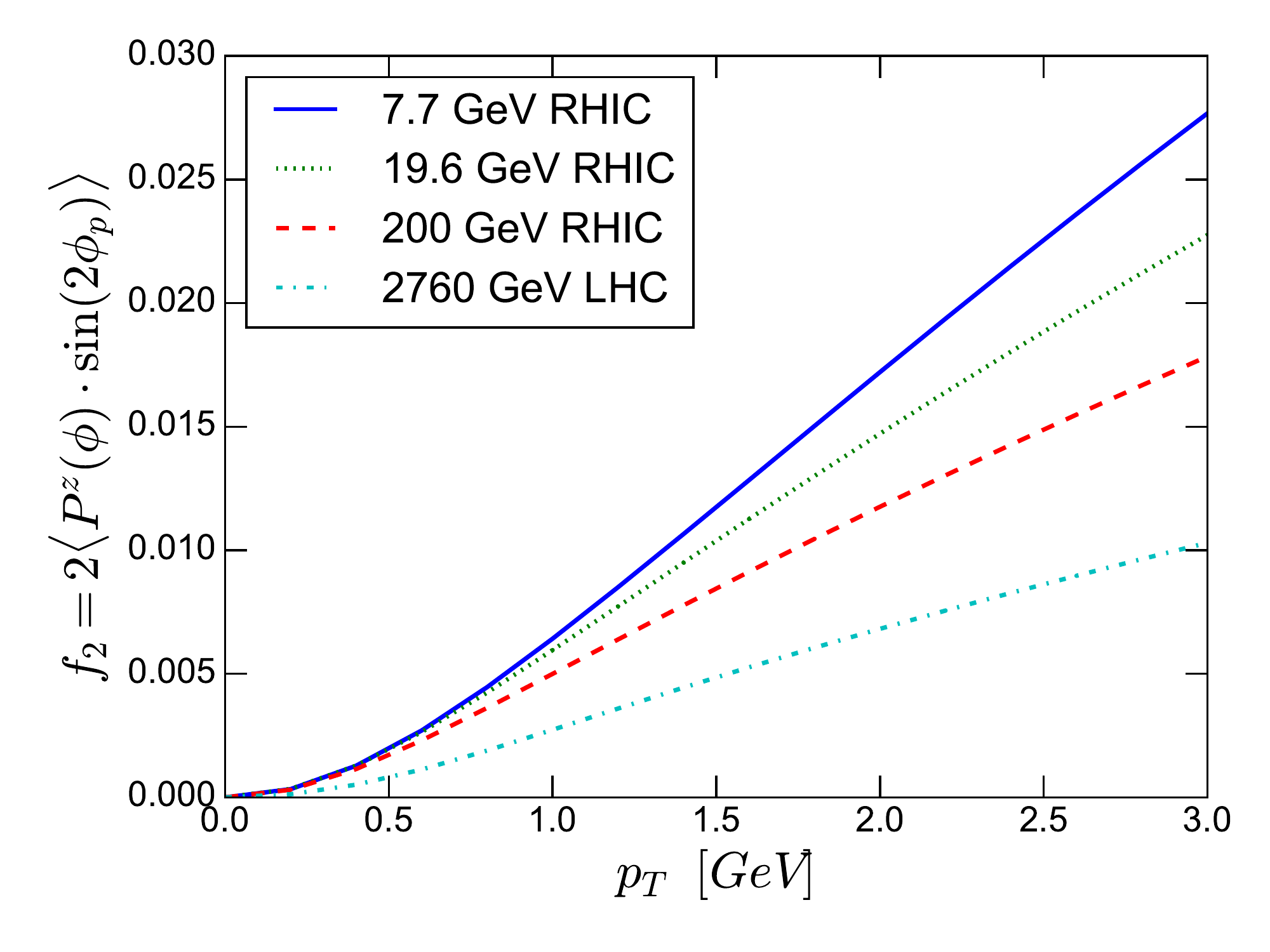}
\caption{Left panel: Global polarization of $\Lambda$ hyperons in 20-50\% central Au-Au (Pb-Pb) collisions at 7.7\dots 200~GeV RHIC (2760 GeV LHC) energies. For the calculations with 3D Glauber IC (initial conditions), the solid one corresponding to longitudinally boost invariant initial flow, and the dashed one corresponding to a small amount of initial shear longitudinal flow as described in \cite{Becattini:2015ska}. The lines connect the points to guide the eye.
Right panel: Second order Fourier harmonic coefficient of polarization component along the beam direction, calculated as a function of $p_T$ for different collision energies; 200 and 2760 GeV points correspond to Monte Carlo Glauber IS.}\label{fig-P2}
\end{figure}

As it was shown in the previous sub-section, the mean, i.e.~momentum averaged, polarization of $\Lambda$ hyperons at mid-rapidity decreases with increasing collision energy. In $\snn=200$~GeV Au-Au collisions at RHIC the mean polarization is less than 0.3\% \cite{Adam:2018ivw}, and at the LHC energies it is presumably smaller, below the accuracy of the experimental measurement. Left panel of Fig.~\ref{fig-P2} demonstrated that, with the same assumptions about the initial state for the hydrodynamic expansion, the spin polarization tends to further decrease between the top RHIC and LHC energies.

However, hydrodynamic results from the previous subsection established not only the global ($\vec{p_T}$-integrated) polarization, but also patterns in local ($\vec{p_T}$-differential) polarization. The magnitude of the longitudinal component $P_\|$ on Fig.~\ref{fig-196gev-4050} was reaching 4\% at high $p_x$ and $p_y$, which is few times larger than the mean polarization, aligned with the total orbital momentum of the fireball.

Indeed, as it has been mentioned in a subsection above, one of the earliest hydrodynamic calculations of $\Lambda$ polarization \cite{Becattini:2015ska} has already demonstrated the existence of local polarization, see Fig.~\ref{fig-pxpypz-echoqgp}.

\begin{figure}
\includegraphics[width=0.98\textwidth]{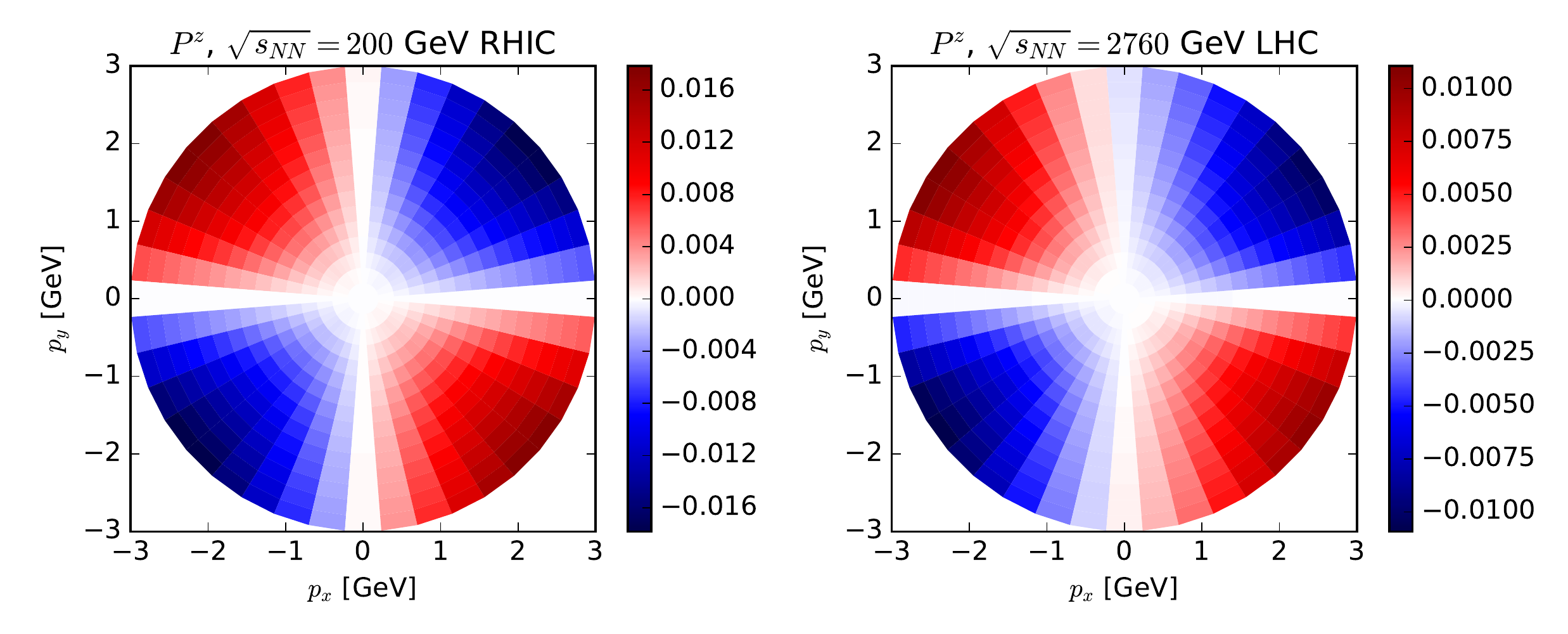}
\caption{Map of longitudinal component of polarization of midrapidity $\Lambda$ from
a hydrodynamic calculation corresponding to 20-50\% central Au-Au collisions at $\snn=200$~GeV 
(left) and 20-50\% central Pb-Pb collisions at $\snn=2760$~GeV (right).}\label{szmap}
\end{figure}

A study \cite{Becattini:2017gcx} took it one step further and computed the local polarization of $\Lambda$ hyperons in a hydrodynamic model using averaged initial state from Monte Carlo Glauber model with its parameters set as in \cite{Bozek:2012fw}. With such initial state, the mid-rapidity slice of the fireball has a small angular momentum. Nevertheless, the quadrupole patterns in the longitudinal polarization persisted at both $\snn=200$~GeV RHIC and $\snn=2760$~GeV LHC energies. The resulting transverse momentum dependence of $P^{*z}$ is shown in Figure~\ref{szmap} for 20-50\% central Au-Au collisions at $\snn = 200$ (RHIC) and 20-50\% Pb-Pb collisions at $\snn = 2760$ GeV (LHC).

Particularly, the rotation-reflection symmetries imply 
that $S^z$ has a Fourier decomposition involving only the sine of even multiples 
of the azimuthal angle $\varphi$:
\be\label{szfour}
 S^z({\bf p}_T,Y=0) = \frac{1}{2} \sum_{k=1}^\infty f_{2k}(p_T) \sin 2 k \varphi
\ee
The corresponding second harmonic coefficients $f_2$ are 
displayed in fig.~\ref{fig-P2} for 4 different collision  energies: 7.7, 19.6~GeV (calculated with initial state from the UrQMD cascade \cite{Karpenko:2015xea}), 200 and 2760~GeV (using averaged initial state from Monte Carlo Glauber 
model). It is worth noting that, whilst the $P^y$ component, along the angular momentum, decreases by about a factor 10 between $\snn=7.7$ and 200 GeV, $f_2$ decreases by only 35\%. We also find that the mean, $p_T$ integrated value of $f_2$ stays around 0.2\% at all collision energies, owing to two compensating effects: decreasing $p_T$ differential $f_2(p_T)$ and increasing mean $p_T$ with increasing collision energy. The $P^y$ component in the UrQMD+vHLLE calculations is produced in the non-central collisions due to anisotropic transverse expansion (elliptic flow) driven by the global geometry of the fireball, whereas in central collisions the initial state fluctuations dominate as shown in \cite{Pang:2016igs}. 

\begin{figure}
\includegraphics[width=0.49\textwidth]{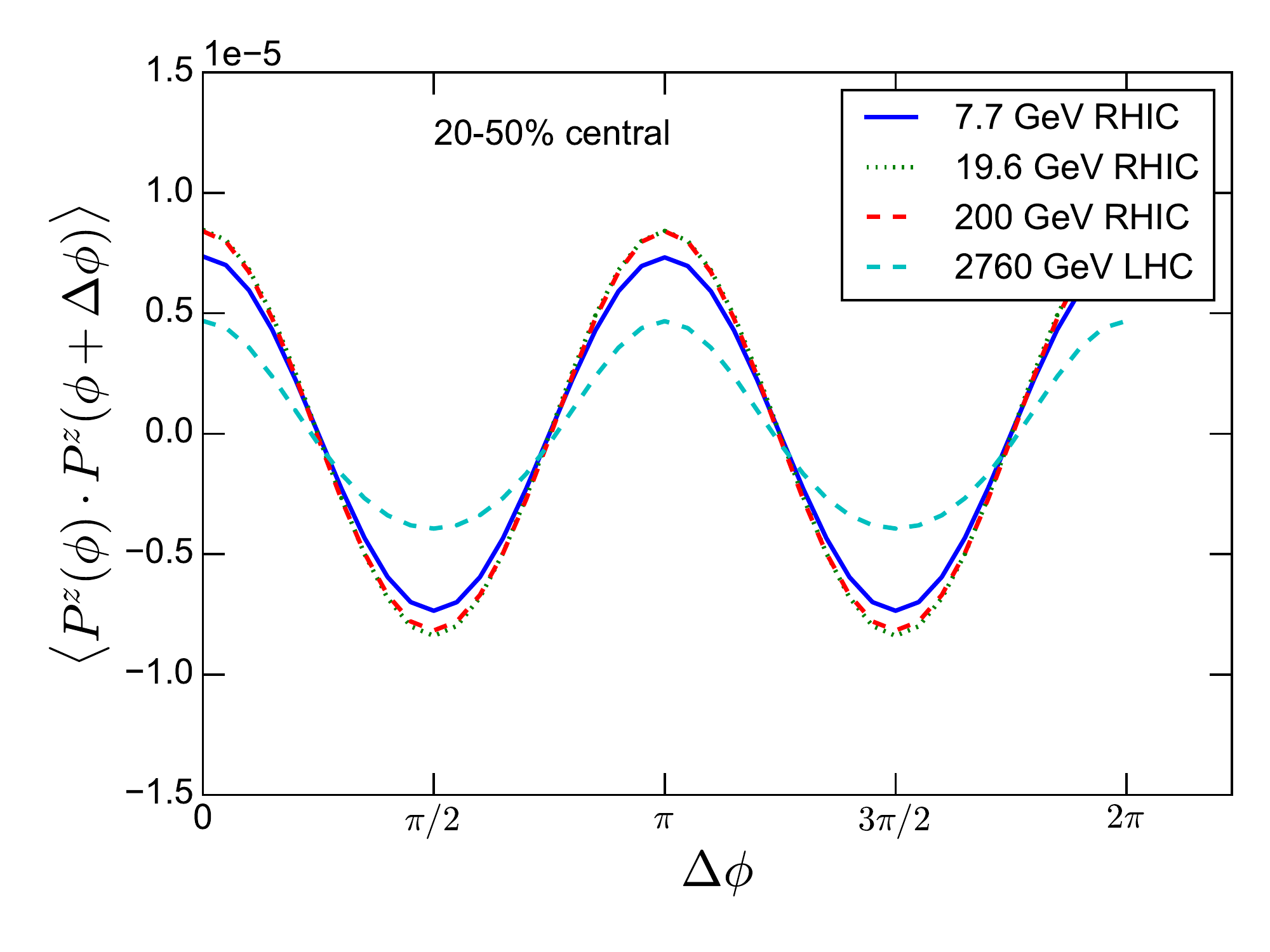}
\includegraphics[width=0.49\textwidth]{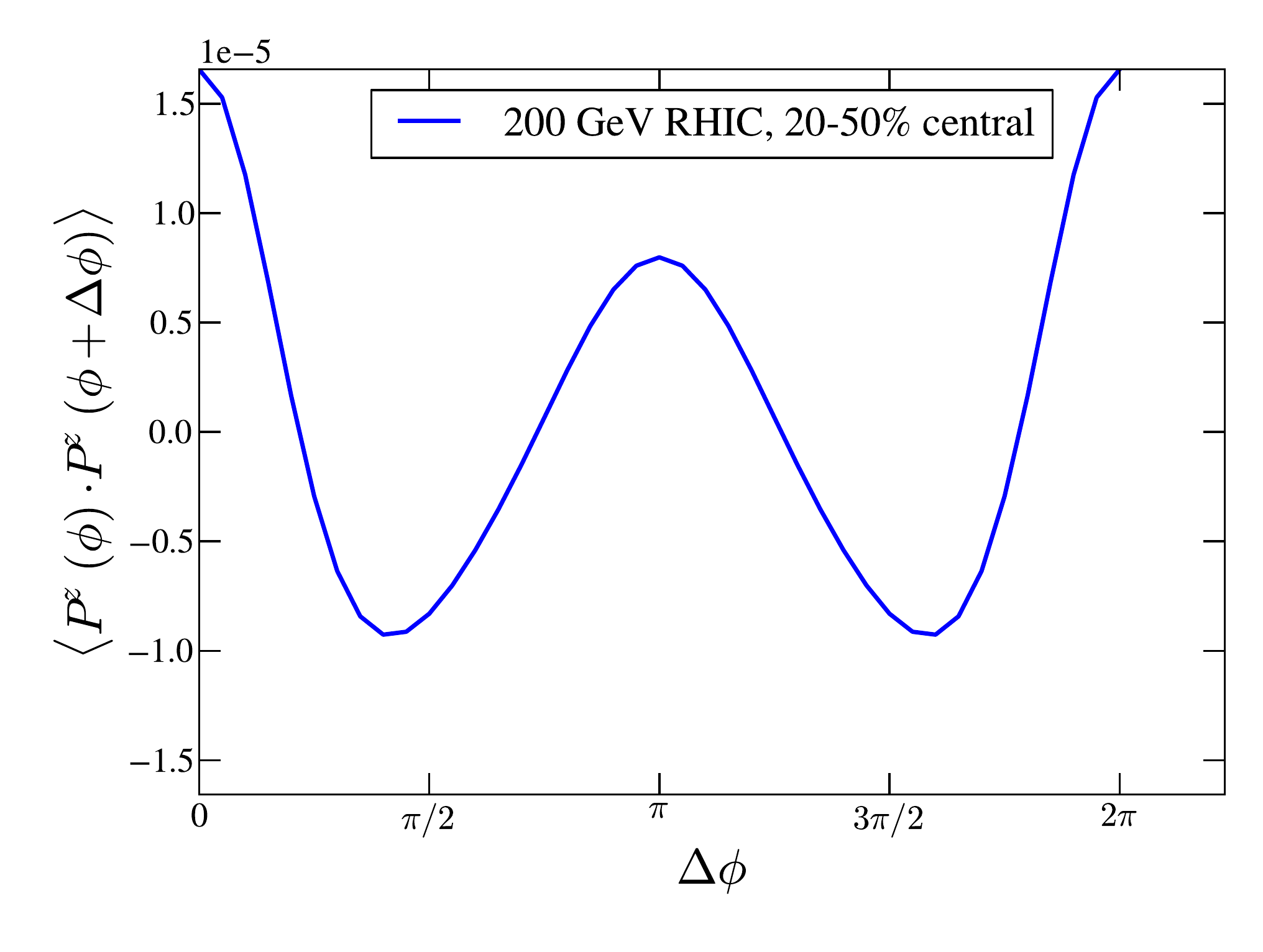}
\caption{Correlation of polarizations of two $\Lambda$ hyperons as a function their opening angle in the transverse plane. Left panel: average initial state, right panel: event-by-event hydrodynamic calculations with a Monte Carlo Glauber initial state.}\label{fig-cosDeltaPhi}
\end{figure}

It is a quick exercise to show that from $P^z=2S^z(\vec{p_T})=f_2(p_T)\sin 2\phi$ it follows that
\be \langle P^z(\phi)P^z(\phi+\Delta\phi)\rangle=\frac{1}{2}f_2^2(p_T)\cos 2\Delta\phi, \ \ee
which means that the correlation function of longitudinal polarization of two $\Lambda$ hyperons, separated by the angle $\Delta\phi$ in the transverse momentum space, behaves as $\cos 2\Delta\phi$. Such behaviour one can see on Fig.~\ref{fig-cosDeltaPhi} left. The right panel of Fig.~\ref{fig-cosDeltaPhi} shows the correlation function from an ensemble of event-by-event hydrodynamic evolutions. In the latter case, the shape of the correlation function deviates from $\cos 2\Delta\phi$ because the underlying azimuthal angle dependence of the $P^z$ after each hydrodynamic evolution in the event-by-event ensemble is randomly fluctuating with respect to the average $\sin 2\phi$ shape.

\begin{figure}
\includegraphics[width=0.52\textwidth]{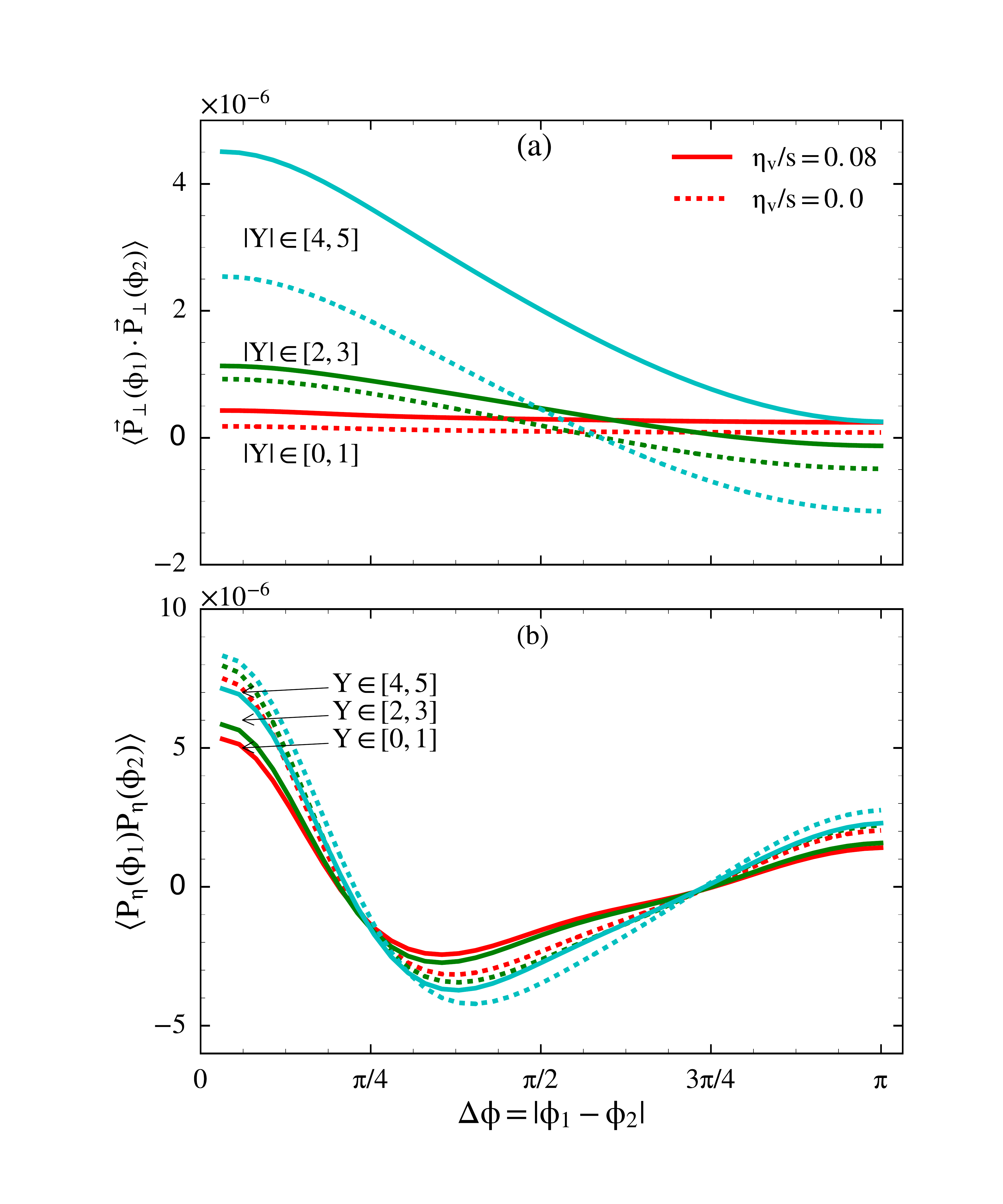}
\includegraphics[width=0.52\textwidth]{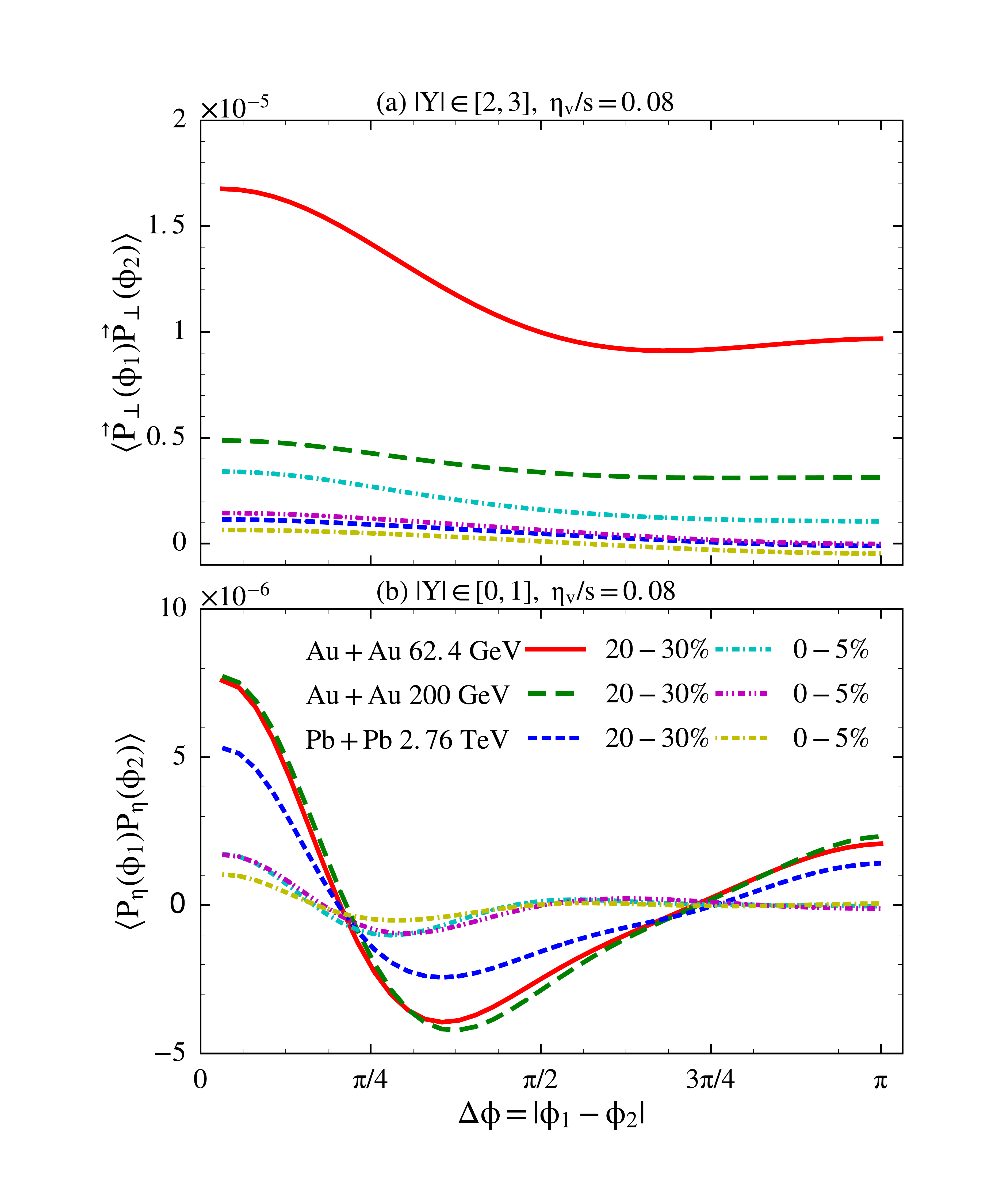}
\caption{Correlation of the transverse (top panel) and longitudinal (bottom panel) components of polarizations of two $\Lambda$ hyperons as a function their opening angle in the transverse plane. Left panel: correlation functions in different rapidity regions from the hydrodynamic calculation for $\snn=2.76$~TeV. Right panel: correlation functions within one rapidity interval but at different collision energies. The plot is taken from \cite{Pang:2016igs}}\label{fig-spincorr-pang}
\end{figure}
In fact, such correlation function of longitudinal, as well as transverse components of $\Lambda$ polarization has been reported in \cite{Pang:2016igs}, see Fig.~\ref{fig-spincorr-pang}.

\subparagraph{Connection between quadrupole longitudinal polarization and elliptic flow}
Indeed, it can be shown that this component does not vanish even in the exact boost 
invariant scenario with no initial state fluctuations and that it decreases slowly 
with increasing center-of-mass energy. 
For the sake of simplicity, let us demonstrate that with an explicit calculation by assuming 
that the fluid is ideal, uncharged and that the {\em initial} transverse velocities 
$u^x,u^y$ vanish. Accumulated evidence in relativistic heavy-ion collisions indicates 
that these are reasonable approximations at very high energy. Under such assumptions, 
it is known that the T-vorticity, 
\be
 \Omega_{\mu\nu} = \partial_\mu (T u_\nu) -\partial_\nu (T u_\mu)
\ee
vanishes at all times \cite{Becattini:2015ska,Deng:2016gyh}, as a consequence of the 
equations of motion. In this case, the thermal vorticity reduces to \cite{Becattini:2015ska}:
\be\label{thvort2}
 \varpi_{\mu\nu} = \frac{1}{T} \left( A_\mu u_\nu - A_\nu u_\mu \right)
\ee
$A$ being the four-acceleration field. This form of the thermal vorticity shows its
entirely relativistic nature, its spatial part being proportional to $({\bf a}\times{\bf v})/c^2$
in the classical units. If we now substitute the eq.~(\ref{thvort2}) in the 
eq.~(\ref{basic}), we get:
\be\label{mainf2}
S^\mu(p) = - \frac{1}{4m} \epsilon^{\mu\rho\sigma\tau} p_\tau \frac{\int_\Sigma \di \Sigma_\lambda 
  p^\lambda A_\rho \beta_\sigma n_F (1-n_F)}{\int_\Sigma \di \Sigma_\lambda p^\lambda n_F} 
\ee
which shows that $S^z(p)$ can get contributions from the vector product of fields
and momenta in the transverse plane, where they are expected to significantly
develop even in the case of longitudinal boost invariance. From this equation on, we use a shortcut $n_F\equiv f(x,p)$. The uncharged perfect fluid 
equations of motion can be written as:
$$
  A_\rho = \frac{1}{T} \nabla_\rho T = \frac{1}{T} \left(\partial_\rho T -
   u^\rho u \cdot \partial T \right)
$$   
If we plug the above acceleration expression in the eq.~(\ref{mainf3}), only
the first term with $\partial_\rho T$ gives a finite contribution as the second 
term vanishes owing to the presence of $\beta_\sigma u_\rho$ factor and the 
Levi-Civita tensor. Furthermore, since:
$$
  \frac{\partial}{\partial p^\sigma} n_F = - \beta_\sigma n_F (1-n_F)
$$
we can rewrite the eq.~(\ref{mainf2}) as:
\be\label{mainf3}
S^\mu(p) = \frac{1}{4mT} \epsilon^{\mu\rho\sigma\tau} p_\tau \frac{\int_\Sigma 
\di \Sigma_\lambda  p^\lambda \frac{\partial n_F}{\partial p^\sigma} \partial_\rho T}
{\int_\Sigma \di \Sigma_\lambda p^\lambda n_F} 
\ee
We can now integrate by parts the numerator in the above equation:
$$
 \int_\Sigma \di \Sigma_\lambda p^\lambda \frac{\partial n_F}{\partial p^\sigma} \partial_\rho T
 = \frac{\partial}{\partial p^\sigma} \int_\Sigma \di \Sigma_\lambda p^\lambda n_F 
 \partial_\rho T - \int_\Sigma \di \Sigma_\sigma n_F \partial_\rho T
 $$
Another very reasonable assumption is that the decoupling hypersurface at high
energy is described by the equation $T=T_{\rm c}$ where $T_{\rm c}$ is the QCD 
pseudo-critical temperature. This entails that the normal vector to the hypersurface 
is the gradient of temperature. 
Then the final expression of the mean spin vector is:
\be\label{mainf4}
S^\mu(p) = \frac{1}{4mT} \epsilon^{\mu\rho\sigma\tau} p_\tau 
 \frac{\frac{\partial}{\partial p^\sigma}\int_\Sigma 
\di \Sigma_\lambda  p^\lambda n_F \partial_\rho T}
{\int_\Sigma \di \Sigma_\lambda p^\lambda n_F} 
\ee
The longitudinal component of the mean spin vector $S^z$ thus depends on the value of 
the temperature gradient on the decoupling hypersurface and its measurement can provide
information thereupon. A simple solution of the above integral appears under 
the assumption of isochronous decoupling hypersurface, with the temperature field
only depending on the Bjorken time $\tau = \sqrt{t^2-z^2}$. In this case the 
parameters describing the hypersurface are $x,y,\eta$ with $\tau=const.$ and 
the only contribution to the numerator of the (\ref{mainf4}) arises from $\rho=0$:
$$
  \int \di \Sigma_\lambda  p^\lambda n_F \frac{\di T}{\di \tau} \cosh \eta
$$
At $Y=0$, the factor $\cosh\eta$ can be approximated with $1$ because of the 
exponential fall-off $\exp[-(m_T/T) \cosh \eta]$ involved in $n_F$, therefore:
\begin{eqnarray*}
S^z({\bf p}_T,Y=0) \hat{\bf k} &\simeq& - \frac{\di T/\di \tau}{4mT} \hat{\bf k} 
\frac{\partial}{\partial \varphi} \log \int_\Sigma \di \Sigma_\lambda p^\lambda n_F  
\end{eqnarray*}
where $\varphi$ is the transverse momentum azimuthal angle, counting from the reaction plane. In the above equation the longitudinal spin component
is a function of the spectrum alone at $Y=0$. By expanding it in Fourier series
in $\varphi$ and retaining only the elliptic flow term, one obtains:
\bea\label{sz1}
S^z({\bf p}_T,Y=0) &\simeq &  - 
\frac{\di T/\di \tau}{4mT} \frac{\partial}{\partial \varphi} 2 v_2(p_T)
\cos 2\varphi \nonumber \\
&=& \frac{\di T}{\di \tau}\frac{1}{mT} v_2(p_T) \sin 2\varphi
\eea
meaning, comparing this result to eq.~(\ref{szfour}) that in this case:
$$
   f_2(p_T) = 2 \frac{\di T}{\di \tau}\frac{1}{mT} v_2(p_T)
$$
This simple formula only applies under special assumptions with regard to the hydrodynamic 
temperature evolution, but it clearly shows the salient features of the longitudinal 
polarization at mid-rapidity as a function of transverse momentum and how it can provide 
direct information on the temperature gradient at hadronization. It also shows,
as has been mentioned - that it is driven by physical quantities related to transverse 
expansion and that it is independent of longitudinal expansion.

\begin{figure}
\begin{center}
\includegraphics[width=0.6\textwidth]{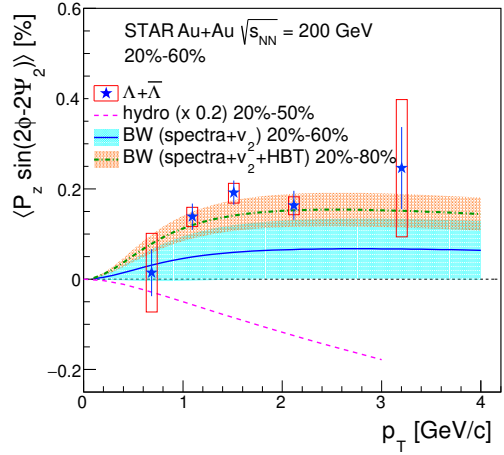}
\caption{The second-order Fourier sine coefficient of the longitudinal $\Lambda$ and $\bar\Lambda$ polarizations as a function of $p_T$, measure by STAR for 20-60\% Au-Au collisions at $\snn=200$~GeV. The curves correspond to a Blast-Wave model calculation (unpublished) and the hydrodynamic calculation from \cite{Becattini:2017gcx}. The plot is taken from \cite{Adam:2019srw}.}\label{fig-Pz-STAR}
\end{center}
\end{figure}

In 2019, STAR collaboration has published a measurement of the $p_T$ and azimuthal angle dependence of the longitudinal polarization of $\Lambda$ and $\bar\Lambda$ hyperons in Au-Au collisions at $\snn=200$~GeV. The same quadrupole structure in the longitudinal polarization has been observed, however its sign is the opposite to the hydrodynamic calculation in \cite{Becattini:2017gcx}.

The same pattern but the opposite sign of the longitudinal $\Lambda$ polarization, consistent with the measurement by STAR \cite{Adam:2019srw} has been reported in a hydrodynamic calculation in PICR model \cite{Xie:2019jun}. Since the origin of the polatization signal, as well as the basic ingredients of PICR model appears to be the similar to the other hydrodynamic calculations, it is not clear why the PICR calculation results in a different sign of the longitudinal polarization.

\begin{figure}
\begin{center}
\includegraphics[width=0.6\textwidth]{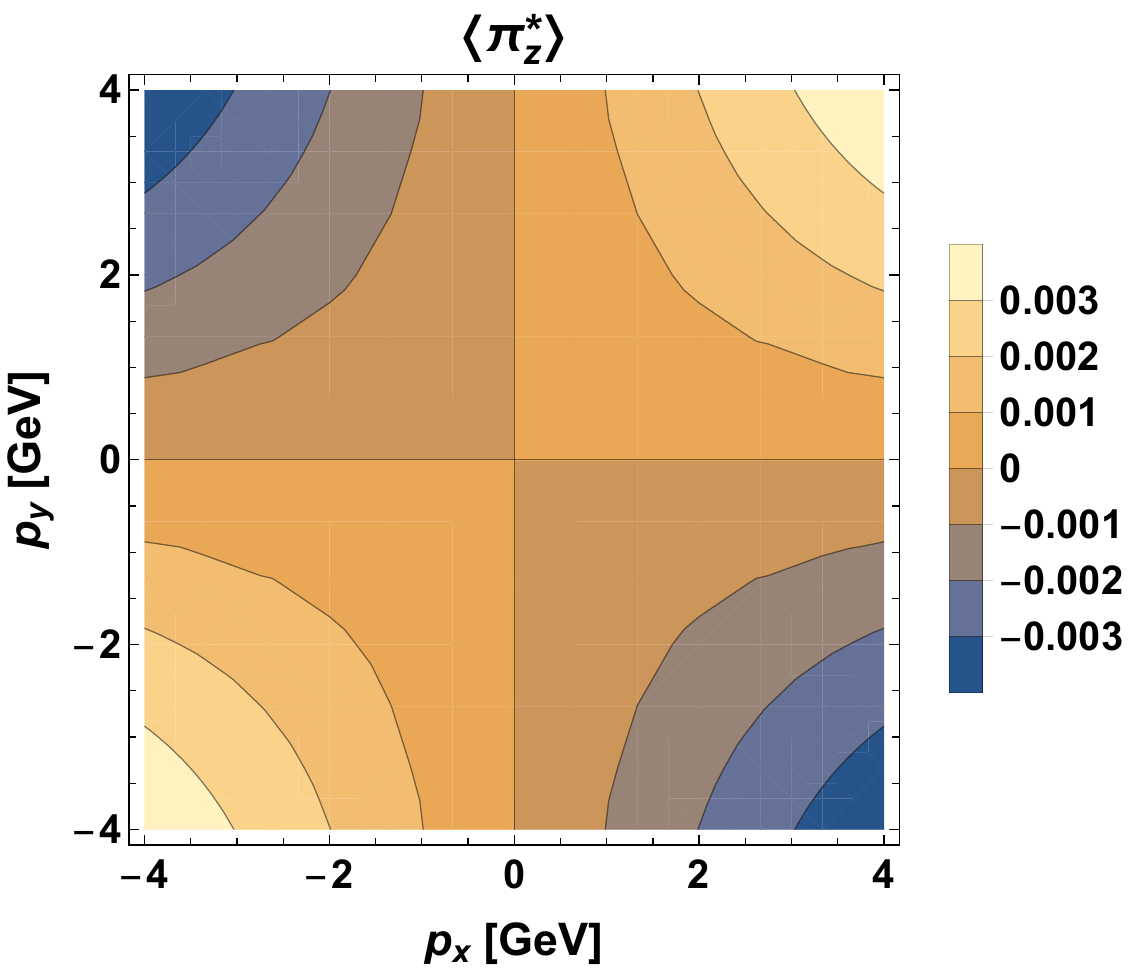}
\end{center}
\caption{Longitudinal component of $\Lambda$ polarization, computed with projected thermal vorticity in Eq.~\ref{eq:Pixp} (for details see text). Note that the sign is compatible with the STAR result in this case. The plot is taken from \cite{Florkowski:2019voj}.}\label{fig-Pz-omega-proj}
\end{figure}

The discrepancy between the hydrodynamic calculations and the experimental measurement by STAR \cite{Adam:2019srw} remains an open question. Few ideas have been proposed to address the question. I.e.\ in \cite{Florkowski:2019voj} it was found that when the thermal vorticity the formula for the spin polarization \ref{eq:Pixp} is replaced by a projected thermal vorticity\footnote{The projection is made on a plane orthogonal to the direction of the collective flow velocity: $\varpi^{\mu\nu}_\text{proj} u_\nu=0$.}: $\varpi^{\mu\nu}_\text{proj}=\varpi_{\alpha\beta}\Delta^\mu_\alpha \Delta^\nu_\beta$, the resulting longitudinal component of polarization flips sign and becomes compatible with the experimental measurement.

\begin{figure}
\begin{center}
\includegraphics[width=0.7\textwidth]{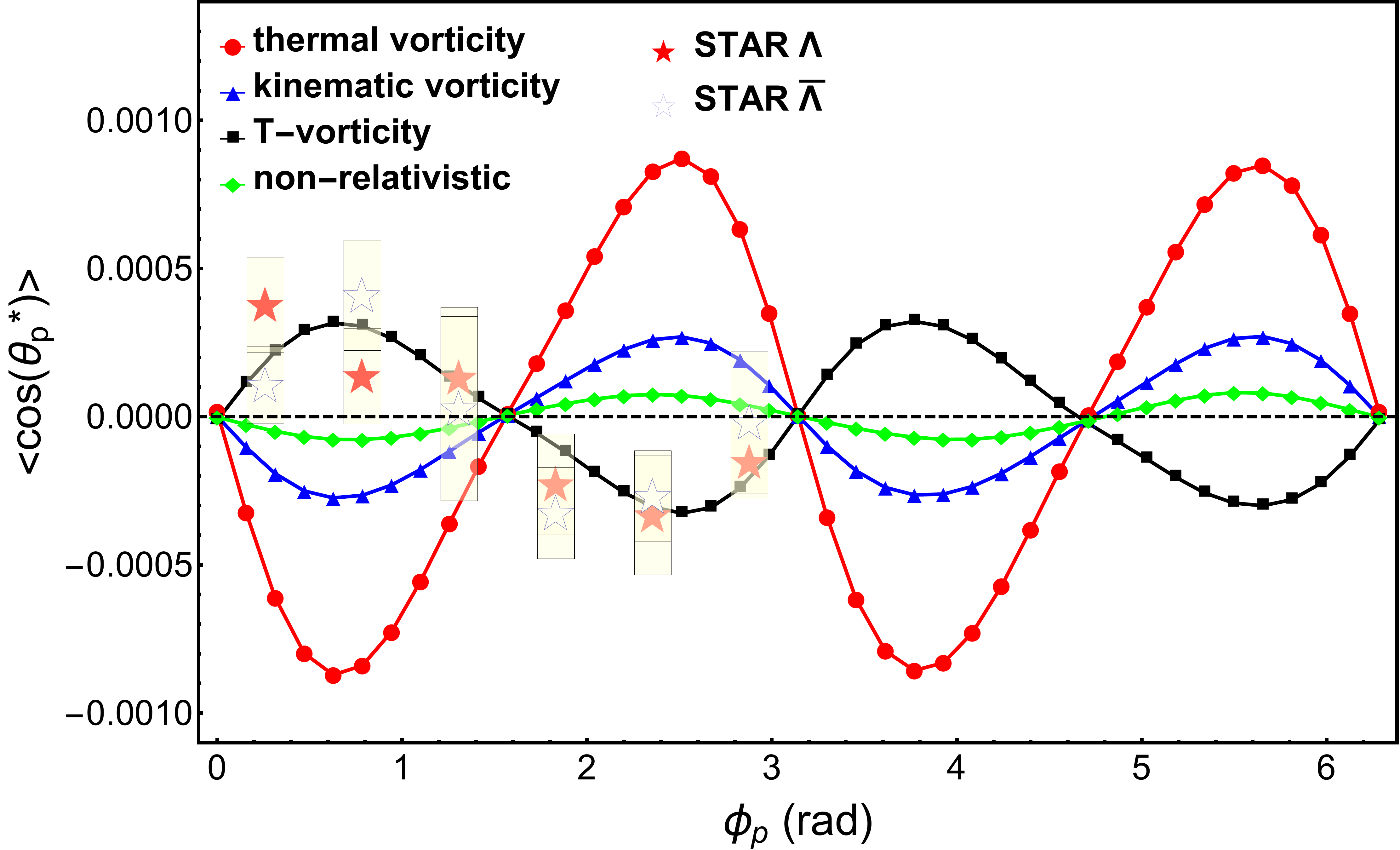}
\end{center}
\caption{Longitudinal component of $\Lambda$ polarization, computed with different definitions of relativistic vorticity in Eq.~\ref{eq:Pixp}. The plot is taken from \cite{Wu:2019eyi}.}\label{fig-Pz-omega-proj}
\end{figure}

Another recent study \cite{Wu:2019eyi} suggests that swapping the thermal vorticity with the T-vorticity \ref{tvort} in the same formula for the spin polarization \ref{eq:Pixp} also results in the sign flip. However, as for now it is not clear whether these suggestions will help solving the ``sign problem'' in the longitudinal polarization component between the hydrodynamic calculations and the experiment, since the basic derivation of the effect leads to the spin polarization to be proportional precisely to thermal vorticity and not to the projected thermal vorticity or the T-vorticity.

\section{Acceleration, grad T and vorticity contributions to polarization}
To gain insight into the physics of polarization in a relativistic fluid, it is 
very useful to decompose the gradients of the four-temperature vector in the 
eq.~(\ref{basic}). We start off with the seperation of the gradients of the comoving 
temperature and four-velocity field:
$$
 \partial_\mu \beta_\nu = \partial_\mu \left(\frac{1}{T}\right) + \frac{1}{T} 
 \partial_\mu u_\nu 
$$
Then, we can introduce the acceleration and the vorticity vector $\omega^\mu$ with
the usual definitions:
\begin{eqnarray*}
A^\mu &= &u \cdot \partial u^\mu  \\
\omega^\mu &=& \frac{1}{2} \epsilon^{\mu\nu\rho\sigma} \partial_\nu u_\rho u_\sigma
\end{eqnarray*}
The antisymmetric part of the tensor $\partial_\mu u_\nu$ can then be expressed as
a function of $A$ and $\omega$:
$$
  \frac{1}{2}\left( \partial_\nu u_\mu - \partial_\mu u_\nu \right) = \frac{1}{2} 
   \left( A_\mu u_\nu - A_\nu u_\mu \right) + \epsilon_{\mu\nu\rho\sigma} \omega^\rho 
  u^\sigma
$$
therafter plugged into the (\ref{basic}) to give:
\begin{eqnarray}\label{spindeco}
S^\mu(x,p)  &=& \frac{1}{8m} (1-n_F)  \epsilon^{\mu\nu\rho\sigma} p_\sigma
   \nabla_\nu (1/T) u_\rho \\
&+&  \frac{1}{8m} (1-n_F) \; 2 \, \frac{\omega^\mu u \cdot p - u^\mu \omega \cdot p}{T} \\
&-& \frac{1}{8m}(1-n_F)\frac{1}{T} \epsilon^{\mu\nu\rho\sigma} p_\sigma  A_\nu u_\rho
\end{eqnarray}
Hence, polarization stems from three contributions: a term proportional to the 
gradient of temperature, a term proportional to the vorticity $\omega$, and a term
proportional to the acceleration. Further insight into the nature of these terms
can be gained by choosing the particle rest frame, where $p=(m,{\bf 0})$ and restoring
the natural units. The eq.~(\ref{spindeco}) then certifies that the spin in the 
rest frame is proportional to the following combination:
\be\label{restframe}
{\bf S}^*(x,p) \propto \frac{\hbar}{KT^2} \gamma {\bf v} \times \nabla T + 
\frac{\hbar}{KT} \gamma (\omegav - (\omegav \cdot {\bf v}) {\bf v}/c^2) 
+ \frac{\hbar}{KT} \gamma {\bf A} \times {\bf v}/c^2
\ee
where $\gamma = 1/\sqrt{1-v^2/c^2}$ and all three-vectors, including vorticity,
acceleration and velocity, are observed in the particle rest frame. 

The three independent contributions are now well discernible in eq.~(\ref{restframe}). 
The second term scales like $\hbar \omega/KT$ and is the one already known from 
non-relativistic physics, proportional to the vorticity vector seen by the particle 
in its motion amid the fluid, with an additional term vanishing in the non-relativistic 
limit. The third term is a purely relativistic one and scales like $\hbar A/KTc^2$;
it is usually overwhelmingly suppressed, except in heavy ion collisions where the 
acceleration of the plasma is huge ($A \sim 10^{30} g$ at the outset of hydrodynamical
stage). The first term, instead, is a new non-relativistic term \cite{becaspin} and 
applies to situations where the velocity field is not parallel to the temperature 
gradient. For ideal uncharged (thus relativistic) fluids, this term is related 
to the acceleration term because the equations of motion reduce to:
\be
 \nabla_\mu T = T A_\mu/c^2 \label{eq-accel-gradT}
\ee
Therefore in the case of ideal uncharged fluid - which QGP is at a very high energy - the grad T and acceleration contributions will be exactly equal to each other.

\begin{figure}
\includegraphics[width=0.5\textwidth]{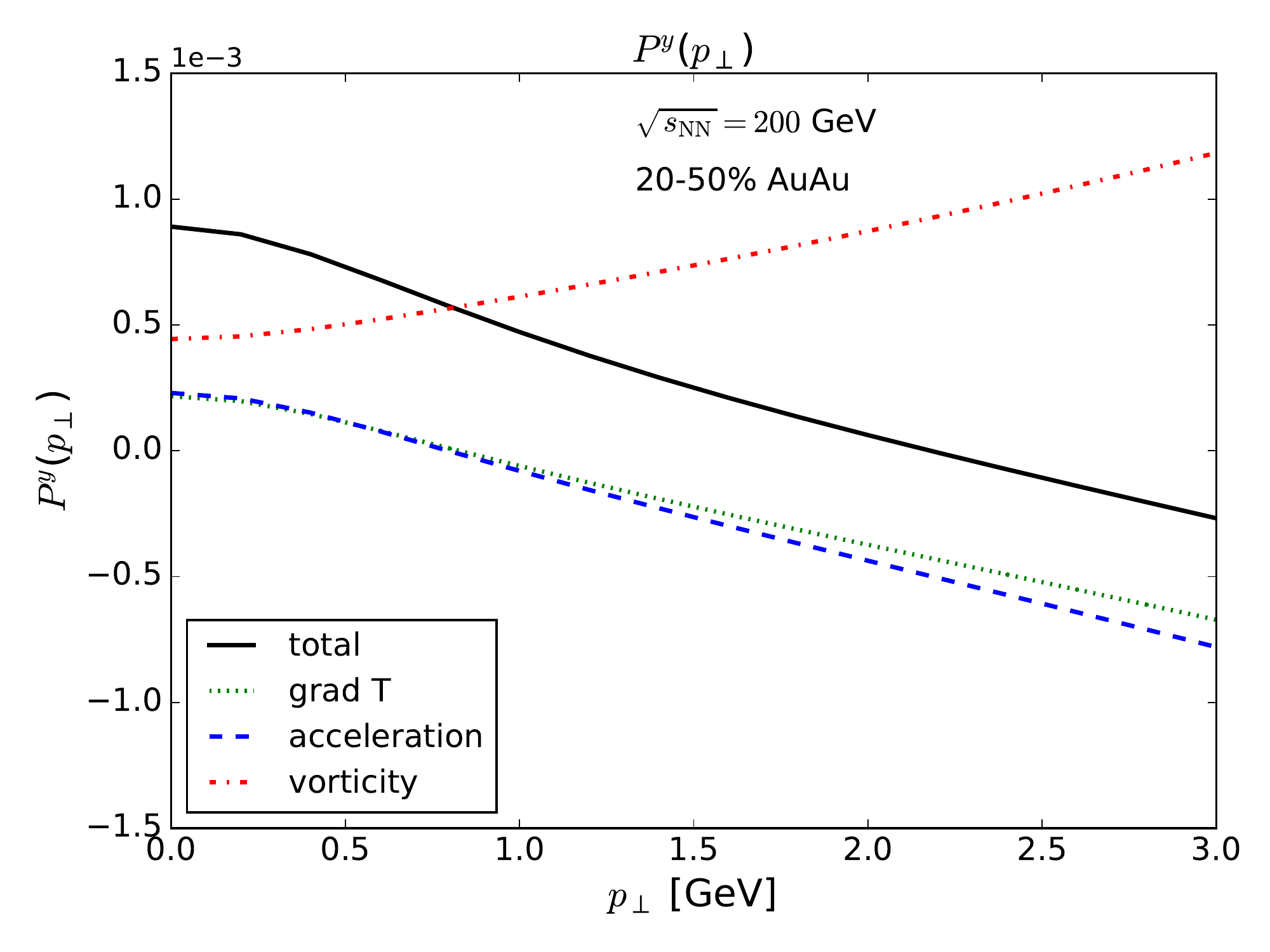}
\includegraphics[width=0.5\textwidth]{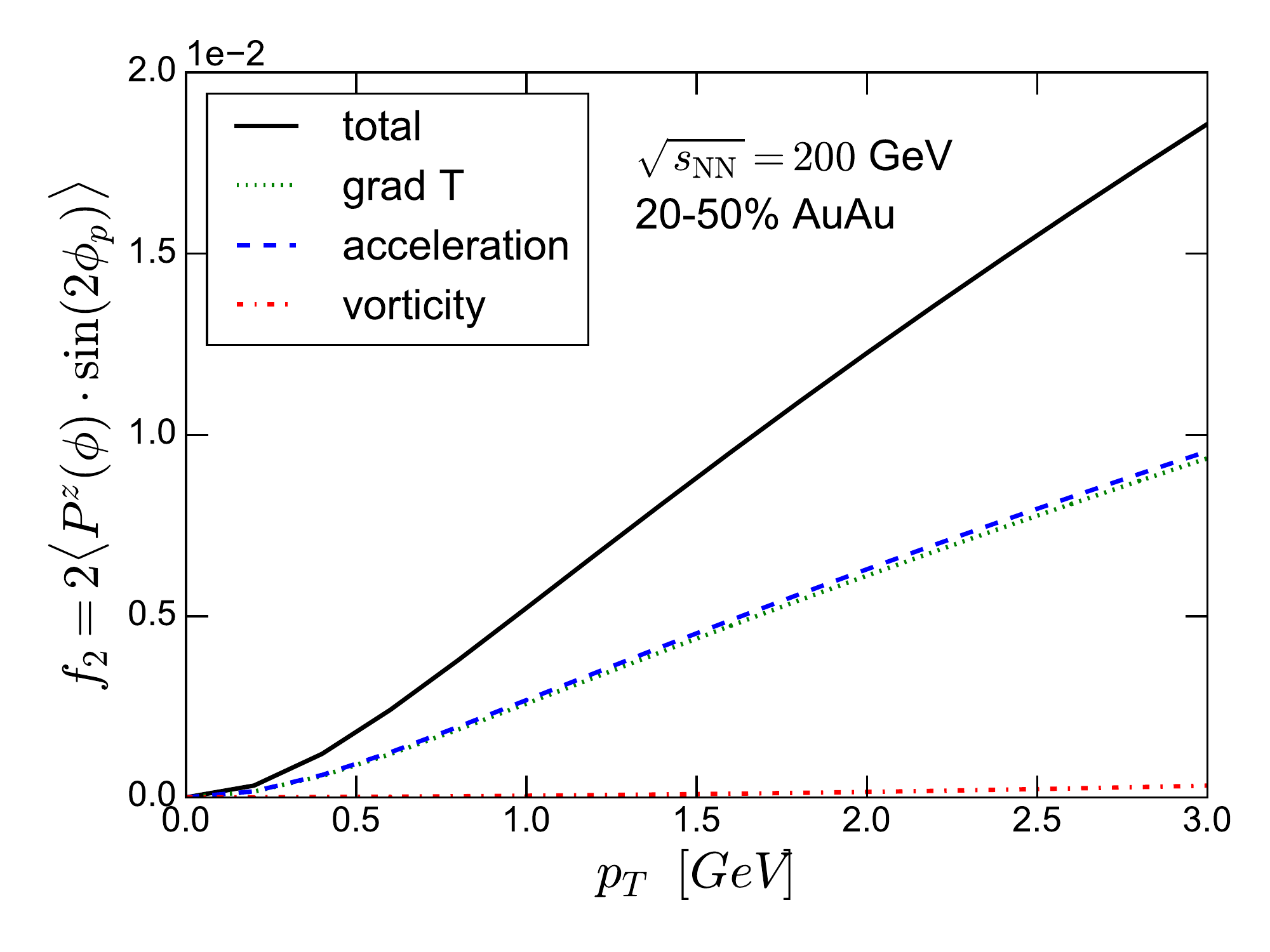}
\caption{Contributions to the global (left panel) and quadrupole longitudinal (right panel) components of $\Lambda$ polarization stemming from gradients of temperature (dotted lines), acceleration (dashed lines) and vorticity (dash-dotted lines). Solid lines show the sums of all 3 contributions. The hydrodynamic calculation with vHLLE is performed with an averaged Monte Carlo Glauber IS corresponding to 20-50\% central Au-Au collisions at $\snn=200$~GeV RHIC energy.}\label{fig-contrib}
\end{figure}
Let's turn to the results from a realistic hydrodynamic calculation \cite{Karpenko:2018erl}. On Fig.~\ref{fig-contrib} we plot the contributions to the global and quadrupole longitudinal polarization components from gradients of temperature, acceleration and vorticity individually, as well as their sum. One can see that the resulting $p_T$-integrated global polarization of $\Lambda$, which is dominated by its low-$p_T$ contributions, has the largest contribution from the classical vorticity term. At the same time, $f_2$ has a negligible contribution from the vorticity term and virtually equal contributions from the grad T and acceleration terms. The latter result is expectable, as in hydrodynamics of ideal uncharged fluid the temperature gradient and acceleration fields are related as follows:
\begin{equation}
 A_\mu = \frac{1}{T}\Delta_{\mu\nu}\partial^\nu T
\end{equation}
Thus the small difference between the grad T and acceleration contributions seen on Fig.~\ref{fig-contrib} shows that, even though the shear viscosity over entropy ratio in the calculations changes between $\eta/s=0.08\dots 0.2$, the resulting hydrodynamic evolution is quantitatively not very different from ideal one.


\newpage{\pagestyle{empty}\cleardoublepage}
\end{document}